\documentclass[aps,prb,amsmath, amssymb, english, twocolumn,reprint, floatfix, superscriptaddress, longbibliography]{revtex4-2}

\usepackage{graphicx} 
\usepackage{hyperref}
\usepackage{amsmath}
\usepackage{todonotes}
\usepackage{subfigure}

\newcommand{\bq}{{\mathbf{q}}}

\newcommand{\bk}{\mathbf{k}}

\newcommand{\bl}{{\boldsymbol\lambda}}

\newcommand{\lmagn}{{\lambda_\mathrm{m}}}
\newcommand{\ldens}{{\lambda_\mathrm{d}}}

\newcommand{\ldga}{\ensuremath{\mathrm{lD}\Gamma\mathrm{A}\,}}
\newcommand{\ldgam}{\ensuremath{\mathrm{lD}\Gamma\mathrm{A}_\text{m}\,}}
\newcommand{\ldgadm}{\ensuremath{\mathrm{lD}\Gamma\mathrm{A}_\text{dm}\,}}

\begin{document}
\title{Suppression of the charge fluctuations by nonlocal correlations close to the Mott transition - Insights from the ladder dynamical vertex approximation}
\author{Irakli Titvinidze}
\author{Julian Stobbe}
\author{Marvin Leusch}
\affiliation{Institute of Theoretical Physics, University of Hamburg, 20355 Hamburg, Germany}
\author{Georg Rohringer}
\affiliation{Theory and Simulation of Condensed Matter, Department of Physics,
King’s College London, The Strand, London WC2R 2LS, United Kingdom}
\affiliation{Institute of Theoretical Physic, University of Hamburg, 20355 Hamburg, Germany}

\date{\today} 

\begin{abstract}
In this paper, we investigate the impact of nonlocal correlations on charge fluctuations in the two-dimensional single-band Hubbard model close to the Mott metal-to-insulator transition, employing the ladder dynamical vertex approximation. At half filling and for interaction strengths and temperatures where the system is in the Mott insulating phase, charge fluctuations are strongly suppressed. Under these conditions, dynamical mean-field theory (DMFT) calculations predict a strong enhancement of the charge susceptibility at small (electron or hole) doping. However, these DMFT results include only the effects of purely local correlations despite the importance of nonlocal correlations in two-dimensional systems. We have, hence, carried out ladder dynamical vertex approximation (lD$\Gamma$A) simulations which allow for the inclusion of such nonlocal correlation effects while retaining the local ones of DMFT. Our lD$\Gamma$A numerical data show that close to half filling the large uniform charge susceptibility of DMFT is strongly suppressed by nonlocal fluctuations but gradually increases with (electron) doping. At a certain doping value, charge fluctuations eventually become  larger in lD$\Gamma$A with respect to DMFT, indicating that the absence of nonlocal correlations underestimates the mobility of the charge carriers in this parameter regime. This metallization effect is also reflected in an enhancement of the lD$\Gamma$A kinetic and potential energies and a corresponding reduction of the (absolute value of the) lD$\Gamma$A Matsubara self-energy with respect to DMFT.
\end{abstract}

\maketitle

\section{Introduction}

Strongly correlated electron systems feature a wide range of interesting physical phenomena, including interaction-driven metal-to-insulator transitions~\cite{Mott_1968,Haule_2017,Moon_2021,Fei_2022}, strange metal behavior~\cite{Chowdhury_2022,Aavishkar_2023}, or Hund's physics~\cite{Georges_2013,Hirjibehedin_2015,Medici2017}.
A particularly interesting effect in this type of material are the collective excitations of the constituent particles. 
In fact, due to their (strong) mutual interactions single electrons can form larger entities whose physical properties are substantially different from the underlying building blocks.
P.~A.~Anderson summarized this emergence of phenomena from well understood fundamental components (i.e., the electrons) in his famous article ``More Is Different''.~\cite{Anderson_1972}
Examples for collective excitations are spin~\cite{Moriya_1985,Das_2013,Schaefer_2021}, pairing~\cite{Bergeal_2008,Maier_2019}, nematic~\cite{Boehmer_2022,Korshunov_2024} or excitonic~\cite{Kaneko_2025} fluctuations.
Another important type of collective behavior in many-electron systems are charge fluctuations, which can be observed, for instance, in high-temperature superconductors~\cite{Karolak_2015}, excitonic insulators~\cite{Volkov_2021} or systems with pronounced long-range Coulomb interactions such as adatoms on Si surfaces~\cite{Hansmann_2013}.

The theoretical description and understanding of such phenomena in realistic strongly correlated electron systems is, however, very challenging due to the strong interaction between the particles which prevents an effective independent-particle description. 
In most cases, an exact solution is unattainable, and one must rely on (i) simplified models and (ii) approximate methods. 
A prime example of the first point is the single-band Hubbard Hamiltonian~\cite{Hubbard_1963,Gutzwiller_1963,Kanamori_1963,Editorial_2013}.
Despite being one of the simplest models for correlated electrons (approximating the Coulomb repulsion as a single scalar value) it entails a rich variety of physical phenomena including collective fluctuations in the spin, pairing and, in particular, the charge sector.
Yet, a rigorous analytical or (at least) numerical solution is available only in very specific cases, for instance in one spatial dimension where Bethe ansatz~\cite{Bethe_1931, Essler_Korepin_2005}, density matrix renormalization group or matrix product states~\cite{White_1992, White_1993, Schollwoeck_2011} provide exact results.
In this respect, dynamical mean-field theory (DMFT)~\cite{Georges_Rosenberg_1996, Metzner_Vollhard_1989, Vollhardt_Kollar_2012} has been a huge step forward as it nonperturbatively captures all purely local correlation effects in the system by means of a local but frequency-dependent dynamical self-energy (which is indeed the exact solution in infinite spatial dimensions~\cite{Metzner_Vollhard_1989}). 
One of the biggest successes of DMFT is the accurate quantitative description of the correlation-driven Mott metal-to-insulator transition in correlated model systems~\cite{Georges_Rosenberg_1996} and real materials~\cite{Imada_1998}.  
While this Mott transition occurs exactly at half filling, fascinating physical phenomena are observed in the (electron or hole) doped Mott insulator, which is believed to capture important aspects of high-temperature superconductors~\cite{Lee_2006}.
In particular, for small deviations of the particle density from unity, a strong increase of the uniform charge susceptibility~\cite{Kotliar_Rozenberg_2002,Nourafkan_Tremblay_2019} (proportional to the compressibility) is observed which eventually diverges at the critical endpoint of the Mott transition within the DMFT framework.
This is quite surprising as, for a one-band model with a purely repulsive local interaction, a suppression of charge fluctuations could be expected for a large coupling strength.
It has been recently demonstrated that this enhancement of charge fluctuations is indeed an emergent phenomenon which is associated with a divergence of the local irreducible charge vertex of DMFT~\cite{Reitner_2020,Kowalski_2024}, which leads to an effective {\em attractive} interaction between the particles, thus enhancing collective charge excitations.

It is, however, not obvious to which extent the enhancement of the charge fluctuations is associated with the local DMFT approximation. 
In fact, DMFT neglects nonlocal correlations, which can certainly play a crucial role in finite dimensions. 
Particularly in low-dimensional systems long-wavelength collective fluctuations become important, especially in the vicinity of transitions to spatially ordered phases.
A good example is the two-dimensional Hubbard model, for which DMFT incorrectly predicts long-range antiferromagnetic order in contradiction with the Mermin-Wagner theorem, which excludes continuous symmetry breaking solutions at finite temperatures. 
Moreover, nonlocal correlations are believed to be essential near the Mott metal-to-insulator transition, which is not prohibited by the Mermin-Wagner theorem and may indeed occur in two-dimensional systems. 
Capturing these nonlocal effects is therefore indispensable for a realistic description of strongly correlated electrons and, in particular, the above mentioned charge fluctuations, in low dimensions.
To overcome these limitations of DMFT, several cluster and diagrammatic extensions of DMFT have been developed. 
These include the cellular DMFT (CDMFT)~\cite{Lichtenstein_Katsnelson_2000, Kotliar_Biroli_2001, Maier_Matthias_2005}, the dynamical cluster approximation (DCA)~\cite{Hettler_Krishnamurthy_1998, Hettler_Krishnamurthy_2000, Maier_Matthias_2005}, the dual fermion (DF)~\cite{Rubtsov_Lichtenstein_2008, Hafermann_Monien_2009, Rohringer_Held_2018}, dual boson (DB)~\cite{Stepanov_Rubtsov_2016, Peters_Stepanov_2019, Rohringer_Held_2018} and one-particle-irreducible (1PI)~\cite{Rohringer_Katanin_2013} approaches, the triply irreducible local expansion (TRILEX)\cite{Ayral2015,Ayral2016a} and quadruply irreducible local expansion (QUADRILEX)\cite{Ayral2016} methods, and the dynamical vertex approximation (D$\Gamma$A)~\cite{Rohringer_Held_2018, Katanin_Held_2009, Toschi_Held_2007,  Rohringer_Toschi_2016, Stobbe_Rohringer_2022, Schaefer_2021}. These methods incorporate nonlocal correlations on top of the local DMFT solution and have led to significant progress in the theoretical understanding of correlated systems in two and three dimensions.

In this paper, our goal is to investigate the impact of nonlocal correlations on the enhancement of charge fluctuation predicted by DMFT for the two-dimensional Hubbard model on a square lattice with nearest-neighbor, next-nearest-neighbor, and next-to-next-nearest-neighbor hopping in the regime close to the Mott transition. 
For this purpose, we employ the dynamical vertex approximation (D$\Gamma$A) in its ladder formulation~\cite{Toschi_Held_2007, Rohringer_Katanin_2013, Rohringer_Toschi_2016}, which systematically includes nonlocal correlations beyond DMFT.
More specifically, we exploit a recently developed improved version of this method~\cite{Stobbe_Rohringer_2022} to analyze the doping dependence of the (uniform) charge susceptibility  $\chi_{\mathrm{d},\bq =\mathbf{0}}^{\omega=0}$. 
This extension of the previous versions of the method allows for a consistent renormalization between both spin and charge fluctuations and is, hence, particularly suitable for this task.
Our D$\Gamma$A calculations predict a substantial suppression of the large charge fluctuations close to half filling and an enhancement at larger doping with respect to the DMFT prediction. 
Hence, DMFT overestimates the dynamics of charge degrees of freedom for densities close to unity while it seems to underestimate it for larger fillings.
This picture is also supported by the study of the spin susceptibility and the momentum-resolved self-energy at the Fermi surface, which offer complementary insights into the interplay between spin and charge degrees of freedom. 
Furthermore, our findings are also confirmed by the behavior of the kinetic and potential energies as a function of doping.

The paper is organized as follows. In Sec.~\ref{sec:model_and_method}, we introduce the Hubbard model and key aspects of the ladder D$\Gamma$A ($\ldga$) method used in this study. 
In Sec.~\ref{sec:results}, we present our numerical results. In particular, Secs.~\ref{sec:charge_susceptibility} and~\ref{sec:spin_susceptibility} discuss the behavior of the charge and spin susceptibilities as a function of filling, respectively. We complete these investigations by an analysis of the correlation length of the charge fluctuations in Sec.~\ref{sec:correlation_length}.
In Sec.~\ref{sec:self-energy}, we examine the self-energy at the Fermi surface, and in Sec.~\ref{sec:potential_kinetic}, we analyze the kinetic and potential energies. We conclude our paper with a summary and outlook in Sec.~\ref{sec:conclusion} and provide more details about the uniform charge susceptibility and the self-energy in the entire Brillouin zone in Appendices~\ref{sec:fitting} and~\ref{sec:self-energy_k}. 
Finally, in Appendix~\ref{sec:codes_and_data} we provide links to the codes employed in our study and to the data underlying the presented results.

\section{Model and Method}
\label{sec:model_and_method}

\subsection{Model}
\label{sec:model}

We study the two-dimensional Hubbard model on a square lattice with nearest-neighbor ($t$), next-nearest-neighbor ($t'$), and next-to-next-nearest-neighbor ($t''$) hopping, along with a local Hubbard interaction $U$. 
The Hamiltonian is given by
\begin{equation} 
\label{equ:hamiltonian} 
\hat{\mathcal{H}} = - \sum_{i,j, \sigma} t_{ij} \hat{c}_{i\sigma}^{\dagger} \hat{c}_{j\sigma}^{\phantom\dagger} + U \sum_i \hat{n}_{i\uparrow}^{\phantom\dagger} \hat{n}_{i\downarrow}^{\phantom\dagger} \,. 
\end{equation}
Here,  $\hat{c}_{i\sigma}^{\dagger}$ ($\hat{c}_{i\sigma}^{\phantom\dagger}$) creates (annihilates) an electron with spin $\sigma$ at lattice site $i$, and  $\hat n_{i\sigma}^{\phantom\dagger} = \hat{c}_{i\sigma}^{\dagger} \hat{c}_{i\sigma}^{\phantom\dagger}$ is the local particle number (or density) operator.
The corresponding dispersion relation is
\begin{align} 
\label{equ:dispersion} 
\varepsilon_\bk = & -2t \left( \cos (\bk_x) + \cos (\bk_y) \right) - 4t' \cos(\bk_x) \cos (\bk_y) \nonumber \\ 
               & - 2 t'' \left( \cos (2 \bk_x) + \cos (2 \bk_y) \right) \, , 
\end{align}

where we set $t\!=\!0.25$eV, which implies that the half bandwidth $D\!=\!4t\!=\!1$eV which we use as unit of energy.
We select hopping parameters commonly used for cuprates, setting $t'=-0.3t$ and $t''=0.2t$ although this choice has no specific significance in the context of this paper since we are conducting a model study without any reference to a specific material. 
In the following we will work in Matsubara frequency space where fermionic and bosonic Matsubara frequencies are denoted by ${\nu=(2n+1)\pi/\beta}$ and ${\omega=2\pi m/\beta}$, respectively, where ${\beta\!=\!1/T}$ is the inverse temperature and $n,m \in \mathbb{Z}$. 
For readability and brevity, integrals over momenta $\bk$ within the Brillouin zone (BZ) are written as ${\frac{1}{V_{\mathrm{BZ}}}  \int\limits_{\mathrm{BZ}}d\bk \equiv \sum\limits_{\bk}}$.

\subsection{Ladder 
\texorpdfstring{D$\mathbf{\Gamma}$A}{DGA}}
\label{sec:method}

The D$\Gamma$A approach was originally developed in Ref.\onlinecite{Toschi_Held_2007}, with its ladder approximation (referred to as $\ldga$ in the following) introduced in Refs.~\onlinecite{Toschi_Held_2007,Katanin_Held_2009,  Rohringer_Toschi_2016, Stobbe_Rohringer_2022}. 
This diagrammatic extension of DMFT constructs a nonlocal self-energy from the DMFT starting point through the (Schwinger-Dyson) equation of motion (EOM)
\begin{align}
\label{equ:self-energy}
\Sigma_{\bk}^{\nu} = \frac{U n}{2} - \frac{U}{\beta^2} \sum_{\nu',\omega,\bk',\bq} F_{\uparrow\downarrow,\bk\bk'\bq}^{\nu\nu'\omega}
G_{\bk'}^{\nu'} G_{\bk' + \bq}^{\nu' + \omega} G_{\bk + \bq}^{\nu + \omega} \,.
\end{align}
where the full vertex function $F_{\uparrow\downarrow,\bk\bk'\bq}^{\nu\nu'\omega}$ is obtained from the two-particle fully irreducible vertex $\Lambda_{\sigma\sigma'}^{\nu\nu'\omega}$ of DMFT via the Bethe-Salpeter and parquet equations in all scattering channels~\cite{Tam_Jarrell_2013,Chen_Bickers_1992,Vali_Toschi_2015,Li_Held_2016}. 
While this method correctly captures the mutual screening effects between charge, spin and particle-particle fluctuations and, hence, fulfills the associated sum rules, it is typically restricted to rather high temperatures  and moderate to intermediate coupling strengths due to its numerical complexity. 
More specifically, the parquet approach to D$\Gamma$A (but also to dual fermion) is limited by two main obstacles: 
(i) In the strong coupling regime parquet equations might converge to an unphysical solution\cite{Kozik2015,Gunnarsson2017} due to the emergence of divergences in the irreducible charge vertex of DMFT\cite{Schaefer_2013,Schaefer_2016,Thunstroem_2018,Chalupa2018,Springer2020}. 
These divergences do not present a mere technical problem as they are in fact responsible for the enhancement of the charge susceptibility close to the Mott transition\cite{Reitner_2020, Kowalski_2024}, which is the central object of our investigation.
Recently\cite{Lihm_2025} a possible solution to this problem has been proposed which, however, has been applied only to the single impurity Anderson model and the atomic limit of the Hubbard model so far.
Let us also stress that a spurious convergence is not observed within lD$\Gamma$A due to the absence of a self-consistent iteration of the corresponding correlation functions.
(ii) Due to their numerical complexity, parquet equations are typically restricted to rather high temperatures due to limited sizes of frequency and momentum grids which are feasible even when modern high-performance computing systems are exploited for the numerical simulations.
In particular, for the parameter regime which we consider in this work we find that $300\times300\times301$ fermionic and bosonic Matsubara frequencies as well as $80\times80$ $\mathbf{k}$ points are required to achieve accurate results.
In spite of many advancements in recent years, which include the improved treatment of the frequency asymptotics of vertex functions\cite{Tagliavini2018,Wentzell2020}, the separation of vertex functions into bosonic and fermionic contributions (single boson exchange formalism\cite{Krien2019,Krien2019a}), where for the less complex bosonic correlation functions larger grid sizes can be considered\cite{Krien_2020,Krien_2020a,Krien_2022}, and the introduction of basis functions for the momentum dependence\cite{Eckhardt2018,Eckhardt2020}, it is still difficult to achieve the grid sizes which are required for the calculations presented in this paper.
In this respect, let us note that the range of applicability of parquet equations might be further extended by exploiting state-of-the-art compression algorithms\cite{Wallerberger2021,Wallerberger2023} or Tensor Train methods\cite{Rohshap_2025}.

Due to the above-mentioned limitations of the self-consistent parquet approach which would indeed affect our numerical analysis, we have exploited the ladder version of D$\Gamma$A which neglects the mutual renormalization between the scattering channels induced by the parquet equations and constructs the generalized susceptibilities $\chi_{r,\mathbf{q}}^{\nu\nu'\omega}$ in the most relevant channels $r$ [which are $r\!=\!d$ density (or charge) and $r\!=\!m$ magnetic (or spin) channels in the case of a repulsive interaction] in a one-shot calculation from the local two-particle vertices $\Gamma_r^{\nu\nu'\omega}$, which are irreducible only in channel $r$, via the respective Bethe-Salpeter equations:
\begin{align}
\label{equ:generalized_susceptibility}
\chi_{r,\bq}^{\nu\nu'\omega} &= \beta\chi_{0,\bq}^{\nu\omega} \delta_{\nu\nu'} - 
\frac{1}{\beta}\chi_{0,\bq}^{\nu\omega}\sum_{\nu_1}  \Gamma_{r}^{\nu\nu_1\omega}\chi_{r,\bq}^{\nu_1\nu'\omega} \, .
\end{align}
Here, the bubble term $\chi_{0,\bq}^{\nu\omega} \!= \!-\sum_{\bk}G_{\bk}^{\nu}G_{\bk+\bq}^{\nu+\omega}$ is just the convolution of two DMFT Green's functions 
\begin{align}
G_{\bk}^{\nu}=\frac{1}{i\nu + \mu_{\mathrm{DMFT}} - \varepsilon_\bk - \Sigma^{\nu}_{\phantom k}}
\end{align} 
in momentum space, where $\Sigma^\nu$ denotes the local DMFT self-energy and $\mu_{\mathrm{DMFT}}$ refers to the chemical potential, which determines the filling $n$.

Within this ladder version of D$\Gamma$A (referred to as lD$\Gamma$A in the following) the EOM~\eqref{equ:self-energy} can be then recast in terms of the physical susceptibilities $\chi_{r,\mathbf{q}}^\omega$ and the triangular vertices $\gamma_{r,\mathbf{q}}^{\nu\omega}$ of DMFT which are obtained from the generalized susceptibilities $\chi_{r,\mathbf{q}}^{\nu\nu'\omega}$ as
\begin{subequations}
\label{equ:triangularphysicalsusc}
\begin{align}
\label{equ:physical_susceptibility}
\chi_{r,\bq}^{\omega} &= \frac{1}{\beta^2}\sum_{\nu,\nu'} \chi_{r,\bq}^{\nu\nu'\omega} \, , \\
\label{equ:triangular_vertex}
\gamma_{r,\bq}^{\nu\omega} &=\left[\chi_{0,\bq}^{\nu\omega}\left(1 \pm U \chi_{r,\bq}^{\omega} \right)\right]^{-1}\frac{1}{\beta}\sum_{\nu'} \chi_{r,\bq}^{\nu\nu'\omega}
\end{align}
\end{subequations}
The EOM for the non-local lD$\Gamma$A self-energy then becomes
\begin{align}
\label{equ:self-emergy_susc}
\Sigma_{\bk}^{\nu} &= \frac{U n}{2} - \frac{U}{\beta} \sum_{\omega,\bq}\left[1+\frac{1}{2}\gamma_{\mathrm{d},\bq}^{\nu\omega}\left(1-U\chi_{\mathrm{d},\bq}^{\omega}\right)\right. 
\\
&\left. -\frac{3}{2}\gamma_{\mathrm{m},\bq}^{\nu\omega}\left(1+U\chi_{\mathrm{m},\bq}^{\omega}\right)  - \frac{1}{\beta}\sum_{\nu'}\chi_{0,\bq}^{\nu'\omega}F_{\mathrm{m}}^{\nu\nu'\omega} \right] G_{\bk +\bq}^{\nu+\omega} \, ,\nonumber
\end{align}
where $F_m^{\nu\nu'\omega}$ is the full local vertex of DMFT in the magnetic (spin) channel.

While the numerical complexity of the lD$\Gamma$A scheme is substantially reduced with respect to the full parquet based D$\Gamma$A, it violates specific consistency relations due to the absence of a self-consistent mutual renormalization of different scattering channels.
In particular, the following two equations for the physical susceptibilities defined in Eq.~\eqref{equ:physical_susceptibility} are typically {\em not} fulfilled within lD$\Gamma$A:
\begin{subequations}
\label{equ:sumrules}
\begin{align}
\label{equ:sumrulePauli}
\frac{1}{2\beta}\sum_{\omega,\bq}\left(\chi^{\omega}_{\mathrm{d},\bq} + \chi^{ \omega}_{\mathrm{m},\bq}\right)&=\frac{n}{2}\left(1-\frac{n}{2}\right)\\
\label{equ:sumruleepot}
\underbrace{\frac{1}{2\beta}\sum_{\omega,\bq}\left(\chi^{\omega}_{\mathrm{d},\bq} - \chi^{ \omega}_{\mathrm{m},\bq}\right)+\frac{Un^2}{4}}_{E^{(2)}_\text{pot}}&=\underbrace{\frac{1}{\beta}\sum_{\nu,\bk} G^{\nu}_\bk \Sigma^{\nu}_{\bk}}_{E^{(1)}_\text{pot}},
\end{align}
\end{subequations}
where $\Sigma_\bk^\nu$ is the lD$\Gamma$A self-energy obtained via Eq.~\eqref{equ:self-emergy_susc} and 
\begin{align}
G_\bk^\nu = \frac{1}{i\nu+\mu_{\mathrm{D}\Gamma\mathrm{A}}-\epsilon_\bk-\Sigma_\bk^\nu}
\end{align}
is the lD$\Gamma$A Green's function.
The relation~\eqref{equ:sumrulePauli} is a consequence of the Pauli principle while the second equation~\eqref{equ:sumruleepot} corresponds to the consistency of the potential energy when it is obtained from one- ($E_\text{pot}^{(1)}$) and two-particle ($E_\text{pot}^{(2)}$) correlation functions, respectively.

\begin{figure}[t]
\centering
\includegraphics[width=0.99\linewidth]{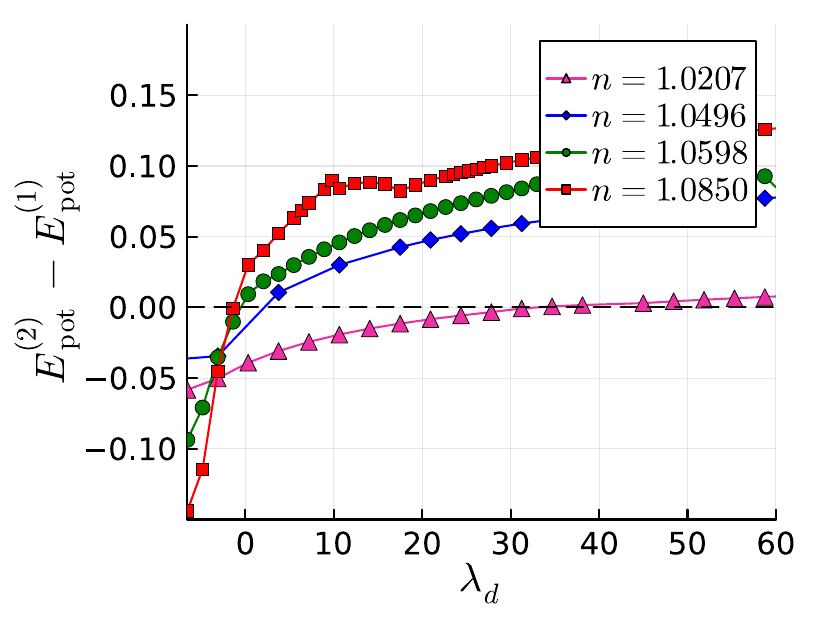}
\caption{Difference between the potential energies obtained from two- and one-particle correlation functions on the left- and right-hand sides of Eq.~\eqref{equ:lambdacondition2}, respectively, as a function of $\lambda_{\mathrm{d}}$ where $\lambda_{\mathrm{m}}$ is determined (for a given $\lambda_{\mathrm{d}}$) from Eq.~\eqref{equ:lambdacondition1}. Crossing with the $x$ axes corresponds to the solution of Eqs.~\eqref{equ:lambdacondition}.
Data are shown for $U=2.64$ and $\beta=44.0$ (see Fig.~\ref{fig:PD} and discussion in Sec.~\ref{sec:charge_susceptibility}) for four different values of the filling $n$.
}
\label{fig:condition_curve}
\end{figure}

\begin{figure}[t]
\centering
\includegraphics[width=0.99\linewidth]{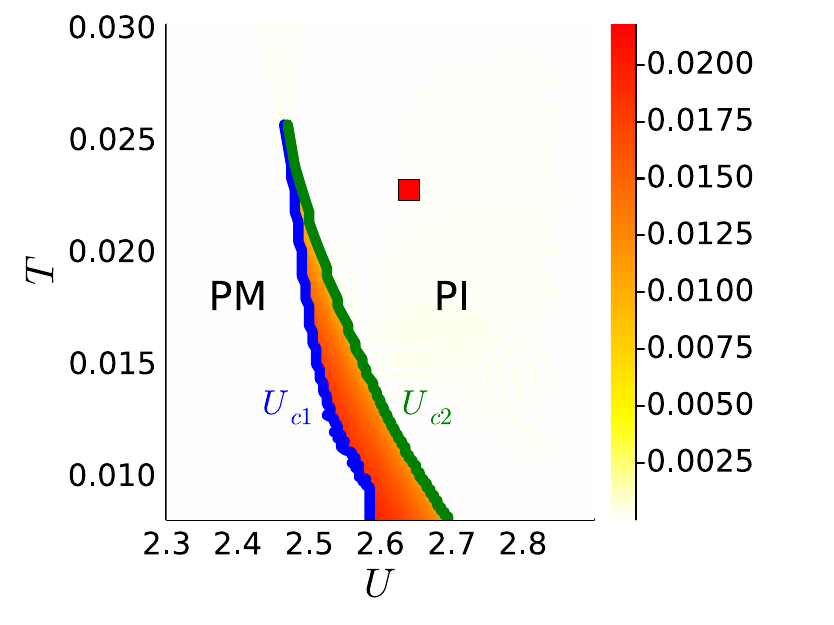}
\caption{DMFT paramagnetic phase diagram of the two-dimensional Hubbard model on a square lattice with nearest ($t$), next-nearest ($t'$), and next-to-next-nearest ($t''$) neighbor hopping, plotted as a function of interaction strength $U$ and temperature $T$ at half filling. Hysteresis is visualized via color coding, representing the difference in double occupancy $d$ obtained from the metallic and insulating solutions, respectively. $U_{c1}$ (blue line) and $U_{c2}$ (green line) are the boundaries between the coexistence region and the paramagnetic metallic (PM) and paramagnetic insulating (PI) phases, respectively. 
The red dot marks the parameter set ($U=2.64$ and $\beta=44$) used to explore the system away from half filling. Model parameters: $t = 0.25$, $t' = -0.075$, and $t'' = 0.05$.
}
\label{fig:PD}
\end{figure}

A restoration of these consistency relations has been demonstrated to reintroduce a re-balancing between the spin and charge channel and improve the predictive power of \ldga in the three-dimensional Hubbard model~\cite{Stobbe_Rohringer_2022}.
This restoration of the consistency relations~\eqref{equ:sumrules} is efficiently possible within the lD$\Gamma$A approach by following Moriya's theory of itinerant magnetism~\cite{Moriya_1985,Katanin_Held_2009,Rohringer_Toschi_2016}.
Here, the physical susceptibilities $\chi_{\mathrm{d},\bq}^\omega$ and $\chi_{\mathrm{m},\bq}^\omega$ are renormalized by static parameters $\lambda_\mathrm{d}$ and $\lambda_\mathrm{m}$:
\begin{align}
\label{equ:lambdacorr}
\chi_{r,\bq}^{\lambda_r,\omega} & = \left( \frac{1}{\chi_{r,\bq}^{\omega}} + \lambda_r \right)^{-1}  \, ,
\end{align}
which corresponds to a renormalization of the correlation length of the corresponding fluctuations [see Ref.~\onlinecite{Stobbe_Rohringer_2022} and Eq.~\eqref{equ:OrnsteinZernicke} in Sec.~\ref{sec:correlation_length}].
The parameters $\lambda_\mathrm{d}$ and $\lambda_\mathrm{m}$ are then chosen in such a way that the consistency relations in Eqs.~\eqref{equ:sumrules} are satisfied, leading to
\begin{subequations}
\label{equ:lambdacondition}
\begin{align}
\label{equ:lambdacondition1}     
\frac{1}{2\beta}\sum_{\omega,\bq}\left(\chi^{\ldens,\omega}_{\mathrm{d},\bq} + \chi^{\lmagn, \omega}_{\mathrm{m},\bq}\right) 
& \overset{!}{=} \frac{n}{2} \left( 1 - \frac{n}{2} \right) 
\\
\label{equ:lambdacondition2}
\frac{U}{2\beta}\sum_{\omega\bq}\left(\chi^{\ldens,\omega}_{\mathrm{d},\bq} - \chi^{\lmagn, \omega}_{\mathrm{m},\bq}\right) + U\frac{n^2}{4} 
& \overset{!}{=} \frac{1}{\beta}\sum_{\nu,\bk} G^{\bl,\nu}_\bk \Sigma^{\bl,\nu}_{\bk}, 
\end{align}
\end{subequations}
which is a system of two nonlinear equations for the two parameters $\lambda_{\mathrm{d}}$ and $\lambda_{\mathrm{m}}$.
These have to be solved together with
\begin{equation}
\label{equ:fixchemicalpotential}
\frac{1}{\beta}\sum_{\bk}G_\bk^{\bl,\nu}\overset{!}{=}\frac{n}{2} \,,
\end{equation}
which determines the chemical potential $\mu_{\mathrm{D}\Gamma\mathrm{A}}$ in 
\begin{align}
G_\bk^{\bl,\nu} = \frac{1}{i\nu+\mu_{\mathrm{D}\Gamma\mathrm{A}}-\epsilon_\bk-\Sigma_\bk^{\bl,\nu}} 
\end{align}
to fix the required density $n$.

In practice, the determination of $\lambda_{\mathrm{d}}$ and $\lambda_{\mathrm{m}}$ is carried out using a specific variant of the Newton method.
Let us, however, illustrate the solution process in a more transparent way which allows for a better understanding of its physical content.
First, we use only the Pauli principle via Eq.~\eqref{equ:lambdacondition1}, to calculate $\lambda_{\mathrm{m}}$ for a given $\lambda_{\mathrm{d}}$~\cite{Stobbe_Rohringer_2022}.
The pairs $[\lambda_{\mathrm{d}},\lambda_{\mathrm{m}}(\lambda_{\mathrm{d}})]$ are then used to calculate the potential energies at the one- and the two-particle level on the right- and left-hand sides of  Eq. \eqref{equ:lambdacondition2}, respectively, as functions of $\lambda_{\mathrm{d}}$.
Furthermore, a chemical potential $\mu_{\text{D}\Gamma\text{A}}$ is obtained for each pair.
A solution of this equation corresponds to the specific value of $\lambda_{\mathrm{d}}$ for which $E_\text{pot}^{(2)}\!-\!E_\text{pot}^{(1)}\!=\!0$.
To illustrate the solution process visually, we show in Fig.~\ref{fig:condition_curve} this difference between the two potential energies as a function of $\lambda_{\mathrm{d}}$ for four different values of the filling at the interaction and temperature indicated by the red square in Fig.~\ref{fig:PD} (for the explicit values, see the caption of Fig.~\ref{fig:condition_curve} and Sec.~\ref{sec:charge_susceptibility}). 
We clearly observe a crossing of all curves with the $\lambda_{\mathrm{d}}$ axis, where the $\lambda_{\mathrm{d}}$ value at this crossing corresponds to the solution of Eqs.~\eqref{equ:lambdacondition}. 
Interestingly, the solutions $\lambda_{\mathrm{d}}$ are monotonously decreasing with increasing filling, indicating that nonlocal correlations affect charge fluctuations in very different ways for different electron occupation numbers.
This will be discussed in more detail in Sec.~\ref{sec:charge_susceptibility}.

Let us finally note that, in the original version of the $\lambda$ corrected $\ldga$\cite{Katanin_Held_2009} method, only Eq.~\eqref{equ:lambdacondition1}, which enforces the Pauli principle, was considered.
Therefore only the spin susceptibility was rescaled, i.e., in our notation, $\lambda_{\mathrm{m}} \neq 0$ and $\lambda_{\mathrm{d}} = 0$, with the charge renormalization only being included recently~\cite{Stobbe_Rohringer_2022}. To distinguish between these two cases, we refer to the approach where both charge and spin susceptibilities are rescaled as $\ldgadm$, and to the version where only the spin susceptibility is rescaled as $\ldgam$.

\begin{figure}[t]
\centering
\includegraphics[width=0.99\linewidth]{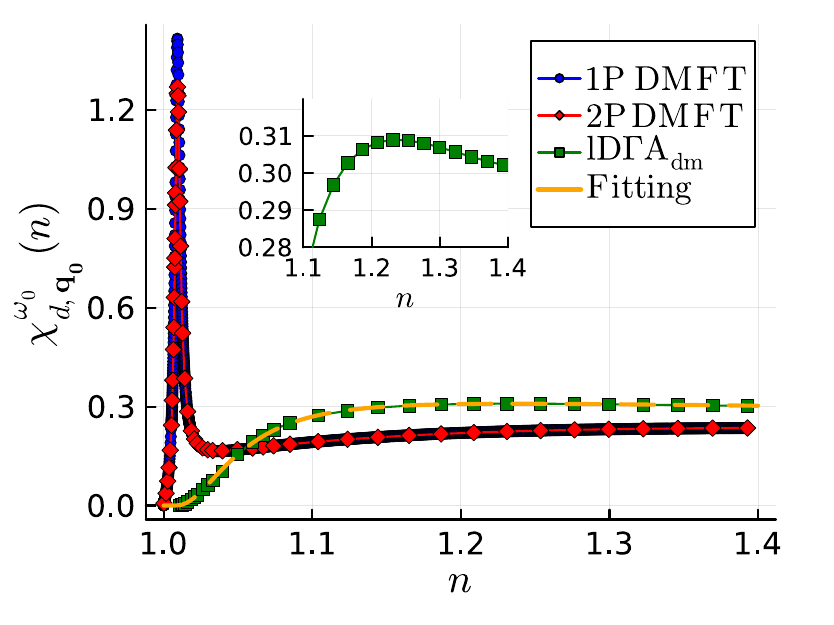}
\caption{Uniform charge susceptibility $\chi_{\mathrm{d},\mathbf{q_0}=\mathbf{0}}^{\omega_0=0}$ for $U\!=\!2.64$ and $\beta\!=\!44.0$ (see red square in Fig.~\ref{fig:PD}) as a function of filling $n$ obtained from DMFT (red diamonds) and $\ldgadm$ (green squares). For comparison also 
$\frac{1}{2}\frac{dn}{d\mu}$ is also depicted (blue circles), which is equivalent to the uniform charge susceptibility within DMFT. The orange curve is obtained by fitting $\ldgadm$ data using Eq.~\eqref{equ:fitting_chid} in Appendix~\ref{sec:fitting} with the fitting parameters are $A = 0.6307$, $B = 0.0658$, and $C = 0.1473$.  
Inset: Enlargement of the $\ldgadm$ uniform charge susceptibility around its maximum value.}
\label{fig:charge_susceptibility}
\end{figure}

\begin{figure}[t]
\centering
\includegraphics[width=0.99\linewidth]{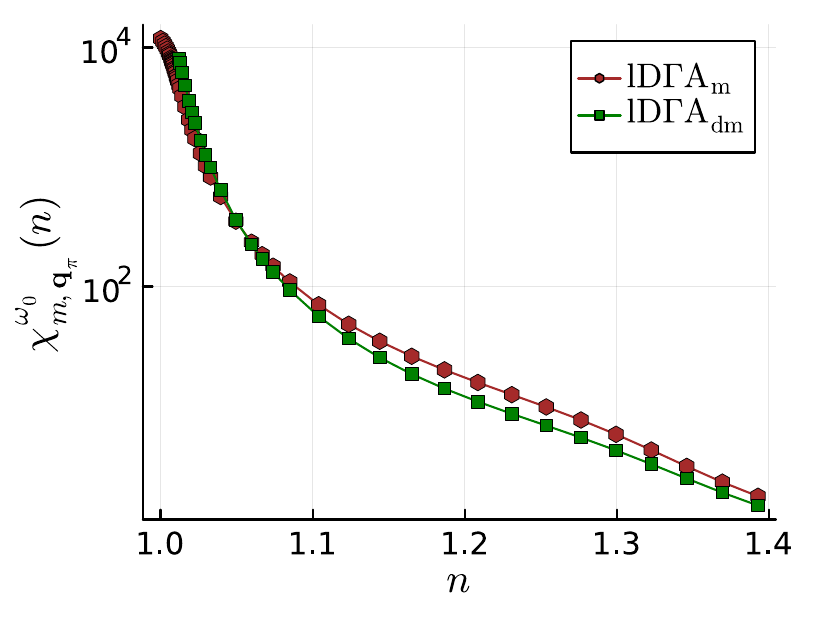}
\caption{Spin susceptibility $\chi_{\mathrm{m},\bq_\pi}^{\omega_0}$ as a function of filling $n$ for $\ldgam$ where only the magnetic channel is corrected ($\lambda_{\mathrm{m}} \neq 0$, $\lambda_{\mathrm{d}} = 0$, brown hexagons) and for $\ldgadm$, where we include both magnetic and charge corrections ($\lambda_{\mathrm{m}} \neq 0$, $\lambda_{\mathrm{d}} \neq 0$, green squares). Here, $\mathbf{q}_\pi = (\pi,\pi)$ denotes the antiferromagnetic ordering vector. Data are shown on a logarithmic scale on the $y$-axis for the same parameters $U$ and $T$ as in Fig.~\ref{fig:charge_susceptibility}. }
\label{fig:mag_susceptibility}
\end{figure}

\begin{center}
\begin{figure*}[t!]
\subfigure[~$n=1.0207$]{
\label{fig:chid_vs_omega_1}
\begin{minipage}[b]{0.235\textwidth}
\centering 
\includegraphics[width=1\textwidth]{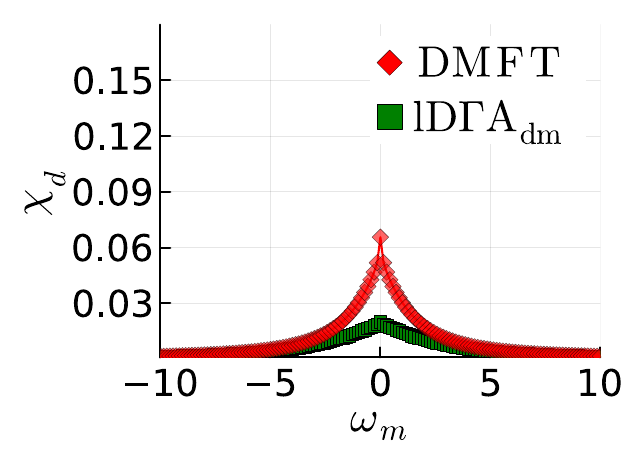}
\end{minipage}}
\subfigure[~$n=1.0496$]{
\label{fig:chid_vs_omega_2}
\begin{minipage}[b]{0.235\textwidth}
\centering 
\includegraphics[width=1\textwidth]{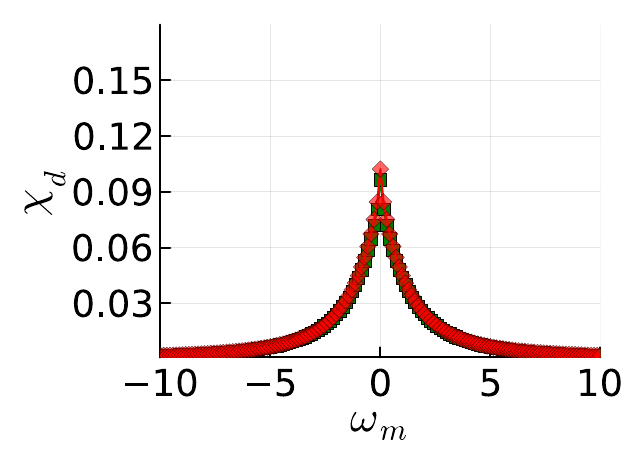}
\end{minipage}}
\subfigure[~$n=1.0598$]{
\label{fig:chid_vs_omega_3}
\begin{minipage}[b]{0.235\textwidth}
\centering 
\includegraphics[width=1\textwidth]{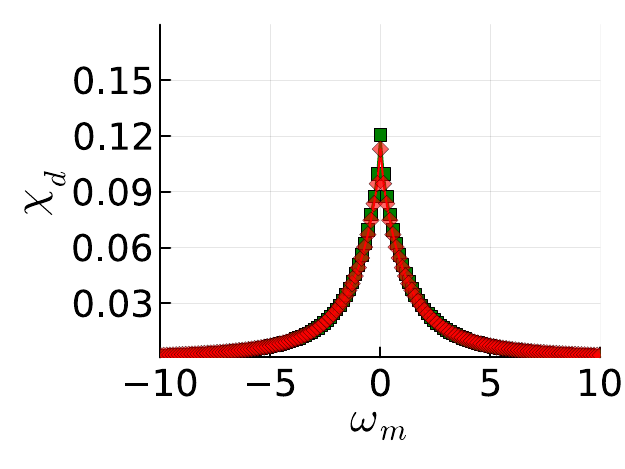}
\end{minipage}}
\subfigure[~$n=1.0850$]{
\label{fig:chid_vs_omega_4}
\begin{minipage}[b]{0.235\textwidth}
\centering 
\includegraphics[width=1\textwidth]{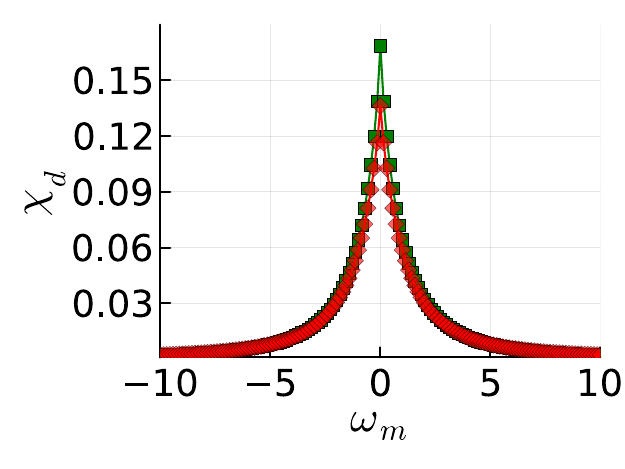}
\end{minipage}}
\\
\subfigure[~$n=1.0207$]{
\label{fig:chid_vs_q_1}
\begin{minipage}[b]{0.235\textwidth}
\centering 
\includegraphics[width=1\textwidth]{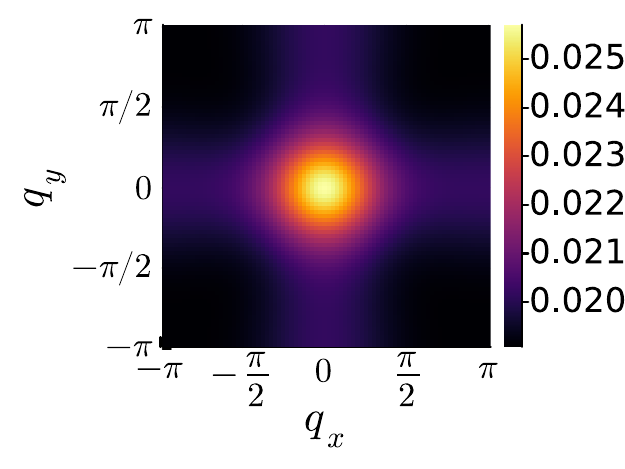}
\end{minipage}}
\subfigure[~$n=1.0496$]{
\label{fig:chid_vs_q_2}
\begin{minipage}[b]{0.235\textwidth}
\centering 
\includegraphics[width=1\textwidth]{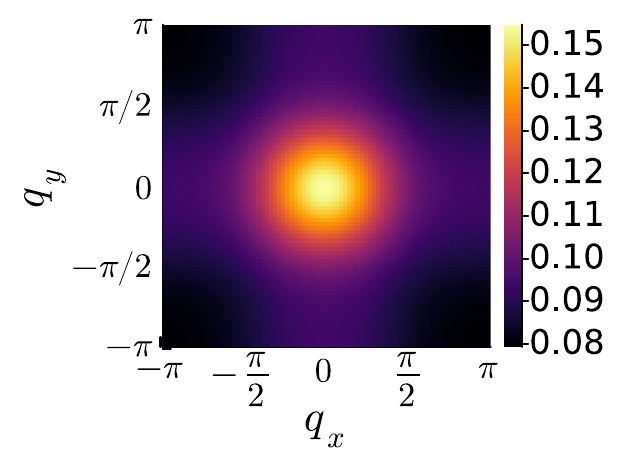}
\end{minipage}}
\subfigure[~$n=1.0598$]{
\label{fig:chid_vs_q_3}
\begin{minipage}[b]{0.235\textwidth}
\centering 
\includegraphics[width=1\textwidth]{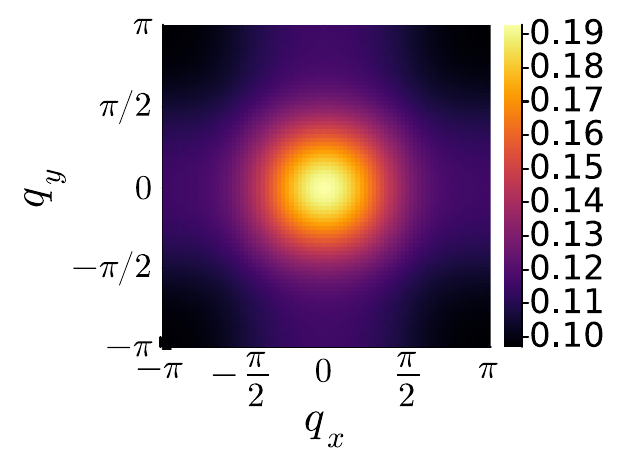}
\end{minipage}}
\subfigure[~$n=1.0850$]{
\label{fig:chid_vs_q_4}
\begin{minipage}[b]{0.235\textwidth}
\centering 
\includegraphics[width=1\textwidth]{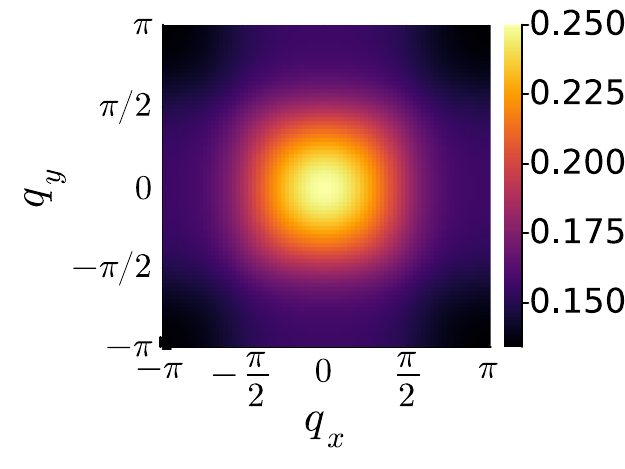}
\end{minipage}}
\\
\subfigure[~$n=1.1652  $]{
\label{fig:chid_vs_q_5}
\begin{minipage}[b]{0.235\textwidth}
\centering 
\includegraphics[width=1\textwidth]{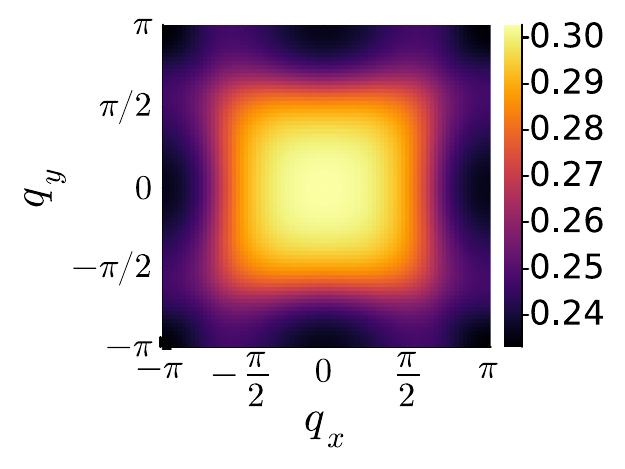}
\end{minipage}}
\subfigure[~$n=1.2087$]{
\label{fig:chid_vs_q_6}
\begin{minipage}[b]{0.235\textwidth}
\centering 
\includegraphics[width=1\textwidth]{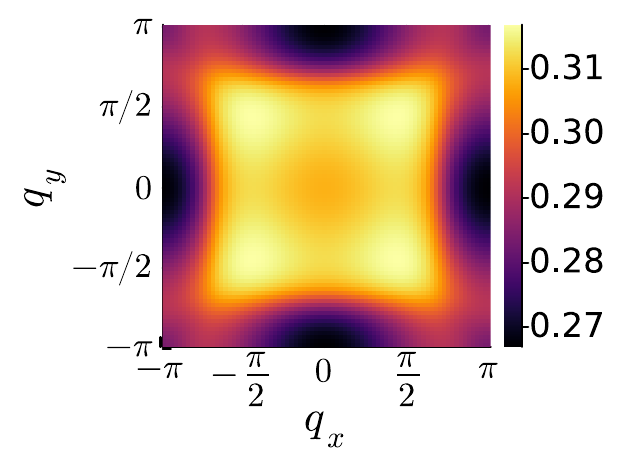}
\end{minipage}}
\subfigure[~$n=1.2995$]{
\label{fig:chid_vs_q_7}
\begin{minipage}[b]{0.235\textwidth}
\centering 
\includegraphics[width=1\textwidth]{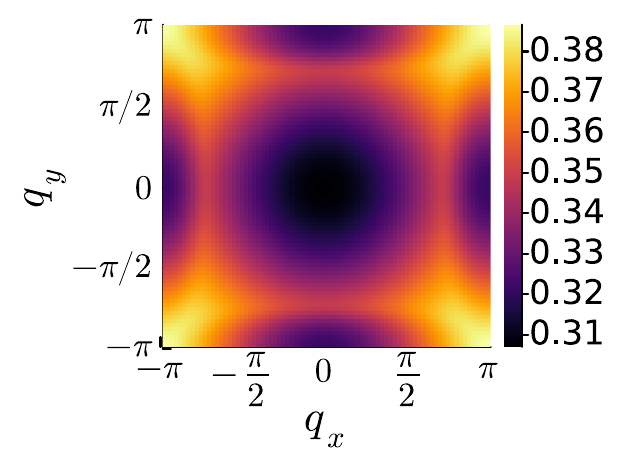}
\end{minipage}}
\subfigure[~$n=1.3928$]{
\label{fig:chid_vs_q_8}
\begin{minipage}[b]{0.235\textwidth}
\centering 
\includegraphics[width=1\textwidth]{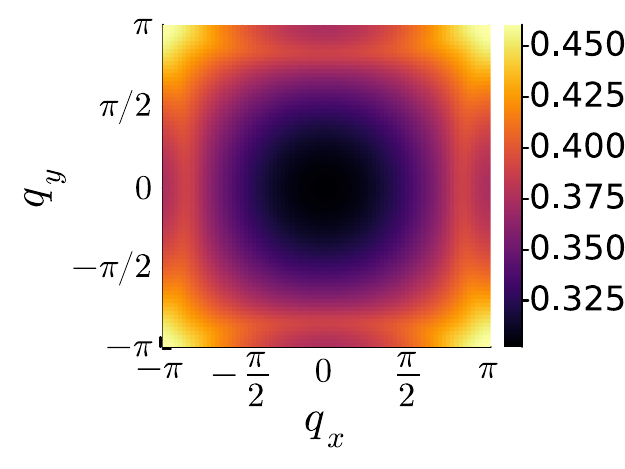}
\end{minipage}}
\caption{Upper row: Local (momentum summed) charge susceptibility $\sum_\bq \chi_{\mathrm{d},\bq}^{\lambda_{\mathrm{d}},\omega}$ of DMFT ($\lambda_{\mathrm{d}}\!=\!0$, red diamonds) vs. $\ldgadm$ ($\lambda_{\mathrm{d}}\!\ne\!0$, green squares) as a function of the bosonic Matsubara frequency $\omega$ for four different values of filling $n$ gradually increasing from (a) to (d); Lower rows: Static $\ldgadm$ charge susceptibility $\chi_{\mathrm{d},\bq}^{\lambda_{\mathrm{d}},\omega=0}$ as a function of momentum $\bq$ for eight different values of $n$ gradually increasing from (e) to (l). Interaction strength and temperature are the same as in Fig.~\ref{fig:charge_susceptibility}.}
\label{fig:charge_susceptibility_omega_q}
\end{figure*}
\end{center}

\section{Results}
\label{sec:results}

\subsection{Charge susceptibility}
\label{sec:charge_susceptibility}

In the following, we will investigate the strong enhancement of charge fluctuations in DMFT at the verge of the Mott transition and, in particular, the impact of nonlocal correlations beyond DMFT on this emerging phenomenon \cite{Reitner_2020,Kowalski_2024,Nourafkan_Tremblay_2019, Sordi_Tremblay_2011}.

To set the stage, we have first mapped out the $U$ vs. $T$ phase diagram of the Hubbard model for the selected hopping parameters $t$, $t'$, and $t''$ detailed in Sec.~\ref{sec:model} at half filling ($n\!=\!1)$.
The corresponding results are shown in Fig.~\ref{fig:PD}, where the difference between the double occupancy $\langle n_\uparrow n_\downarrow\rangle$ of the metallic and insulating solution is depicted as a heat map.
This difference is obviously zero if only one of the phases exists [which is the paramagnetic metal (PM) for small values of $U$ and the paramagnetic insulator (PI) for large values of $U$] and takes finite values in the {\em coexistence region} where the self-consistent DMFT iteration has two fixed points, i.e., a metallic and an insulating one which obviously have different double occupancies.
Note that exactly at half filling, charge fluctuations are typically strongly suppressed in the insulating phase on the right-hand side of the coexistence region.

We now aim at finding a point in the phase diagram where the uniform charge susceptibility of DMFT is continuous as a function of the filling $n$ and shows a pronounced peak close to $n\!=\!1$.
To this end, we have to avoid parameter sets $(U, T)$ where the filling $n$ as a function of the chemical potential $\mu_{\mathrm{DMFT}}$ displays a discontinuity. 
As is well known~\cite{Nourafkan_Tremblay_2019, Werner_Millis_2007, Liebsch_2008, Steinbauer_Biermann_2019, Sordi_Tremblay_2011}, such behavior occurs just at the first-order metal–Mott insulator transition where the system features a finite jump in the filling $n$ as a function of $\mu_\text{DMFT}$.
Therefore, we have considered an interaction and a temperature slightly away from the metal–insulator transition in the half-filled model which is indicated by the red square in Fig.~\ref{fig:PD} and corresponds to $U\!=\!2.64$ and $\beta\!=\!44$.
Let us, however, note that also other points on the Mott insulating side of the phase diagram are also possible for our analysis as long as the charge susceptibility is a continuous function of the chemical potential $\mu_{\mathrm{DMFT}}$.

We now investigate how nonlocal correlations affect charge fluctuations for different fillings $n$.
To this end, we analyze the uniform charge susceptibility $\chi_{\mathrm{d},\bq_0}^{\omega_0}$, where $\omega_0\!=\!0$ and $\bq_0\!=\!(0,0)$, which has been obtained in three different ways in Fig.~\ref{fig:charge_susceptibility}:
First we have evaluated the DMFT charge susceptibility obtained via Eqs.~\eqref{equ:generalized_susceptibility} and \eqref{equ:physical_susceptibility} at $\omega\!=\!\omega_0$ and $\mathbf{q}\!=\!\mathbf{q}_0$ as a function of $n$, which is shown by red diamonds in Fig.~\ref{fig:charge_susceptibility}.
While at and (very) close to half filling this correlation function is very small, it gets strongly enhanced for finite (but still small) doping, reaching a maximum around $n\!\approx 1.01$.
For larger values of the doping, it rapidly decays and starts to depend only weakly on $n$ for $n\!>\!1.2$ where $\chi_{\mathrm{d},\bq_0}^{\omega_0}\!\approx\!0.3$.
To confirm these results, we have also calculated the compressibility $\kappa\!=\!\frac{1}{n^2}\frac{dn}{d\mu}$, which is proportional to the uniform charge susceptibility ($\chi_{\mathrm{d},\mathbf{q_0}=\mathbf{0}}^{\omega_0=0}=\frac{1}{2}n^2 \kappa$) for a conserving theory such as DMFT.
The corresponding results are given by blue circles in Fig.~\ref{fig:charge_susceptibility} and indeed show excellent agreement with the results obtained from the charge susceptibility (except for very small deviations at the peak, which can be attributed to the finite-difference approximation used in the numerical evaluation of the derivative $\frac{dn}{d\mu}$).

As discussed in the Introduction, DMFT captures local fluctuations accurately but neglects nonlocal correlations (or, more precisely, treats them only on a mean field level). 
To account for the latter, we performed $\ldgadm$ calculations for $\chi_{\mathrm{d},\bq}^{\lambda_{\mathrm{d}},\omega}$ as a function of the density $n$.
Our results, shown as green squares in Fig.~\ref{fig:charge_susceptibility}, demonstrate that nonlocal correlations have a pronounced effect on charge fluctuations. 
Close to half filling, the peak in the uniform charge susceptibility $\chi^{\lambda_{\mathrm{d}},\omega_0}_{\mathrm{d},\bq_0}$ observed within DMFT is strongly suppressed in $\ldgadm$. 
As the filling increases, the rescaled susceptibility $\chi^{\lambda_{\mathrm{d}},\omega_0}_{\mathrm{d},\bq_0}$ becomes larger and eventually exceeds the corresponding DMFT value for $n \gtrsim 1.0544$. 
It reaches a maximum at around $n_\text{max}\!\approx\!1.24$ (see inset of Fig.~\ref{fig:charge_susceptibility}) and decays for larger $n\!>\!n_\text{max}$ eventually approaching the DMFT results. 
This is indeed the expected behavior, as correlation effects generally become weaker with increasing filling.
Our results have a clear-cut physical interpretation: Close to half filling, the neglect of nonlocal correlations in DMFT leads to a substantial overestimation of uniform charge fluctuation whereas (far) away from $n\!=\!1$ DMFT underestimates them.
Within our $\ldgadm$ approach this effect can be interpreted as a rebalancing of charge and spin fluctuations induced by consistency relations~\eqref{equ:lambdacondition}, which mimics a fully self-consistent parquet treatment at substantially lower numerical cost.
More specifically, the dominance of antiferromagnetic spin fluctuations in this filling regime, which is also indicated by a very large spin susceptibility in Fig.~\ref{fig:mag_susceptibility},
leads to a localization of electrons by the formation of antiferromagnetic domains where charge fluctuations are strongly suppressed.
On the contrary, for larger values of $n$, the electron (or rather hole) mobility is increased as more free lattice sites (without holes) are available.

To get further insights into the impact of nonlocal correlations on charge fluctuations we have also investigated the frequency and momentum dependence of the charge susceptibility $\chi^{\lambda_{\mathrm{d}},\omega}_{\mathrm{d},\bq}$ in Fig.~\ref{fig:charge_susceptibility_omega_q}.
In Figs.~\ref{fig:chid_vs_omega_1}-\ref{fig:chid_vs_omega_4},
we analyze the momentum-summed susceptibility $\chi^{\lambda_{\mathrm{d}},\omega}_{\mathrm{d}} =\sum\limits_{\bq} \chi^{\lambda_{\mathrm{d}},\omega}_{\mathrm{d},\bq}$ as a function of the bosonic Matsubara frequency $\omega$ obtained by DMFT ($\lambda_{\mathrm{d}}\!=\!0$, red diamonds) and $\ldgadm$ ($\lambda_{\mathrm{d}}\!\ne\!0$, green squares) for four different fillings.
As expected, all curves show a pronounced peak at $\omega = 0$ as classical fluctuations are typically dominant at finite temperatures. 
The doping dependence of the local (momentum summed) susceptibility follows the one observed for the uniform susceptibility in Fig.~\ref{fig:charge_susceptibility}:
Close to half filling [Fig.~\ref{fig:chid_vs_omega_1}, $n\!=\!1.0207$] nonlocal correlations substantially reduce the DMFT susceptibilities which is also indicated by the large positive value of $\lambda_{\mathrm{d}}$ at which $E_\text{pot}^{(2)}\!-\!E_\text{pot}^{(1)}\!=\!0$ for this filling in Fig.~\ref{fig:condition_curve}.
At fillings $n\!=\!1.0496$ [Fig.~\ref{fig:chid_vs_omega_2}] and $n\!=\!1.0598$ [Fig.~\ref{fig:chid_vs_omega_3}], which lie close to $n\!\simeq\!1.0544$, where the DMFT and $\ldgadm$ uniform charge susceptibilities coincide in Fig.~\ref{fig:charge_susceptibility}, we observe a similar agreement for their local (momentum-summed) counterparts. 
For $n\!=\!1.0496$, the maximum value obtained from $\ldgadm$ is slightly lower than that from DMFT, while for $n\!=\!1.0598$, the situation is reversed. At even higher filling, $n\!=\!1.085$ [Fig.~\ref{fig:chid_vs_omega_4}], the local $\ldgadm$ susceptibility becomes significantly larger than the DMFT result. This behavior for $n\!=\!1.0598$ and $n\!=\!1.085$, where the local $\ldgadm$ susceptibility exceeds the DMFT one, is consistent with the negative values of $\lambda_{\mathrm{d}}$ at which $E_\text{pot}^{(2)}\!-\!E_\text{pot}^{(1)}$ crosses zero in Fig.\ref{fig:condition_curve}.

The momentum dependence of the $\ldgadm$ charge susceptibility at $\omega\!=\!0$ is shown in the middle and lower rows of Fig.~\ref{fig:charge_susceptibility_omega_q} [Fig.~\ref{fig:chid_vs_q_1}-\ref{fig:chid_vs_q_8}]. 
At low fillings, we observe a sharp maximum at $\bq = (0,0)$ which indicates that uniform charge fluctuations are dominant in this parameter regime, although the overall size of the charge susceptibility is strongly suppressed due to the rebalancing between the charge and spin channels discussed above.
Upon increasing the filling, this sharp feature broadens, becomes completely flat around $n\!=\!1.1652$ [Fig.~\ref{fig:chid_vs_q_5}], shifts away from $\bq_0$ to some incommensurate wave vectors $\mathbf{q}$ at $n=1.2087$ [Fig.~\ref{fig:chid_vs_q_6}] and eventually moves to $\mathbf{q}\!=\!(\pi,\pi)$ corresponding to a checkerboard structure for the two largest fillings considered here [Figs.~\ref{fig:chid_vs_q_7} and \ref{fig:chid_vs_q_8}].
This behavior can be understood as an effect similar to the superexchange entailing anitferromagnetic spin order at half filling:
For fillings substantially larger than $n\!=\!1$, the system features a lot of double occupancies.
Its total energy can be lowered by hopping processes which are, however, only possible if doubly occupied sites are next to singly occupied sites  which requires an alternating (checkerboard)-type distribution of these two types of occupancies.

Let us point out that the consistency condition in Eq.~\eqref{equ:lambdacondition1} forbids any charge or spin order at finite temperatures in two spatial dimensions when the charge and spin susceptibilities are obtained from DMFT.
In fact, a DMFT susceptibility at a (finite) transition temperature is given by $\chi_{r,\bq}^{\omega=0}\!\sim\!\frac{1}{q^2}$ in the vicinity of $\omega\!=\!\omega_0$ and $\mathbf{q}\!=\!\mathbf{q}_0$.
This, however, would lead to a divergence of the corresponding momentum integral, which is incompatible with the finite value on the right-hand side of sum rule~\eqref{equ:lambdacondition1}.
While this is in agreement with the Mermin-Wagner theorem for the three dimensional magnetic order parameter in two dimensions\cite{Mermin1966}, for which any ordered state at finite temperature is destroyed by fluctuations, a finite-temperature divergence of the charge susceptibility is, in principle, possible in two dimensions.
This can be compatible with condition~\eqref{equ:lambdacondition1} only if one replaces DMFT by a more accurate starting point which takes into account a non-mean-field critical exponents $\eta\!>\!0$ for which the integral over $\sim\!\frac{1}{q^{2-\eta}}$ is finite.
In our case, we are sufficiently far away from the critical endpoint of DMFT (see phase diagram in Fig.~\ref{fig:PD}), where the mean field exponent $\eta\!=\!0$ is justified, which entails the above-described reduction of the charge susceptibility.
It is, however, not excluded that sizable charge fluctuations can be found in other parameter regimes closer to the critical endpoint of the Mott transition or at very low temperatures which cannot be unambiguously identified within the combination of DMFT and $\ldgadm$.
Future studies of this question can also be relevant to clarify a possible correlation-driven enhancement of electron-phonon coupling in strongly correlated electron systems \cite{Huang_2003,Coulter_2025,Moghadas_2025}.

\begin{figure}
\centering
\includegraphics[width=0.99\linewidth]{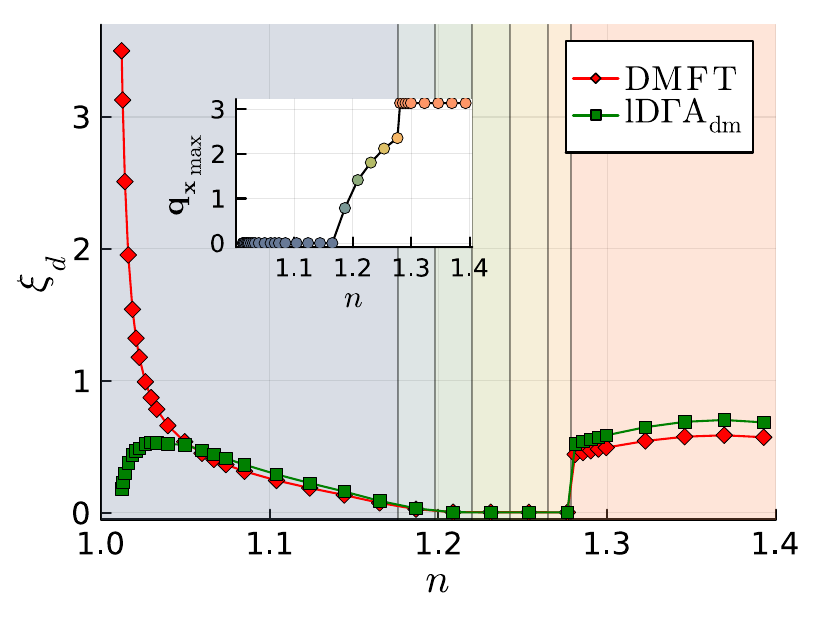}
\caption{The charge correlation length $\xi_{\mathrm{d}}$ as a function of filling $n$ obtained from DMFT (red diamonds) 
and $\ldgadm$ (green squares). Different $\mathbf{q}_{\mathrm{max}}$ regions are shaded by different colors. Inset: The component $q_x$ of $\mathbf{q}_{\mathrm{max}}$ (where $q_x\!=\!q_y$) as a function of the filling $n$. 
Parameters are the same as in Fig.~\ref{fig:charge_susceptibility}. 
}
\label{fig:correlation_length}
\end{figure}

\begin{center}
\begin{figure*}[t!]
\subfigure[~$n=1.0207$]{
\label{fig:chim_omega_1}
\begin{minipage}[b]{0.235\textwidth}
\centering 
\includegraphics[width=1\textwidth]{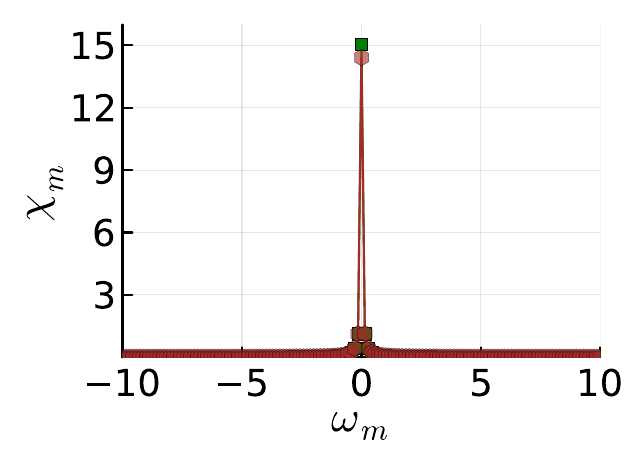}
\end{minipage}}
\subfigure[~$n=1.0496$]{
\label{fig:chim_omega_2}
\begin{minipage}[b]{0.235\textwidth}
\centering 
\includegraphics[width=1\textwidth]{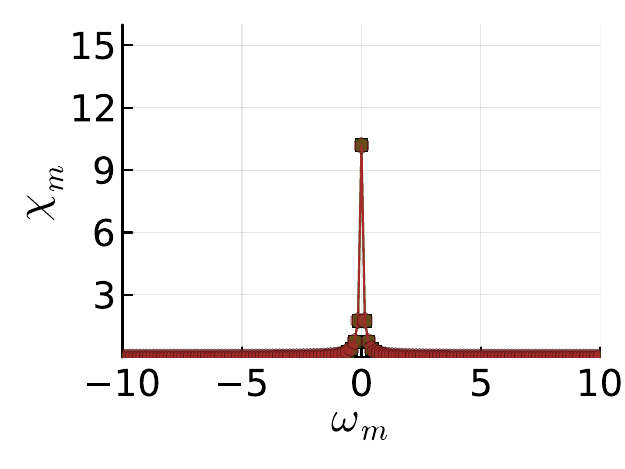}
\end{minipage}}
\subfigure[~$n=1.0598$]{
\label{fig:chim_omega_3}
\begin{minipage}[b]{0.235\textwidth}
\centering 
\includegraphics[width=1\textwidth]{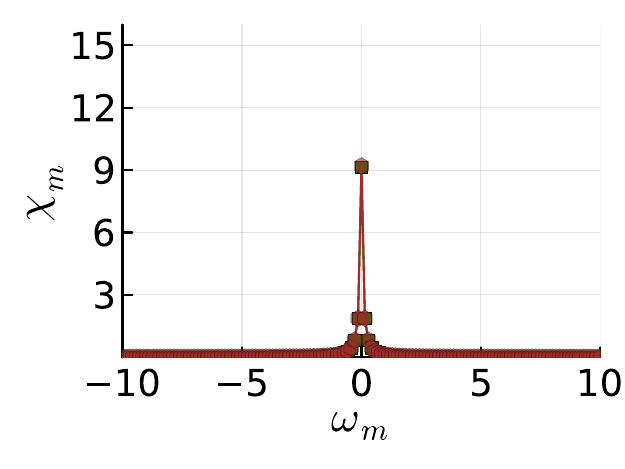}
\end{minipage}}
\subfigure[~$n=1.0850$]{
\label{fig:chim_omega_4}
\begin{minipage}[b]{0.235\textwidth}
\centering 
\includegraphics[width=1\textwidth]{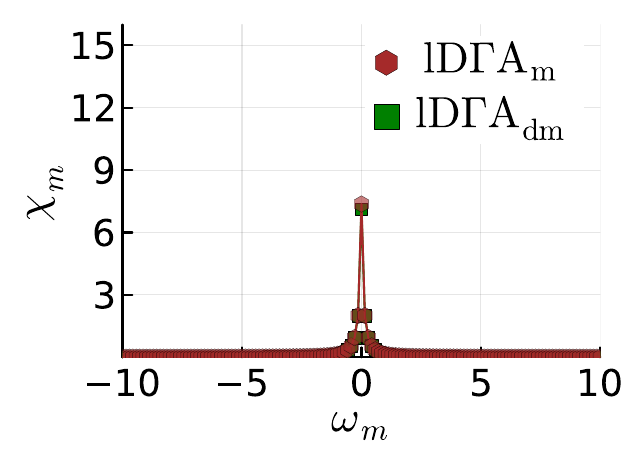}
\end{minipage}}
\\
\subfigure[~$n=1.0207$]{
\label{fig:chim_q_1}
\begin{minipage}[b]{0.235\textwidth}
\centering 
\includegraphics[width=1\textwidth]{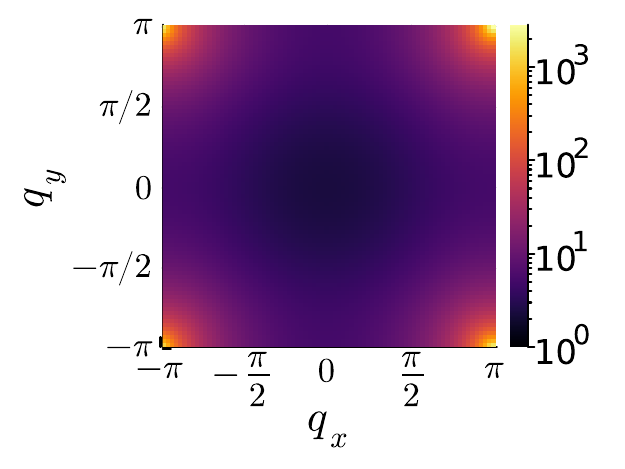}
\end{minipage}}
\subfigure[~$n=1.0496$]{
\label{fig:chim_q_2}
\begin{minipage}[b]{0.235\textwidth}
\centering 
\includegraphics[width=1\textwidth]{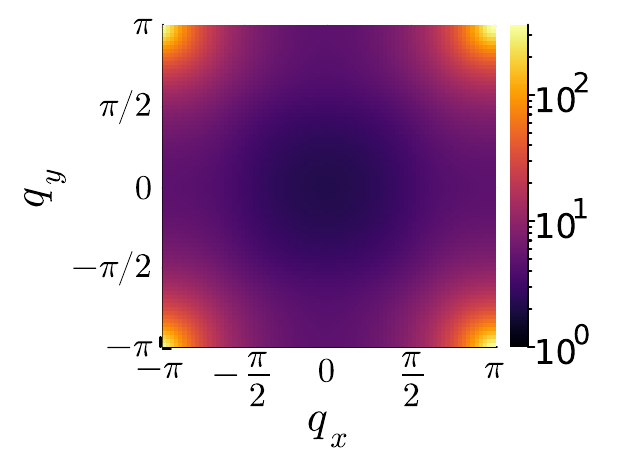}
\end{minipage}}
\subfigure[~$n=1.0598$]{
\label{fig:chim_q_3}
\begin{minipage}[b]{0.235\textwidth}
\centering 
\includegraphics[width=1\textwidth]{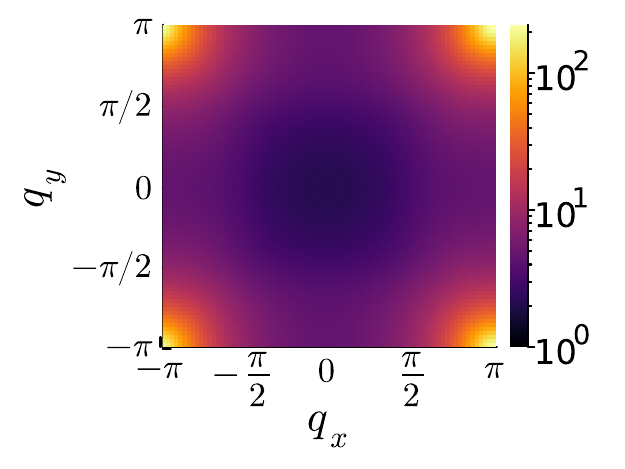}
\end{minipage}}
\subfigure[~$n=1.0850$]{
\label{fig:chim_q_4}
\begin{minipage}[b]{0.235\textwidth}
\centering 
\includegraphics[width=1\textwidth]{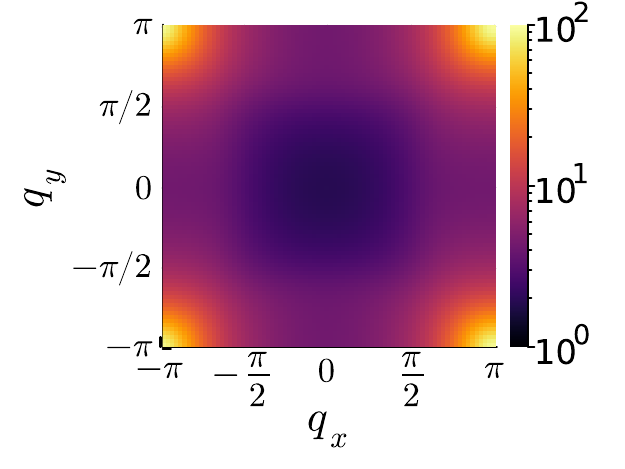}
\end{minipage}}
\\
\subfigure[~$n=1.1652$]{
\label{fig:chim_q_5}
\begin{minipage}[b]{0.235\textwidth}
\centering 
\includegraphics[width=1\textwidth]{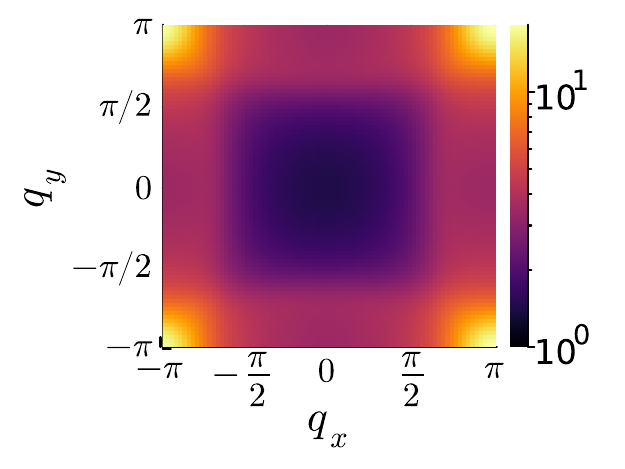}
\end{minipage}}
\subfigure[~$n=1.2087$]{
\label{fig:chim_q_6}
\begin{minipage}[b]{0.235\textwidth}
\centering 
\includegraphics[width=1\textwidth]{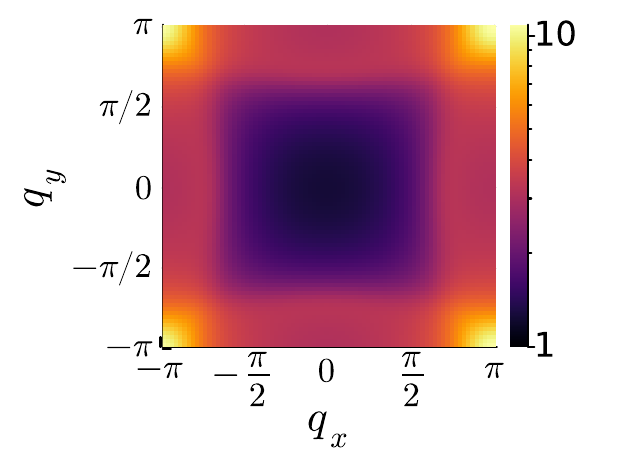}
\end{minipage}}
\subfigure[~$n=1.2995$]{
\label{fig:chim_q_7}
\begin{minipage}[b]{0.235\textwidth}
\centering 
\includegraphics[width=1\textwidth]{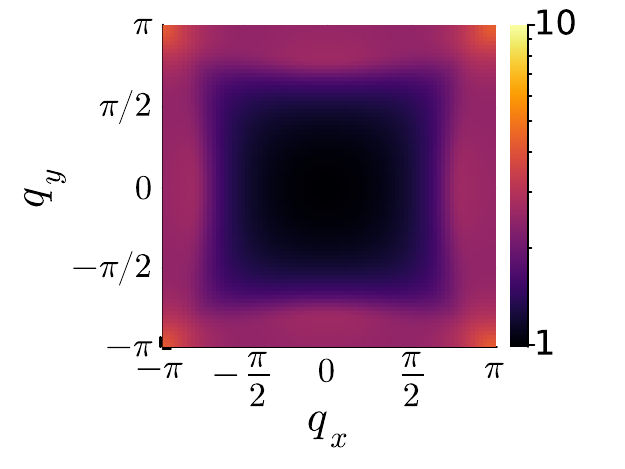}
\end{minipage}}
\subfigure[~$n=1.3928$]{
\label{fig:chim_q_8}
\begin{minipage}[b]{0.235\textwidth}
\centering 
\includegraphics[width=1\textwidth]{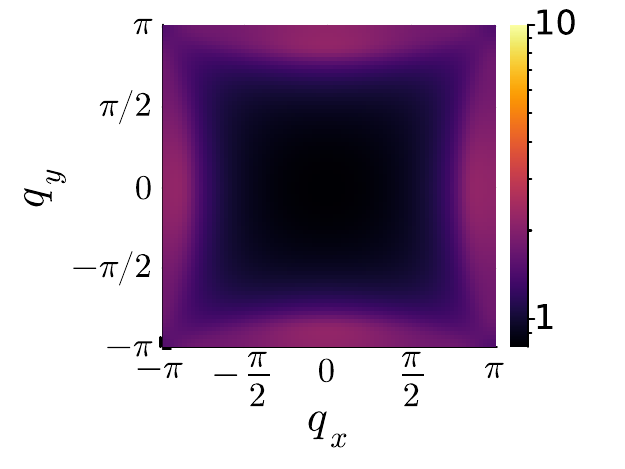}
\end{minipage}}
\caption{Same as in Fig.~\ref{fig:charge_susceptibility_omega_q} but for the spin susceptibility $\chi^{\lambda_{\mathrm{m}},\omega}_{\mathrm{m},\bq}$ of $\ldgadm$ (green squares). In addition, we show data for the $\ldgam$ method (brown hexagon) where {\em only} the spin susceptibility is corrected by $\lambda_{\mathrm{m}}$ while $\lambda_{\mathrm{d}}\!=\!0$.}
\label{fig:magnetic_susceptibility_omega_q}
\end{figure*}
\end{center}

\subsection{Correlation length}
\label{sec:correlation_length}

In this section, we investigate the correlation length $\xi_{\mathrm{d}}$ of charge fluctuations as a function of the filling $n$.
$\xi_{\mathrm{d}}$ is obtained from a fit of our numerical data for $\chi_{\mathrm{d},\bq}^{\omega=0}$ to the following Ornstein-Zernicke form\cite{Ornstein1914}:
\begin{equation}
\label{equ:OrnsteinZernicke}
\chi_{r,\mathbf{q}}^{\omega=0}= \frac{A_d}{{(\bq-\bq_\text{max})^{2}+\xi_d^{-2}}} \,,
\end{equation}
where $\bq_\text{max}$ is the momentum at which the charge susceptibility takes its maximum.
In Fig.~\ref{fig:correlation_length} results are shown both for DMFT (red diamonds) and for the $\ldgadm$ approach (green squares).
The different background colors indicate different values of $\bq_\text{max}$ where $\chi_{\mathrm{d},\bq}^{\omega=0}$ assumes its maximum. 
The specific values of $\bq_\text{max}$ (or, more precisely, of the $q_x$ component of $\bq_\text{max}$ for $q_y\!=\!q_x$) is depicted in the inset of Fig.~\ref{fig:correlation_length}  as a function of the filling $n$.
In the case of DMFT the charge correlation length of uniform charge fluctuations $\bq_\text{max}\!=\!(0,0)$ is largest close to half filling, where the corresponding susceptibility is also strongly enhanced. 
Upon increasing the doping, the correlation length rapidly decreases and eventually becomes negligible around $n\!\approx\!1.18$.
This is consistent with the broadening of the peak around $\bq\!=\!(0,0)$ in $\chi_{\mathrm{d},\bq}^{\omega=0}$ with increasing doping observed in Figs.~\ref{fig:chid_vs_q_1}-\ref{fig:chid_vs_q_5}.
As already mentioned, for the filling $n \simeq 1.1652$ the sharp maximum becomes essentially flat around $\bq\!=\!(0,0)$ [Fig.~\ref{fig:chid_vs_q_5}]. For this filling, the fitting results are less precise and reliable compared to cases where a pronounced maximum of the charge susceptibility as a function of $\bq$ is obtained, but they still capture qualitatively the suppression of the correlation length.
In fact, fitting a weakly momentum dependent data set for $\chi_{d,\mathbf{q}}^{\omega=0}$ to the Ornstein-Zernicke Eq. \eqref{equ:OrnsteinZernicke} requires a value for $\xi^{-2}$ which is much larger than $(\mathbf{q}-\mathbf{q}_\text{max})^2$, which leads to an almost vanishing correlation length.
At $n\!\approx\!1.18$, the maximum of $\chi_{\mathrm{d},\bq}^{\omega=0}$ changes from $\bq\!=\!(0,0)$ to an incommensurate value.  After this shift, the peak around $\bq_{\mathrm{max}}$ is broadened and remains nearly flat, which again implies a strong reduction of the correlation length as discussed above.
With a further increase of the filling $n$, the maximum of $\chi_{\mathrm{d},\bq}^{\omega=0}$ eventually reaches $\bq\!=\!(\pi,\pi)$ at around $n\!\approx\!1.28$, indicating the dominance of a kind of checkerboard type order as discussed in the previous section. For these fillings, we again obtain a pronounced maximum of $\chi_{\mathrm{d},\bq}^{\omega=0}$ around $\bq_{\mathrm{max}}$, and therefore our fit is reliable once more. This could explain the apparent discontinuous behavior around $n\!\approx\!1.28$.

Let us now compare these DMFT data to corresponding $\ldgadm$ results.
Close to half filling the $\ldgadm$ correlation length is significantly suppressed with respect to DMFT in accordance with the very small value of the associated susceptibility due to the mutual renormalization between spin and charge channels. 
Upon increasing $n$ the correlation length also gets larger, reaches a maximum of $\xi \simeq 0.5173$ at $n \simeq 1.0330$, and eventually becomes slightly larger than the DMFT correlation length for fillings $n > 1.0598$, which has also been observed for the susceptibility in Fig.~\ref{fig:charge_susceptibility}.
For even larger values of $n$ the $\ldgadm$ data closely follow the DMFT results where also the momentum vector at which the charge susceptibility attains its maximum changes at the same fillings as in DMFT.
This can be easily understood by the fact a momentum-independent $\lambda$ correction can only rescale the entire  DMFT charge susceptibility but is not able to invert its relative magnitude at different $\bq$ points.

\begin{center}
\begin{figure*}[t!]
\subfigure[~$n=1.0207$]{
\label{fig:ReSigma_1}
\begin{minipage}[b]{0.235\textwidth}
\centering 
\includegraphics[width=1\textwidth]{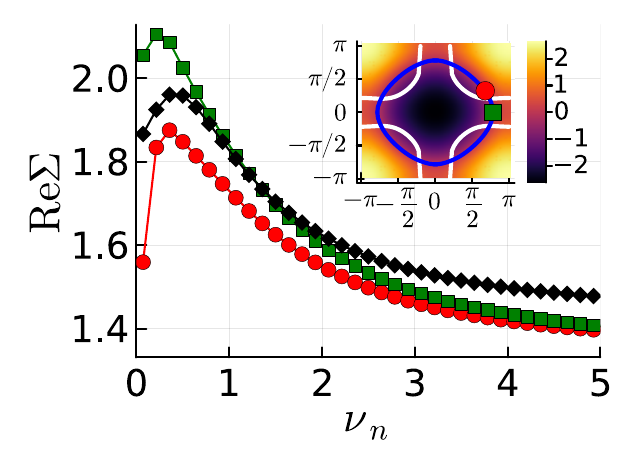}
\end{minipage}}
\subfigure[~$n=1.0496$]{
\label{fig:ReSigma_2}
\begin{minipage}[b]{0.235\textwidth}
\centering 
\includegraphics[width=1\textwidth]{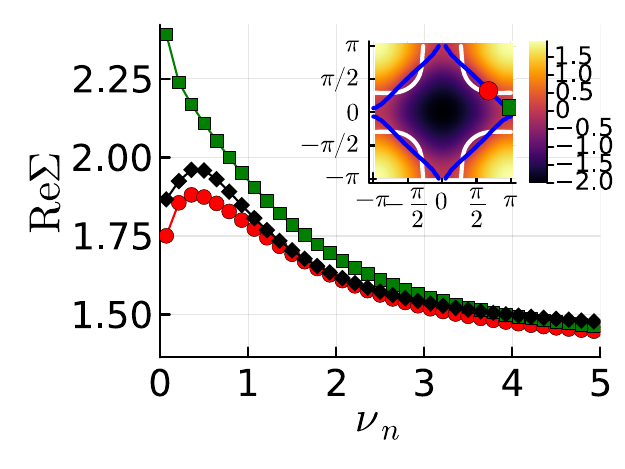}
\end{minipage}}
\subfigure[~$n=1.0598$]{
\label{fig:ReSigma_3}
\begin{minipage}[b]{0.235\textwidth}
\centering 
\includegraphics[width=1\textwidth]{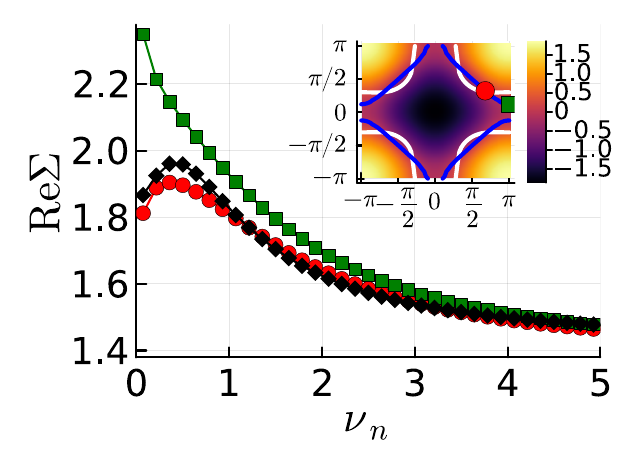}
\end{minipage}}
\subfigure[~$n=1.0850$]{
\label{fig:ReSigma_4}
\begin{minipage}[b]{0.235\textwidth}
\centering 
\includegraphics[width=1\textwidth]{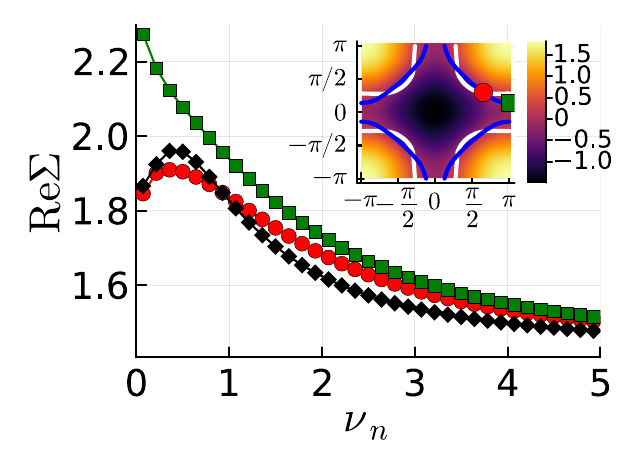}
\end{minipage}}
\\
\subfigure[~$n=1.0207$]{
\label{fig:ImSigma_1}
\begin{minipage}[b]{0.235\textwidth}
\centering 
\includegraphics[width=1\textwidth]{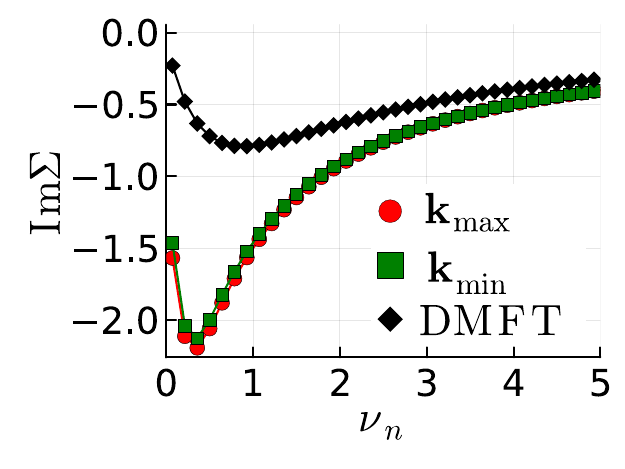}
\end{minipage}}
\subfigure[~$n=1.0496$]{
\label{fig:ImSigma_2}
\begin{minipage}[b]{0.235\textwidth}
\centering 
\includegraphics[width=1\textwidth]{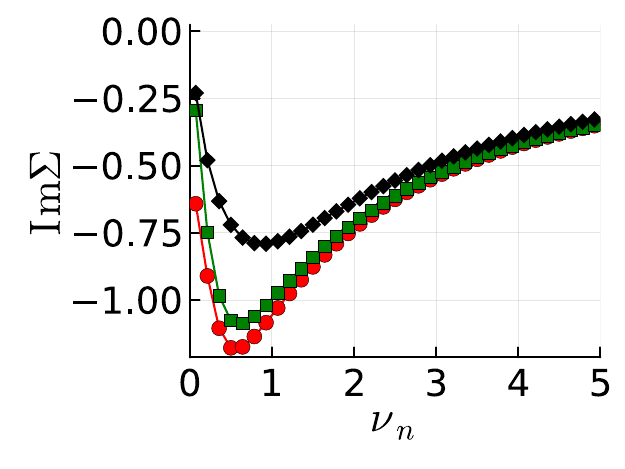}
\end{minipage}}
\subfigure[~$n=1.0598$]{
\label{fig:ImSigma_3}
\begin{minipage}[b]{0.235\textwidth}
\centering 
\includegraphics[width=1\textwidth]{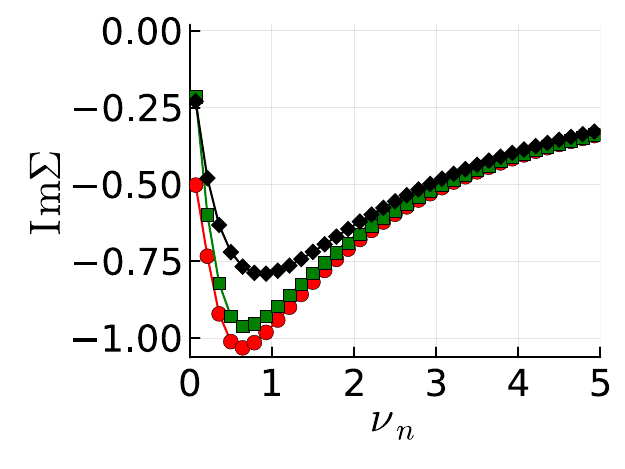}
\end{minipage}}
\subfigure[~$n=1.0850$]{
\label{fig:ImSigma_4}
\begin{minipage}[b]{0.235\textwidth}
\centering 
\includegraphics[width=1\textwidth]{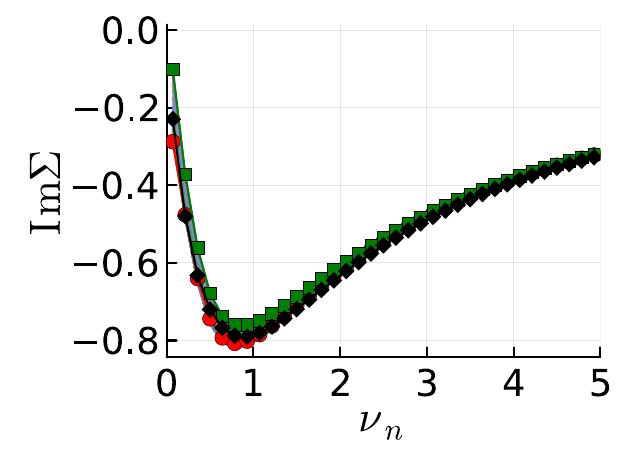}
\end{minipage}}
\caption{Real (a)–(d) and imaginary (e)–(h) parts of the $\ldgadm$ self-energy for $\bk$-points on the Fermi surface compared to the local DMFT self-energy (black diamonds}). In particular, we present results for $\bk_\text{max}$ and $\bk_\text{min}$, highlighted with red circles and green squares, respectively. 
Inset [upper panels, (a)–(d)]: Color-coded interacting spectrum $E_\bk - \mu_{\mathrm{D}\Gamma\mathrm{A}}$, as defined in Eq.~\eqref{equ:interacting_Spectrum}.  
$\bk_\text{min}$ and $\bk_\text{max}$ points, which lie close to the antinodal and nodal points, respectively, are marked using the same symbols as in the self-energy plots. Calculation parameters are identical to those in Fig.~\ref{fig:charge_susceptibility}.
\label{fig:self-energy}
\end{figure*}
\end{center}

\subsection{Spin susceptibility}
\label{sec:spin_susceptibility}

We now turn our attention to the spin susceptibility $\chi^{\lambda_{\mathrm{m}},\omega}_{\mathrm{m},\bq}$, which probes magnetic fluctuations in the system. 
Although long-range antiferromagnetic order is absent in two dimensions at finite temperature due to the Mermin-Wagner theorem, antiferromagnetic correlations remain the dominant mechanism in the repulsive Hubbard model. 
We therefore show in Fig.~\ref{fig:mag_susceptibility} the static antiferromagnetic susceptibility $\chi^{\lambda_{\mathrm{m}},\omega=0}_{\mathrm{m},\bq_\pi}$ as a function of filling $n$ where $\bq_\pi = (\pi,\pi)$ is the antiferromagnetic ordering vector.
Data are presented for $\ldgam$, where $\lambda_{\mathrm{d}}\!=\!0$ (brown hexagons) and $\ldgadm$, where both spin and charge fluctuations are renormalized (green squares).
In both cases, the antiferromagnetic spin susceptibility exhibits a maximum at half filling and decreases monotonously with $n$.
For $n\!<\!1.0544$, we observe larger values of $\chi^{\lambda_{\mathrm{m}},\omega=0}_{\mathrm{m},\bq_\pi}$ for $\ldgadm$ with respect to $\ldgam$.
This is consistent with the fact that charge fluctuations are suppressed in this parameter regime via a positive $\lambda_{\mathrm{d}}$.
In fact, if $\chi^{\lambda_{\mathrm{d}},\omega=0}_{\mathrm{d},\bq}$ is reduced in Eq.~\eqref{equ:lambdacondition1} by $\lambda_{\mathrm{d}}\!>\!0$, we need a larger $\chi^{\lambda_{\mathrm{m}},\omega=0}_{\mathrm{m},\bq}$ to fulfill this consistency relation compared to the case of $\ldgam$, where $\lambda_{\mathrm{d}}\!=\!0$, which requires a stronger renormalization in the spin channel.
The situation is reversed for $n\!>\!1.0544$ where $\lambda_{\mathrm{d}}\!<\!0$.
In this case, charge fluctuations are enhanced with respect to DMFT (see Fig.~\ref{fig:charge_susceptibility}), which requires an even larger value of $\lambda_{\mathrm{m}}$ compared to $\ldgam$ to fulfill Eq.~\eqref{equ:lambdacondition1}.
Hence, in this parameter regime $\ldgadm$ predicts smaller spin fluctuations than the older $\ldgam$ approach.
This behavior nicely illustrates the rebalancing of spin and charge fluctuations which is achieved by enforcing the sum rule in Eq.~\eqref{equ:lambdacondition2} in addition to Eq.~\eqref{equ:lambdacondition1} in the framework of $\ldgadm$.
This picture is also reflected in the frequency dependence of the local (momentum summed) antiferromagnetic spin susceptibility $\chi^{\lambda_r,\omega}_{\mathrm{m}} =\sum\limits_\bq \chi^{\lambda_r,\omega}_{\mathrm{m},\bq}$ shown in the upper row of Fig.~\ref{fig:magnetic_susceptibility_omega_q} [Figs.~\ref{fig:chim_omega_1}-\ref{fig:chim_omega_4}], where for $n\!<\!1.0544$ [Figs.~\ref{fig:chim_omega_1}-\ref{fig:chim_omega_2}] the peak at $\omega\!=\!0$ is larger for $\ldgadm$ with respect to $\ldgam$, while the opposite is true for $n\!>\!1.0544$ [Figs.~\ref{fig:chim_omega_3}-\ref{fig:chim_omega_4}].
In the lower row of Fig.~\ref{fig:magnetic_susceptibility_omega_q} we depict the spin susceptibility at $\omega\!=\!0$ for four selected densities as a function of $\bq\!=\!(q_x,q_y$). 
We observe a maximum at the antiferromagnetic wave vector $\bq = \bq_{\boldsymbol{\pi}}$, confirming that antiferromagnetic spin fluctuations dominate near half filling. For fillings beyond $n > 1.35$, the maximum shifts to $\bq = (\pi, 0)$ and $\bq = (0, \pi)$, indicating a tendency toward stripe order.

Let us finally remark that we do not show any DMFT results for the (antiferromagnetic) spin susceptibility as this correlation function is already negative at $\beta\!=\!44$, which is an agreement with the antiferromagnetically ordered phase which DMFT predicts in this parameter regime.

\subsection{Self-energy}
\label{sec:self-energy}

Next, we discuss the momentum-dependent self-energy of $\ldgadm$ and compare it to the local one obtained from DMFT at four selected densities for which the charge and spin susceptibilities are also shown in Figs.~\ref{fig:charge_susceptibility_omega_q} and \ref{fig:magnetic_susceptibility_omega_q}. 
To provide an overview over the momentum dependence of this correlation function we show its imaginary part at the first positive Matsubara frequency $\nu_0\!=\!\frac{\pi}{\beta}$, $\mathrm{Im}\Sigma_\mathbf{q}^{\nu_0}$, 
as a heat map in the $(k_x,k_y)$ plane in Fig.~\ref{fig:self-energy_bk} of Appendix~\ref{sec:self-energy_k}. 
We are, however, particularly interested in this correlation function for $\bk$ points on the Fermi surface where the largest phase space for scattering between the electrons is available.
In interacting systems, the Fermi surface is defined by the zeros of the real part of the retarded Green's function $\mathrm{Re}G_{\bk}^\omega$ at zero (real) frequency $\omega\!=\!0$. 
We, hence, define a renormalized dispersion as
\begin{align}
\label{equ:interacting_Spectrum}
E_\bk = \epsilon_\bk + \mathrm{Re}\Sigma_{\bk}^{\omega=0}\,,
\end{align}
where $\mathrm{Re}\Sigma_{\bk}^{\omega}$ is the retarded self-energy as a function of real frequencies $\omega$, which can be obtained from the associated Matsubara self-energy $\Sigma_\bk^\nu$ via an analytic continuation $i\nu\!\rightarrow\!\omega$.
The Fermi surface is then determined by the condition $E_\bk = \mu_{\mathrm{D}\Gamma\mathrm{A}}$, where the chemical potential is determined via Eq.~\eqref{equ:fixchemicalpotential} to fix the density within $\ldgadm$ to the desired value.
Note that in the nonbipartite case, $\mu_{\mathrm{D}\Gamma\mathrm{A}}$ typically differs from the corresponding chemical potential in DMFT as nonlocal correlations renormalize the energy scales in the system.

In the upper row of Fig.~\ref{fig:self-energy}, we show the real part of the Matsubara self-energy as a function of the Matsubara frequency $\nu$. 
We have extrapolated these curves to zero frequency using a quadratic fit to obtain $E_\bk$ defined in Eq.~\eqref{equ:interacting_Spectrum}.
The insets show the function $E_\bk\!-\!\mu_{\mathrm{D}\Gamma\mathrm{A}}$ as a heat map in the $(k_x,k_y)$ plane where we have indicated the interacting Fermi surface of $\ldgadm$, defined by $E_\bk\!-\!\mu_{\mathrm{D}\Gamma\mathrm{A}}\!=\!0$, with a blue solid line (for reference the Fermi surface of the noninteracting system is shown by a white solid line).
We observe that while the Fermi surfaces for larger doping [Figs.~\ref{fig:ReSigma_2}-\ref{fig:ReSigma_4}] are all hole like (as the ones for the noninteracting system), an electronlike Fermi surface is found for the lowest doping $n\!=\!1.0207$ in Fig.~\ref{fig:ReSigma_1}.
Such a Lifshitz-type transition is purely driven by electronic correlations and has also been found in studies of the Hall conductivity of the Hubbard model in Refs.~\onlinecite{Rojo_1993,Markov_2019}, where a change of majority charge carriers is signaled by a sign change of $\sigma_{xy}$.

To proceed with our discussion of the self-energy, we have identified the two $\bk$ points on the Fermi surface, $\bk_{\mathrm{max}}$ and $\bk_{\mathrm{min}}$, where the absolute value of the imaginary part of the Matsubara self-energy $\Sigma_\bk^\nu$ takes its maximum (red squares) and its minimum (green circles), respectively.
Note that for the electron-doped case considered here, the maximum is closer to the so-called nodal point $\bk\!=\!\left(\frac{\pi}{2},\frac{\pi}{2}\right)$ while the minimum is in the vicinity of the antinodal point $\bk\!=\!(\pi,0)$ which is the opposite behavior compared to the hole-doped case \cite{Senechal_Tremblay_2004,Worm_Toschi_2024}.
In the lower row of Fig.~\ref{fig:self-energy}, we depict the imaginary part of the self-energy as a function of the Matsubara frequency $\nu$ for different momenta located on the Fermi surface between the maximum (red square) and the minimum (green circle).
We clearly observe that close to half filling [Fig.~\ref{fig:ImSigma_1}], the absolute value of the imaginary part of the self-energy at the Fermi surface obtained from the $\ldgadm$ calculations is significantly larger than that from DMFT. 
This can be attributed to the scattering of electrons at strong antiferromagnetic spin fluctuations which are observed at this filling (see Fig.~\ref{fig:mag_susceptibility}).
Upon increasing the doping to $n\!=\!1.0496$ [Fig.~\ref{fig:ImSigma_2}], we observe a substantial increase in momentum differentiation which is consistent with the emergence of a pseudogap in this parameter regime (although for the definite identification of this feature the calculation of the full spectral function would be required).
For an even larger doping of $n\!=\!1.0598$ in Fig.~\ref{fig:ImSigma_3}, we observe a very interesting effect:
At the first Matsubara frequency, the minimal $\ldgadm$ self-energy (green circle) becomes even {\em smaller} than the DMFT self-energy which indicates an enhancement of metallicity at this specific $\bk$ point.
This is consistent with the fact that charge fluctuations start to get enhanced in $\ldgadm$ with respect to DMFT at exactly this filling (see Fig.~\ref{fig:charge_susceptibility}).
This trend continues to the largest doping in Fig.~\ref{fig:ImSigma_4}, where the minimal self-energy of $\ldgadm$ becomes smaller than the DMFT one at all Matsubara frequencies.
Let us, however, note that for this doping value the maximal self-energy is also already rather close to its DMFT counterpart indicating that the effect of nonlocal correlations gradually decreases with increasing particle number.

\begin{figure}[t!]
\centering
\includegraphics[width=0.99\linewidth]{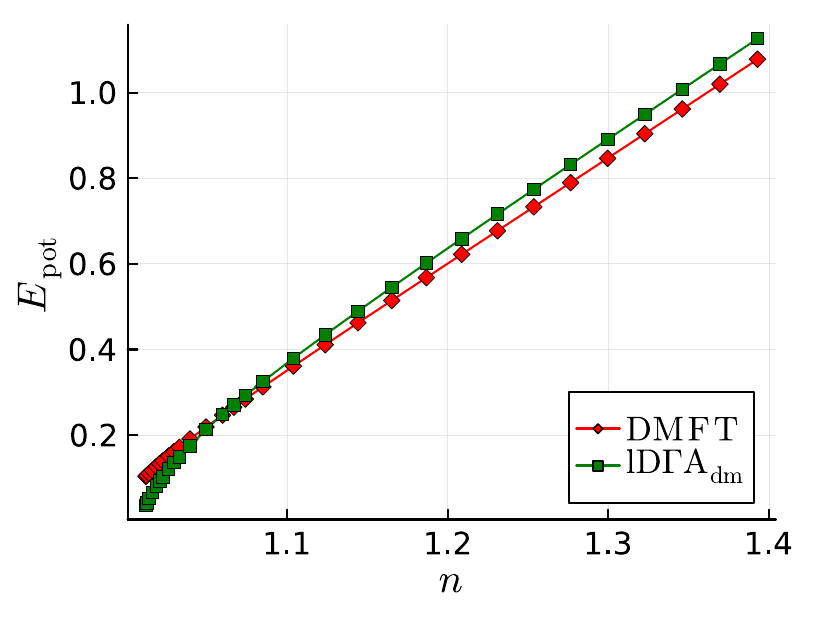}
\caption{Potential energy $E_\text{pot}$ as a function of filling $n$ obtained from DMFT (red diamonds) and $\ldgadm$ (green squares).
DMFT data are calculated at the one-particle level ($E_\text{pot}^{(1)}$) while for $\ldgadm$ the potential energies at the one- and two-particle levels are equivalent, which is enforced by Eq.~\eqref{equ:lambdacondition2}.
Parameters are the same as in Fig.~\ref{fig:charge_susceptibility}. 
}
\label{fig:potential_energy}
\end{figure}

\begin{figure}[t!]
\centering
\includegraphics[width=0.99\linewidth]{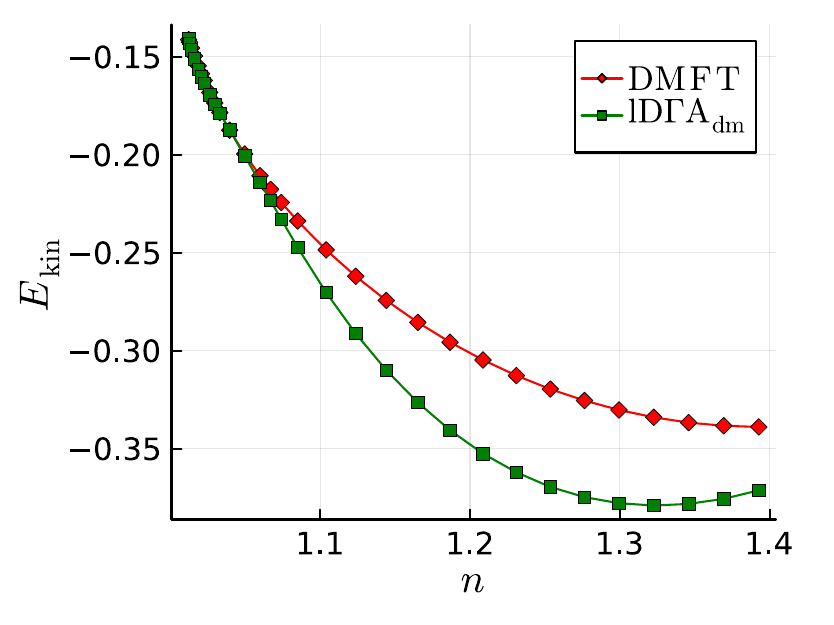}
\caption{Kinetic energy $E_\text{kin}$ as a function of filling $n$ obtained from DMFT (red diamonds) 
and $\ldgadm$ (green squares). Parameters are the same as in Fig.~\ref{fig:charge_susceptibility}. 
}
\label{fig:kinetic_energy}
\end{figure}

Let us finally remark, that the $\ldga$ self-energy obtained via Eq.~\eqref{equ:self-emergy_susc} can feature tiny violations of the correct high-frequency asymptotic behavior which makes some kind of tail correction necessary.
While all qualitative (and also almost all quantitative) results of our calculations are unaffected by the concrete choice of the tail correction scheme, the height of the maximum of the uniform $\ldgadm$ charge susceptibility as a function of $n$ shown in the inset of Fig.~\ref{fig:charge_susceptibility} can change depending on the  method used (while its location remains roughly at the same density $n$). More technical details about the tail corrections of the self-energy will be provided in 
Ref.~\cite{Stobbe_2025prep}.

\subsection{Potential and kinetic energies}
\label{sec:potential_kinetic}

We finally discuss the behavior of the potential and kinetic energies of DMFT and $\ldgadm$ 
as a function of filling, shown in Figs.~\ref{fig:potential_energy} and \ref{fig:kinetic_energy}, respectively.
As for the potential energy in Fig.~\ref{fig:potential_energy}, let us note that this thermodynamic observable can be calculated from the one-particle self-energy and Green's function ($E_\text{pot}^{(1)}$) or from the two-particle spin and charge susceptibilities ($E_\text{pot}^{(2)}$) [see Eq.~\eqref{equ:sumruleepot}].
For DMFT, we present only $E_\text{pot}^{(1)}$ since $E_\text{pot}^{(2)}$ is not meaningful as we are already in the symmetry-broken phase of this method where the spin susceptibility is negative, which makes it unsuitable as an input for the calculation of other thermodynamic observables \cite{vanLoon_2016, Rohringer_Toschi_2016}.
For the $\ldgadm$ approach, on the other hand, the equality of both potential energies is enforced by our $\lambda$ correction method.
We observe that for low values of the doping, the potential energy of $\ldgadm$ is smaller than the corresponding DMFT result.
This can be easily understood by the fact that strong spin fluctuations make the system more insulating in this parameter regime and, hence, suppress double occupancies of two electrons at one lattice site.
Upon increasing the filling, the $\ldgadm$ potential energy gradually approaches the corresponding DMFT from below and eventually becomes larger at $n\!=\!1.0598$, where also larger charge fluctuations are predicted by $\ldgadm$ with respect to DMFT.
This result provides further evidence that at this filling nonlocal fluctuations lead to a delocalization of electrons which is not captured within the local DMFT approach.
Such an interpretation is also supported by the behavior of the kinetic energy in Fig.~\ref{fig:kinetic_energy}.
This thermodynamic variable is small (in absolute value) close to half filling, where correlation effects suppress the mobility of the particles and increases with increasing doping.
This increase is stronger in $\ldgadm$ compared to DMFT, which again signals a delocalization of electrons due to nonlocal correlations.

\section{Conclusion and outlook}
\label{sec:conclusion}

Employing the ladder dynamical vertex approximation, we have investigated the effect of nonlocal correlations in the two-dimensional Hubbard model as a function of (electron) doping for an interaction parameter and a temperature at which the system is a Mott insulator at half filling and DMFT predicts a substantial enhancement of uniform charge fluctuations for small doping. 
Our $\ldgadm$ approach, which enforces a consistent evaluation of the potential energy from one- and two-particle quantities by means of an effective renormalization of DMFT charge and spin susceptibilities, predicts a strong suppression of these charge fluctuations by strong nonlocal antiferromagnetic spin fluctuations.
This is a consequence of the rebalancing between these two types of fluctuations by the enforcement of sum rules for the Pauli principle and the potential energy within the $\ldgadm$ approach.
For larger fillings, we observe the opposite trend, where nonlocal correlations induced by $\ldgadm$ enhance the uniform charge fluctuations with respect to DMFT indicating an increase of the electron mobility.
This scenario is also consistent with the large value of the antiferromagnetic susceptibility at half filling and its strong decay with increasing doping.
The picture outlined above is also reflected in the correlation length which is strongly reduced in $\ldgadm$ with respect to DMFT at small fillings $n$.

We also investigated the behavior of the self-energy at the Fermi surface. 
Close to half filling, we find that the absolute value of the imaginary part of the self-energy $\Sigma_{\bk}^{\nu}$ obtained from $\ldgadm$ is larger than that from DMFT for all $\bk$ points on the Fermi surface reflecting the impact of the strong nonlocal (spin) fluctuations on this correlation function. 
On the contrary, for $n > 1.0544$ the imaginary part of the DMFT self-energy becomes larger than that from $\ldgadm$ for some $\bk$ points. 
This again reflects the trend observed in the charge susceptibility where a metallization due to nonlocal correlations is found by $\ldgadm$.
Finally, this picture is also confirmed by the analysis of the doping dependence of the potential and kinetic energies, which also shows a suppression of electron mobility at low doping and a corresponding enhancement at larger fillings induced by nonlocal correlation.

It would be indeed very interesting to understand how nonlocal correlation effects and the consistent treatment of charge and spin fluctuations affect the physics of the two-dimensional Hubbard model in a broader parameter regime.
A particularly important question is how the consistent charge renormalization might affect the superconducting state at optimal doping, as this can potentially guide the way larger critical temperatures can be achieved.
This is certainly a very interesting future research direction.

{\sl Acknowledgements.} We thank A. Toschi and E. Moghadas for insightful discussions. We acknowledge financial support from the Deutsche Forschungsgemeinschaft (DFG) through
Projects No. 407372336 and No. 449872909. The authors gratefully acknowledge the computing time granted by the Resource Allocation Board and provided on the supercomputers Lise and Emmy/Grete at NHR@ZIB and NHR@Göttingen as part of the NHR infrastructure. The calculations for this research were conducted with computing resources under Project No. hhp00048.

\begin{center}
\begin{figure*}[tb!]
\subfigure[~$n=1.0207$]{
\label{fig:ImSigma_1k}
\begin{minipage}[b]{0.235\textwidth}
\centering 
\includegraphics[width=1\textwidth]{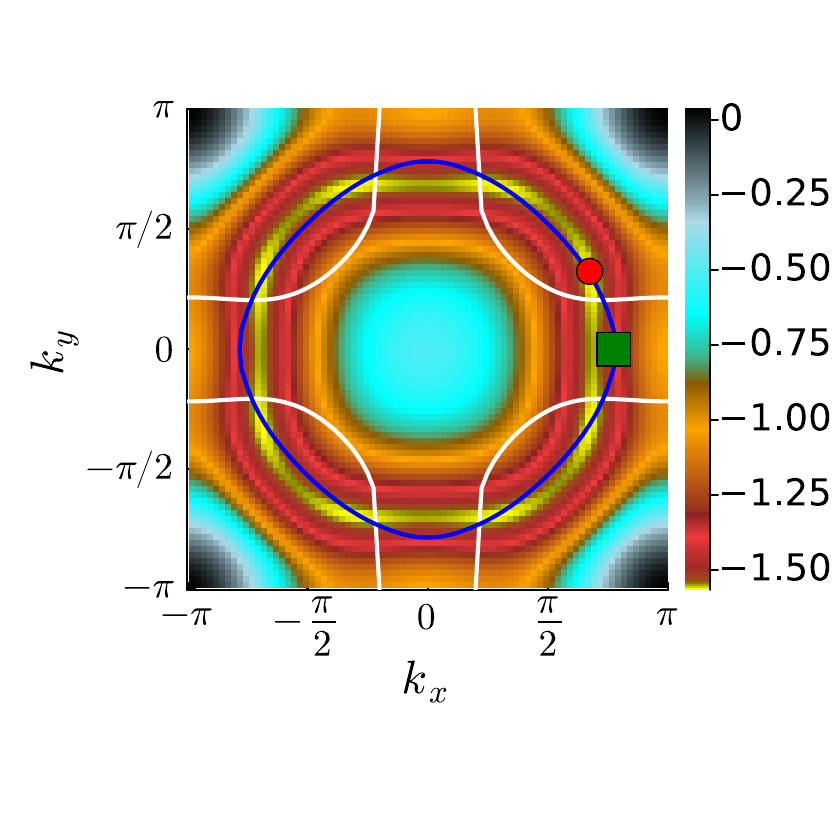}
\end{minipage}}
\subfigure[~$n=1.0496$]{
\label{fig:ImSigma_2k}
\begin{minipage}[b]{0.235\textwidth}
\centering 
\includegraphics[width=1\textwidth]{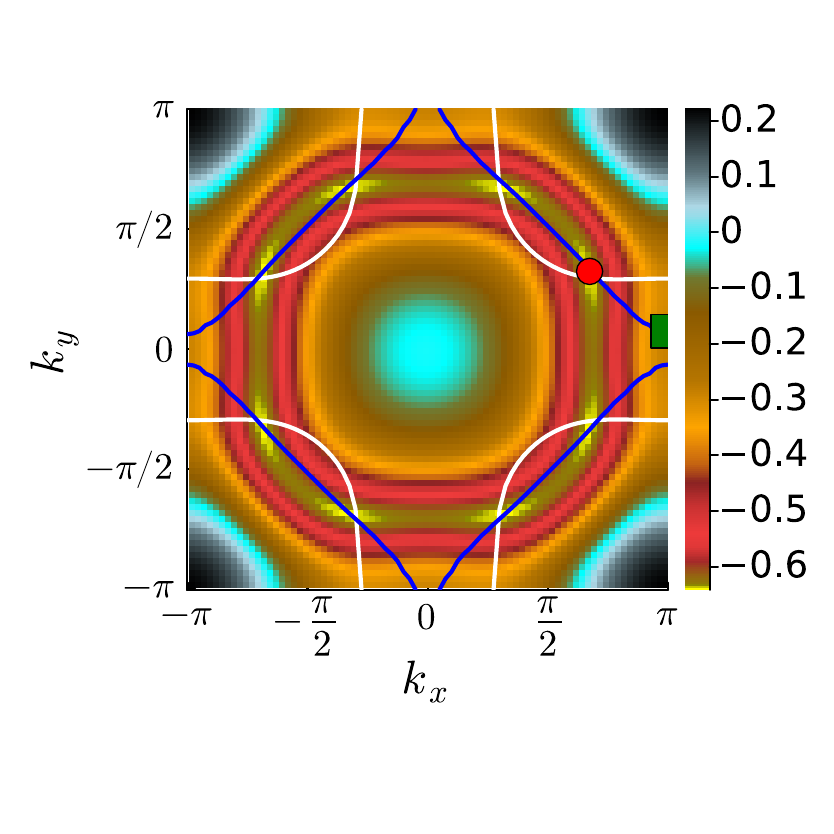}
\end{minipage}}
\subfigure[~$n=1.0598$]{
\label{fig:ImSigma_3k}
\begin{minipage}[b]{0.235\textwidth}
\centering 
\includegraphics[width=1\textwidth]{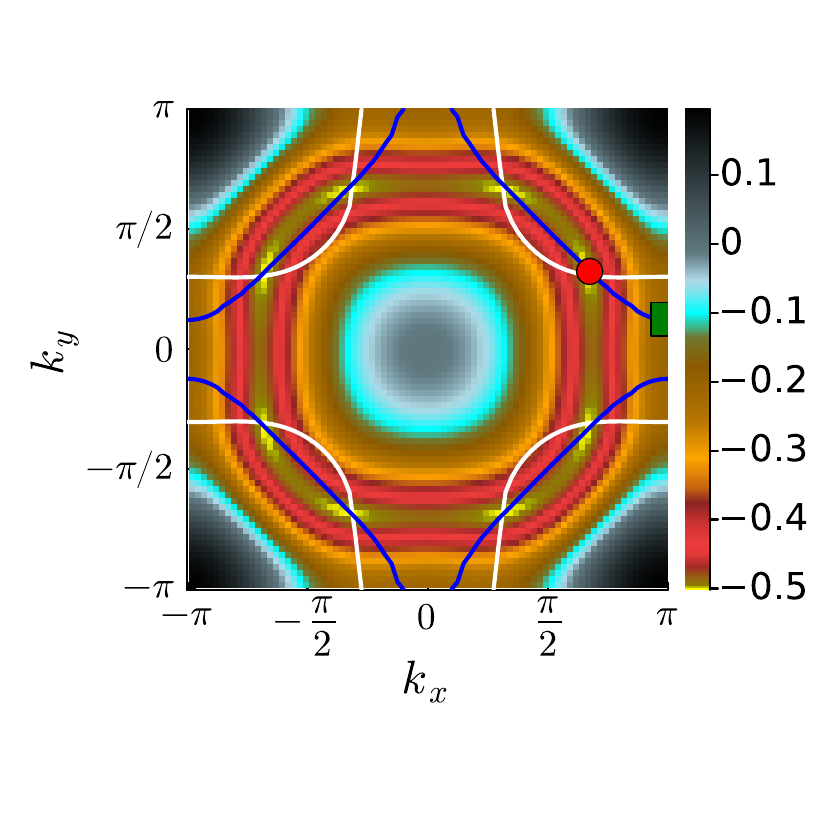}
\end{minipage}}
\subfigure[~$n=1.0850$]{
\label{fig:ImSigma_4k}
\begin{minipage}[b]{0.235\textwidth}
\centering 
\includegraphics[width=1\textwidth]{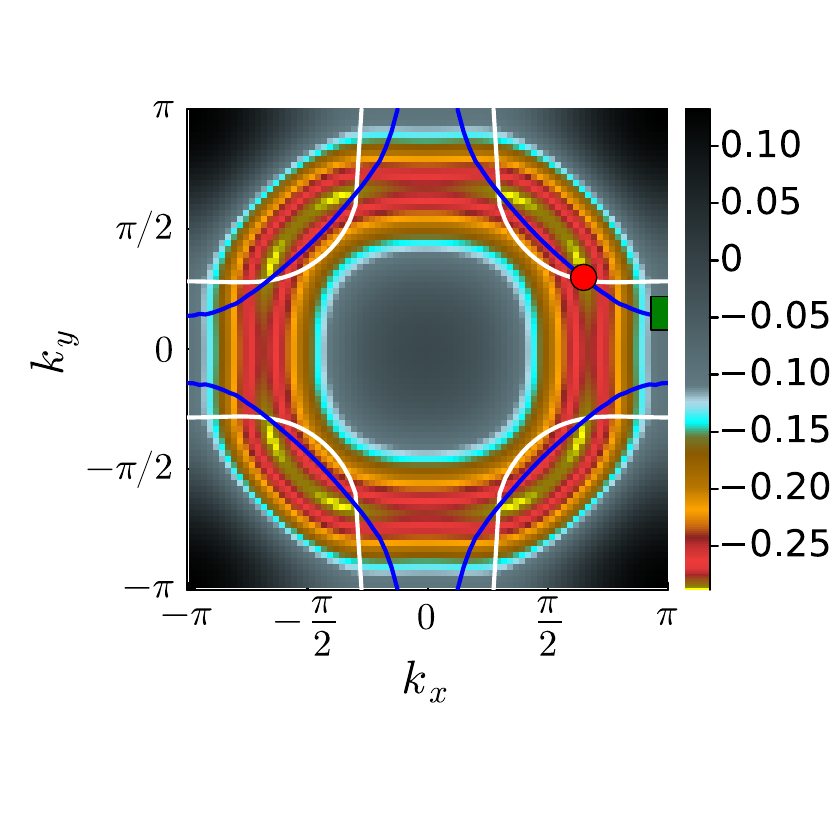}
\end{minipage}}
\caption{
Imaginary part of the self-energy for $\nu\!=\!\nu_0\!=\!\pi/\beta$ as a heat map in the $(k_x,k_y)$ plane for four different fillings $n$ in (a) to (d).
$\bk_\text{min}$ and $\bk_\text{max}$ points, which lie close to the antinodal and nodal points, respectively,  are marked using the same symbols as in the self-energy plots in Fig.~\ref{fig:self-energy}. The blue lines indicate the Fermi surface predicted by $\ldgadm$ while the noninteracting Fermi surface is shown by white lines. Calculation parameters are identical to those in Fig. \ref{fig:charge_susceptibility}.}
\label{fig:self-energy_bk}
\end{figure*}
\end{center}

\appendix

\section{Fitting function for \texorpdfstring{$\chi^{\lambda_{\mathrm{d}},\omega_0}_{\mathrm{d},\bq_0}$}{Chid(omega=0,q=0)}}
\label{sec:fitting}

The behavior of the susceptibility $\chi^{\lambda_{\mathrm{d}},\omega_0}_{\mathrm{d},\bq_0}$ as a function of filling $n$ is well described by the following fit function:
\begin{align}
\label{equ:fitting_chid}
\chi^{\lambda_{\mathrm{d}},\omega_0}_{\mathrm{d},\bq_0} &= \frac{A}{e^{\frac{B}{(n-1)(2-n)}} + e^{\frac{C}{(1-n)n}}} \,,
\end{align}
where $A = 0.6307$, $B = 0.0658$, and $C = 0.1473$.

This somewhat elaborate functional form is motivated by the distinct behavior of the susceptibility near the limiting cases of the electronic filling $n$. 
Close to half filling ($|n-1| \ll 1$), the susceptibility is exponentially suppressed. In this regime, a functional form of the type $\chi^{\lambda_{\mathrm{d}},\omega_0}_{\mathrm{d},\bq_0} \sim e^{-B/|n-1|}$ provides a reasonable approximation up to $n \simeq 1.05$.
However, this simple form fails to capture the correct physical behavior further away from half filling and, in particular, for the limits $n\!\rightarrow\!0$ and $n\!\rightarrow\!2$ where the susceptibility is expected to vanish. 
To account for this, we have modified the fitting function in the following way:
To capture the suppression of the susceptibility near full filling ($n \to 2$) a term of the form $e^{-B/(n-1)(2-n)}$ is required. 
On the other hand, to correctly reproduce the behavior near empty filling ($n \to 0$), a term proportional to $e^{-C/(1-n)n}$ becomes necessary.
Equation \eqref{equ:fitting_chid} effectively combines both limiting behaviors, providing an accurate fit across the full range of fillings which accurately describes our numerical data within $n\!\in\![1,1.4]$.

\section{Self-energy}
\label{sec:self-energy_k}

In Fig.~\ref{fig:self-energy_bk}, we present the imaginary part of the self-energy, $\Sigma_\bk^\nu$ at the first positive Matsubara frequency $\nu = \nu_0 = \pi/\beta$ for different electron fillings $n$ as a heat map in the $(k_x,k_y)$ plane. 
As expected, we observe that the absolute value of the imaginary part of the self-energy is largest at and near the Fermi surface where a relatively large phase space is available for scattering. 
This behavior is most pronounced close to half filling for $n \simeq 1.0207$ [Fig.~\ref{fig:ImSigma_1k}], where the Fermi surface is electron-like (see also discussion in the main text). 
For this small deviation of the electron density from $n\!=\!1$, the imaginary part of the self-energy remains negative for all momenta $\bk$. 
Its absolute value is minimal at the edges of the Brillouin zone $\bk = (\pm \pi, \pm \pi)$ and increases upon approaching the Fermi surface, where it takes its maximum values. 
Moving further to the center of the Brillouin zone $\bk = (0, 0)$, the magnitude of the self-energy again decreases which is compatible with the reduction of available phase space for scattering.

Upon increasing the filling $n$, the (absolute value of the) imaginary part of the self-energy becomes smaller, which is consistent with the fact that the system becomes gradually less correlated for increased doping [Figs.~\ref{fig:ImSigma_2k}-\ref{fig:ImSigma_4k}].
At the Brillouin zone boundaries around $\bq\!~\sim\!(\pi,\pi)$, we even observe an unphysical overshooting where the imaginary part of the Matsubara self-energy becomes positive corresponding to a violation of causality.
This behavior, which indicates that D$\Gamma$A is less reliable far away from the Fermi surface has been observed in previous works even for the full parquet based version of the method \cite{Vali_Toschi_2015}.

\section{Codes and Data}
\label{sec:codes_and_data}

The main $\ldga$ code requires different one- and two-particle quantities as input, all condensed into a single file with the help of Ref.~\cite{VertexPostprocessing}

\begin{itemize}
    \item $U, \beta, \mu$
    \item Anderson Parameters from exact diagonalization (we use Ref.~\cite{jED}) 
    \item Two-Particle Green's function (we use an in-house ED solver)
    \item Matsubara frequency grid (Ref.~\cite{EquivalenceClassesConstructor}
    generates symmetry reduced grids)
\end{itemize}

To run  $\ldga$ \cite{LadderDGA}  
two additional modules are required: Refs. \cite{Dispersions} and \cite{BSE_Asymptotics}
With all packages and data in place, the $\ldga$ code provides functions to calculate all the desired $\lambda$-corrections and the resulting non-local self energies.

All data can be found at Ref.~\cite{Our_Data}. 


\begin{thebibliography}{103}%
\makeatletter
\providecommand \@ifxundefined [1]{%
 \@ifx{#1\undefined}
}%
\providecommand \@ifnum [1]{%
 \ifnum #1\expandafter \@firstoftwo
 \else \expandafter \@secondoftwo
 \fi
}%
\providecommand \@ifx [1]{%
 \ifx #1\expandafter \@firstoftwo
 \else \expandafter \@secondoftwo
 \fi
}%
\providecommand \natexlab [1]{#1}%
\providecommand \enquote  [1]{``#1''}%
\providecommand \bibnamefont  [1]{#1}%
\providecommand \bibfnamefont [1]{#1}%
\providecommand \citenamefont [1]{#1}%
\providecommand \href@noop [0]{\@secondoftwo}%
\providecommand \href [0]{\begingroup \@sanitize@url \@href}%
\providecommand \@href[1]{\@@startlink{#1}\@@href}%
\providecommand \@@href[1]{\endgroup#1\@@endlink}%
\providecommand \@sanitize@url [0]{\catcode `\\12\catcode `\$12\catcode
  `\&12\catcode `\#12\catcode `\^12\catcode `\_12\catcode `\%12\relax}%
\providecommand \@@startlink[1]{}%
\providecommand \@@endlink[0]{}%
\providecommand \url  [0]{\begingroup\@sanitize@url \@url }%
\providecommand \@url [1]{\endgroup\@href {#1}{\urlprefix }}%
\providecommand \urlprefix  [0]{URL }%
\providecommand \Eprint [0]{\href }%
\providecommand \doibase [0]{http://dx.doi.org/}%
\providecommand \selectlanguage [0]{\@gobble}%
\providecommand \bibinfo  [0]{\@secondoftwo}%
\providecommand \bibfield  [0]{\@secondoftwo}%
\providecommand \translation [1]{[#1]}%
\providecommand \BibitemOpen [0]{}%
\providecommand \bibitemStop [0]{}%
\providecommand \bibitemNoStop [0]{.\EOS\space}%
\providecommand \EOS [0]{\spacefactor3000\relax}%
\providecommand \BibitemShut  [1]{\csname bibitem#1\endcsname}%
\let\auto@bib@innerbib\@empty
\bibitem [{\citenamefont {Mott}(1968)}]{Mott_1968}%
  \BibitemOpen
  \bibfield  {author} {\bibinfo {author} {\bibfnamefont {N.~F.}\ \bibnamefont
  {Mott}},\ }\bibfield  {title} {\emph {\enquote {\bibinfo {title}
  {Metal-Insulator Transition},}\ }}\href {\doibase 10.1103/RevModPhys.40.677}
  {\bibfield  {journal} {\bibinfo  {journal} {Rev. Mod. Phys.}\ }\textbf
  {\bibinfo {volume} {40}},\ \bibinfo {pages} {677} (\bibinfo {year}
  {1968})}\BibitemShut {NoStop}%
\bibitem [{\citenamefont {Haule}\ and\ \citenamefont
  {Pascut}(2017)}]{Haule_2017}%
  \BibitemOpen
  \bibfield  {author} {\bibinfo {author} {\bibfnamefont {K.}~\bibnamefont
  {Haule}}\ and\ \bibinfo {author} {\bibfnamefont {G.~L.}\ \bibnamefont
  {Pascut}},\ }\bibfield  {title} {\emph {\enquote {\bibinfo {title} {Mott
  Transition and Magnetism in Rare Earth Nickelates and its Fingerprint on the
  X-ray Scattering},}\ }}\href {\doibase 10.1038/s41598-017-10374-2} {\bibfield
   {journal} {\bibinfo  {journal} {Scientific Reports}\ }\textbf {\bibinfo
  {volume} {7}},\ \bibinfo {pages} {10375} (\bibinfo {year}
  {2017})}\BibitemShut {NoStop}%
\bibitem [{\citenamefont {Moon}(2021)}]{Moon_2021}%
  \BibitemOpen
  \bibfield  {author} {\bibinfo {author} {\bibfnamefont {B.~H.}\ \bibnamefont
  {Moon}},\ }\bibfield  {title} {\emph {\enquote {\bibinfo {title}
  {Metal-insulator transition in two-dimensional transition metal
  dichalcogenides},}\ }}\href {\doibase 10.1007/s42247-021-00202-9} {\bibfield
  {journal} {\bibinfo  {journal} {Emergent Materials}\ }\textbf {\bibinfo
  {volume} {4}},\ \bibinfo {pages} {989} (\bibinfo {year} {2021})}\BibitemShut
  {NoStop}%
\bibitem [{\citenamefont {Fei}\ \emph {et~al.}(2022)\citenamefont {Fei},
  \citenamefont {Wu}, \citenamefont {Zhang},\ and\ \citenamefont
  {Yin}}]{Fei_2022}%
  \BibitemOpen
  \bibfield  {author} {\bibinfo {author} {\bibfnamefont {Y.}~\bibnamefont
  {Fei}}, \bibinfo {author} {\bibfnamefont {Z.}~\bibnamefont {Wu}}, \bibinfo
  {author} {\bibfnamefont {W.}~\bibnamefont {Zhang}}, \ and\ \bibinfo {author}
  {\bibfnamefont {Y.}~\bibnamefont {Yin}},\ }\bibfield  {title} {\emph
  {\enquote {\bibinfo {title} {Understanding the Mott insulating state in
  1T-TaS2 and 1T-TaSe2},}\ }}\href {\doibase 10.1007/s43673-022-00049-0}
  {\bibfield  {journal} {\bibinfo  {journal} {AAPPS Bulletin}\ }\textbf
  {\bibinfo {volume} {32}},\ \bibinfo {pages} {20} (\bibinfo {year}
  {2022})}\BibitemShut {NoStop}%
\bibitem [{\citenamefont {Chowdhury}\ \emph {et~al.}(2022)\citenamefont
  {Chowdhury}, \citenamefont {Georges}, \citenamefont {Parcollet},\ and\
  \citenamefont {Sachdev}}]{Chowdhury_2022}%
  \BibitemOpen
  \bibfield  {author} {\bibinfo {author} {\bibfnamefont {D.}~\bibnamefont
  {Chowdhury}}, \bibinfo {author} {\bibfnamefont {A.}~\bibnamefont {Georges}},
  \bibinfo {author} {\bibfnamefont {O.}~\bibnamefont {Parcollet}}, \ and\
  \bibinfo {author} {\bibfnamefont {S.}~\bibnamefont {Sachdev}},\ }\bibfield
  {title} {\emph {\enquote {\bibinfo {title} {Sachdev-Ye-Kitaev models and
  beyond: Window into non-Fermi liquids},}\ }}\href {\doibase
  10.1103/RevModPhys.94.035004} {\bibfield  {journal} {\bibinfo  {journal}
  {Rev. Mod. Phys.}\ }\textbf {\bibinfo {volume} {94}},\ \bibinfo {pages}
  {035004} (\bibinfo {year} {2022})}\BibitemShut {NoStop}%
\bibitem [{\citenamefont {Patel}\ \emph {et~al.}(2023)\citenamefont {Patel},
  \citenamefont {Guo}, \citenamefont {Esterlis},\ and\ \citenamefont
  {Sachdev}}]{Aavishkar_2023}%
  \BibitemOpen
  \bibfield  {author} {\bibinfo {author} {\bibfnamefont {A.~A.}\ \bibnamefont
  {Patel}}, \bibinfo {author} {\bibfnamefont {H.}~\bibnamefont {Guo}}, \bibinfo
  {author} {\bibfnamefont {I.}~\bibnamefont {Esterlis}}, \ and\ \bibinfo
  {author} {\bibfnamefont {S.}~\bibnamefont {Sachdev}},\ }\bibfield  {title}
  {\emph {\enquote {\bibinfo {title} {Universal theory of strange metals from
  spatially random interactions},}\ }}\href {\doibase 10.1126/science.abq6011}
  {\bibfield  {journal} {\bibinfo  {journal} {Science}\ }\textbf {\bibinfo
  {volume} {381}},\ \bibinfo {pages} {790} (\bibinfo {year} {2023})},\ \Eprint
  {http://arxiv.org/abs/https://www.science.org/doi/pdf/10.1126/science.abq6011}{https://www.science.org/doi/pdf/10.1126/science.abq6011}\BibitemShut
  {NoStop}%
\bibitem [{\citenamefont {Georges}\ \emph {et~al.}(2013)\citenamefont
  {Georges}, \citenamefont {Medici},\ and\ \citenamefont
  {Mravlje}}]{Georges_2013}%
  \BibitemOpen
  \bibfield  {author} {\bibinfo {author} {\bibfnamefont {A.}~\bibnamefont
  {Georges}}, \bibinfo {author} {\bibfnamefont {L.~d.}\ \bibnamefont {Medici}},
  \ and\ \bibinfo {author} {\bibfnamefont {J.}~\bibnamefont {Mravlje}},\
  }\bibfield  {title} {\emph {\enquote {\bibinfo {title} {Strong Correlations
  from Hund’s Coupling},}\ }}\href {\doibase
  https://doi.org/10.1146/annurev-conmatphys-020911-125045} {\bibfield
  {journal} {\bibinfo  {journal} {Annual Review of Condensed Matter Physics}\
  }\textbf {\bibinfo {volume} {4}},\ \bibinfo {pages} {137} (\bibinfo {year}
  {2013})}\BibitemShut {NoStop}%
\bibitem [{\citenamefont {Hirjibehedin}(2015)}]{Hirjibehedin_2015}%
  \BibitemOpen
  \bibfield  {author} {\bibinfo {author} {\bibfnamefont {C.~F.}\ \bibnamefont
  {Hirjibehedin}},\ }\bibfield  {title} {\emph {\enquote {\bibinfo {title} {The
  makings of a Hund's metal},}\ }}\href {\doibase 10.1038/nnano.2015.225}
  {\bibfield  {journal} {\bibinfo  {journal} {Nature Nanotechnology}\ }\textbf
  {\bibinfo {volume} {10}},\ \bibinfo {pages} {914} (\bibinfo {year}
  {2015})}\BibitemShut {NoStop}%
\bibitem [{\citenamefont {de' Medici}(2017)}]{Medici2017}%
  \BibitemOpen
  \bibfield  {author} {\bibinfo {author} {\bibfnamefont {L.}~\bibnamefont {de'
  Medici}},\ }\enquote {\bibinfo {title} {The physics of Correlated Insulator,
  Metals and Superconductors},}\ \ (\bibinfo  {publisher} {Forschungszentrum
  J\"ulich GmbH, Institute for Advanced Simulations},\ \bibinfo {year} {2017})\
  Chap.\ \bibinfo {chapter} {Hund's metals explained}, pp.\ \bibinfo {pages}
  {14.1--14.22}\BibitemShut {NoStop}%
\bibitem [{\citenamefont {Anderson}(1972)}]{Anderson_1972}%
  \BibitemOpen
  \bibfield  {author} {\bibinfo {author} {\bibfnamefont {P.~W.}\ \bibnamefont
  {Anderson}},\ }\bibfield  {title} {\emph {\enquote {\bibinfo {title} {More Is
  Different},}\ }}\href {\doibase 10.1126/science.177.4047.393} {\bibfield
  {journal} {\bibinfo  {journal} {Science}\ }\textbf {\bibinfo {volume}
  {177}},\ \bibinfo {pages} {393} (\bibinfo {year} {1972})}\BibitemShut
  {NoStop}%
\bibitem [{\citenamefont {Moriya}(1985)}]{Moriya_1985}%
  \BibitemOpen
  \bibfield  {author} {\bibinfo {author} {\bibfnamefont {T.}~\bibnamefont
  {Moriya}},\ }\href {\doibase 10.1007/978-3-642-82499-9} {\emph {\bibinfo
  {title} {Spin Fluctuations in Itinerant Electron Magnetism}}},\ \bibinfo
  {series} {Springer Series in Solid-State Sciences}, Vol.~\bibinfo {volume}
  {56}\ (\bibinfo  {publisher} {Springer-Verlag},\ \bibinfo {address} {Berlin,
  Heidelberg},\ \bibinfo {year} {1985})\BibitemShut {NoStop}%
\bibitem [{\citenamefont {Das}\ \emph {et~al.}(2013)\citenamefont {Das},
  \citenamefont {Zhu},\ and\ \citenamefont {Graf}}]{Das_2013}%
  \BibitemOpen
  \bibfield  {author} {\bibinfo {author} {\bibfnamefont {T.}~\bibnamefont
  {Das}}, \bibinfo {author} {\bibfnamefont {J.-X.}\ \bibnamefont {Zhu}}, \ and\
  \bibinfo {author} {\bibfnamefont {M.~J.}\ \bibnamefont {Graf}},\ }\bibfield
  {title} {\emph {\enquote {\bibinfo {title} {Self-consistent spin fluctuation
  spectrum and correlated electronic structure of actinides},}\ }}\href
  {\doibase 10.1557/jmr.2012.423} {\bibfield  {journal} {\bibinfo  {journal}
  {Journal of Materials Research}\ }\textbf {\bibinfo {volume} {28}},\ \bibinfo
  {pages} {659} (\bibinfo {year} {2013})}\BibitemShut {NoStop}%
\bibitem [{\citenamefont {Sch\"afer}\ \emph {et~al.}(2021)\citenamefont
  {Sch\"afer}, \citenamefont {Wentzell}, \citenamefont {\ifmmode~\check{S}\else
  \v{S}\fi{}imkovic}, \citenamefont {He}, \citenamefont {Hille}, \citenamefont
  {Klett}, \citenamefont {Eckhardt}, \citenamefont {Arzhang}, \citenamefont
  {Harkov}, \citenamefont {Le~R\'egent}, \citenamefont {Kirsch}, \citenamefont
  {Wang}, \citenamefont {Kim}, \citenamefont {Kozik}, \citenamefont {Stepanov},
  \citenamefont {Kauch}, \citenamefont {Andergassen}, \citenamefont {Hansmann},
  \citenamefont {Rohe}, \citenamefont {Vilk}, \citenamefont {LeBlanc},
  \citenamefont {Zhang}, \citenamefont {Tremblay}, \citenamefont {Ferrero},
  \citenamefont {Parcollet},\ and\ \citenamefont {Georges}}]{Schaefer_2021}%
  \BibitemOpen
  \bibfield  {author} {\bibinfo {author} {\bibfnamefont {T.}~\bibnamefont
  {Sch\"afer}}, \bibinfo {author} {\bibfnamefont {N.}~\bibnamefont {Wentzell}},
  \bibinfo {author} {\bibfnamefont {F.}~\bibnamefont {\ifmmode~\check{S}\else
  \v{S}\fi{}imkovic}}, \bibinfo {author} {\bibfnamefont {Y.-Y.}\ \bibnamefont
  {He}}, \bibinfo {author} {\bibfnamefont {C.}~\bibnamefont {Hille}}, \bibinfo
  {author} {\bibfnamefont {M.}~\bibnamefont {Klett}}, \bibinfo {author}
  {\bibfnamefont {C.~J.}\ \bibnamefont {Eckhardt}}, \bibinfo {author}
  {\bibfnamefont {B.}~\bibnamefont {Arzhang}}, \bibinfo {author} {\bibfnamefont
  {V.}~\bibnamefont {Harkov}}, \bibinfo {author} {\bibfnamefont {F.~m. c.-M.}\
  \bibnamefont {Le~R\'egent}}, \bibinfo {author} {\bibfnamefont
  {A.}~\bibnamefont {Kirsch}}, \bibinfo {author} {\bibfnamefont
  {Y.}~\bibnamefont {Wang}}, \bibinfo {author} {\bibfnamefont {A.~J.}\
  \bibnamefont {Kim}}, \bibinfo {author} {\bibfnamefont {E.}~\bibnamefont
  {Kozik}}, \bibinfo {author} {\bibfnamefont {E.~A.}\ \bibnamefont {Stepanov}},
  \bibinfo {author} {\bibfnamefont {A.}~\bibnamefont {Kauch}}, \bibinfo
  {author} {\bibfnamefont {S.}~\bibnamefont {Andergassen}}, \bibinfo {author}
  {\bibfnamefont {P.}~\bibnamefont {Hansmann}}, \bibinfo {author}
  {\bibfnamefont {D.}~\bibnamefont {Rohe}}, \bibinfo {author} {\bibfnamefont
  {Y.~M.}\ \bibnamefont {Vilk}}, \bibinfo {author} {\bibfnamefont {J.~P.~F.}\
  \bibnamefont {LeBlanc}}, \bibinfo {author} {\bibfnamefont {S.}~\bibnamefont
  {Zhang}}, \bibinfo {author} {\bibfnamefont {A.-M.~S.}\ \bibnamefont
  {Tremblay}}, \bibinfo {author} {\bibfnamefont {M.}~\bibnamefont {Ferrero}},
  \bibinfo {author} {\bibfnamefont {O.}~\bibnamefont {Parcollet}}, \ and\
  \bibinfo {author} {\bibfnamefont {A.}~\bibnamefont {Georges}},\ }\bibfield
  {title} {\emph {\enquote {\bibinfo {title} {Tracking the Footprints of Spin
  Fluctuations: A MultiMethod, MultiMessenger Study of the Two-Dimensional
  Hubbard Model},}\ }}\href {\doibase 10.1103/PhysRevX.11.011058} {\bibfield
  {journal} {\bibinfo  {journal} {Phys. Rev. X}\ }\textbf {\bibinfo {volume}
  {11}},\ \bibinfo {pages} {011058} (\bibinfo {year} {2021})}\BibitemShut
  {NoStop}%
\bibitem [{\citenamefont {Bergeal}\ \emph {et~al.}(2008)\citenamefont
  {Bergeal}, \citenamefont {Lesueur}, \citenamefont {Aprili}, \citenamefont
  {Faini}, \citenamefont {Contour},\ and\ \citenamefont
  {Leridon}}]{Bergeal_2008}%
  \BibitemOpen
  \bibfield  {author} {\bibinfo {author} {\bibfnamefont {N.}~\bibnamefont
  {Bergeal}}, \bibinfo {author} {\bibfnamefont {J.}~\bibnamefont {Lesueur}},
  \bibinfo {author} {\bibfnamefont {M.}~\bibnamefont {Aprili}}, \bibinfo
  {author} {\bibfnamefont {G.}~\bibnamefont {Faini}}, \bibinfo {author}
  {\bibfnamefont {J.~P.}\ \bibnamefont {Contour}}, \ and\ \bibinfo {author}
  {\bibfnamefont {B.}~\bibnamefont {Leridon}},\ }\bibfield  {title} {\emph
  {\enquote {\bibinfo {title} {Pairing fluctuations in the pseudogap state of
  copper-oxide superconductors probed by the Josephson effect},}\ }}\href
  {\doibase 10.1038/nphys1017} {\bibfield  {journal} {\bibinfo  {journal}
  {Nature Physics}\ }\textbf {\bibinfo {volume} {4}},\ \bibinfo {pages} {608}
  (\bibinfo {year} {2008})}\BibitemShut {NoStop}%
\bibitem [{\citenamefont {Maier}\ and\ \citenamefont
  {Scalapino}(2019)}]{Maier_2019}%
  \BibitemOpen
  \bibfield  {author} {\bibinfo {author} {\bibfnamefont {T.~A.}\ \bibnamefont
  {Maier}}\ and\ \bibinfo {author} {\bibfnamefont {D.~J.}\ \bibnamefont
  {Scalapino}},\ }\bibfield  {title} {\emph {\enquote {\bibinfo {title}
  {Pairfield fluctuations of a 2D Hubbard model},}\ }}\href {\doibase
  10.1038/s41535-019-0169-9} {\bibfield  {journal} {\bibinfo  {journal} {npj
  Quantum Materials}\ }\textbf {\bibinfo {volume} {4}},\ \bibinfo {pages} {30}
  (\bibinfo {year} {2019})}\BibitemShut {NoStop}%
\bibitem [{\citenamefont {Böhmer}\ \emph {et~al.}(2022)\citenamefont
  {Böhmer}, \citenamefont {Chu}, \citenamefont {Lederer},\ and\ \citenamefont
  {Yi}}]{Boehmer_2022}%
  \BibitemOpen
  \bibfield  {author} {\bibinfo {author} {\bibfnamefont {A.~E.}\ \bibnamefont
  {Böhmer}}, \bibinfo {author} {\bibfnamefont {J.-H.}\ \bibnamefont {Chu}},
  \bibinfo {author} {\bibfnamefont {S.}~\bibnamefont {Lederer}}, \ and\
  \bibinfo {author} {\bibfnamefont {M.}~\bibnamefont {Yi}},\ }\bibfield
  {title} {\emph {\enquote {\bibinfo {title} {Nematicity and nematic
  fluctuations in iron-based superconductors},}\ }}\href {\doibase
  10.1038/s41567-022-01833-3} {\bibfield  {journal} {\bibinfo  {journal}
  {Nature Physics}\ }\textbf {\bibinfo {volume} {18}},\ \bibinfo {pages} {1412}
  (\bibinfo {year} {2022})}\BibitemShut {NoStop}%
\bibitem [{\citenamefont {Korshunov}\ and\ \citenamefont
  {Togushova}(2024)}]{Korshunov_2024}%
  \BibitemOpen
  \bibfield  {author} {\bibinfo {author} {\bibfnamefont {M.~M.}\ \bibnamefont
  {Korshunov}}\ and\ \bibinfo {author} {\bibfnamefont {Y.~N.}\ \bibnamefont
  {Togushova}},\ }\bibfield  {title} {\emph {\enquote {\bibinfo {title}
  {Superconducting Order Parameter Structure in the Nematic Phase of Iron-based
  Materials},}\ }}\href {\doibase 10.1134/S0021364024600046} {\bibfield
  {journal} {\bibinfo  {journal} {JETP Letters}\ }\textbf {\bibinfo {volume}
  {119}},\ \bibinfo {pages} {310} (\bibinfo {year} {2024})}\BibitemShut
  {NoStop}%
\bibitem [{\citenamefont {Kaneko}\ and\ \citenamefont
  {Ohta}(2025)}]{Kaneko_2025}%
  \BibitemOpen
  \bibfield  {author} {\bibinfo {author} {\bibfnamefont {T.}~\bibnamefont
  {Kaneko}}\ and\ \bibinfo {author} {\bibfnamefont {Y.}~\bibnamefont {Ohta}},\
  }\bibfield  {title} {\emph {\enquote {\bibinfo {title} {A New Era of
  Excitonic Insulators},}\ }}\href {\doibase 10.7566/JPSJ.94.012001} {\bibfield
   {journal} {\bibinfo  {journal} {Journal of the Physical Society of Japan}\
  }\textbf {\bibinfo {volume} {94}},\ \bibinfo {pages} {012001} (\bibinfo
  {year} {2025})},\ \Eprint
  {http://arxiv.org/abs/https://doi.org/10.7566/JPSJ.94.012001}{https://doi.org/10.7566/JPSJ.94.012001}\BibitemShut
  {NoStop}%
\bibitem [{\citenamefont {Karolak}\ \emph {et~al.}(2015)\citenamefont
  {Karolak}, \citenamefont {Izquierdo}, \citenamefont {Molodtsov},\ and\
  \citenamefont {Lichtenstein}}]{Karolak_2015}%
  \BibitemOpen
  \bibfield  {author} {\bibinfo {author} {\bibfnamefont {M.}~\bibnamefont
  {Karolak}}, \bibinfo {author} {\bibfnamefont {M.}~\bibnamefont {Izquierdo}},
  \bibinfo {author} {\bibfnamefont {S.~L.}\ \bibnamefont {Molodtsov}}, \ and\
  \bibinfo {author} {\bibfnamefont {A.~I.}\ \bibnamefont {Lichtenstein}},\
  }\bibfield  {title} {\emph {\enquote {\bibinfo {title} {Correlation-Driven
  Charge and Spin Fluctuations in ${\mathrm{LaCoO}}_{3}$},}\ }}\href {\doibase
  10.1103/PhysRevLett.115.046401} {\bibfield  {journal} {\bibinfo  {journal}
  {Phys. Rev. Lett.}\ }\textbf {\bibinfo {volume} {115}},\ \bibinfo {pages}
  {046401} (\bibinfo {year} {2015})}\BibitemShut {NoStop}%
\bibitem [{\citenamefont {Volkov}\ \emph {et~al.}(2021)\citenamefont {Volkov},
  \citenamefont {Ye}, \citenamefont {Lohani}, \citenamefont {Feldman},
  \citenamefont {Kanigel},\ and\ \citenamefont {Blumberg}}]{Volkov_2021}%
  \BibitemOpen
  \bibfield  {author} {\bibinfo {author} {\bibfnamefont {P.~A.}\ \bibnamefont
  {Volkov}}, \bibinfo {author} {\bibfnamefont {M.}~\bibnamefont {Ye}}, \bibinfo
  {author} {\bibfnamefont {H.}~\bibnamefont {Lohani}}, \bibinfo {author}
  {\bibfnamefont {I.}~\bibnamefont {Feldman}}, \bibinfo {author} {\bibfnamefont
  {A.}~\bibnamefont {Kanigel}}, \ and\ \bibinfo {author} {\bibfnamefont
  {G.}~\bibnamefont {Blumberg}},\ }\bibfield  {title} {\emph {\enquote
  {\bibinfo {title} {Critical charge fluctuations and emergent coherence in a
  strongly correlated excitonic insulator},}\ }}\href {\doibase
  10.1038/s41535-021-00351-4} {\bibfield  {journal} {\bibinfo  {journal} {npj
  Quantum Materials}\ }\textbf {\bibinfo {volume} {6}},\ \bibinfo {pages} {52}
  (\bibinfo {year} {2021})}\BibitemShut {NoStop}%
\bibitem [{\citenamefont {Hansmann}\ \emph {et~al.}(2013)\citenamefont
  {Hansmann}, \citenamefont {Ayral}, \citenamefont {Vaugier}, \citenamefont
  {Werner},\ and\ \citenamefont {Biermann}}]{Hansmann_2013}%
  \BibitemOpen
  \bibfield  {author} {\bibinfo {author} {\bibfnamefont {P.}~\bibnamefont
  {Hansmann}}, \bibinfo {author} {\bibfnamefont {T.}~\bibnamefont {Ayral}},
  \bibinfo {author} {\bibfnamefont {L.}~\bibnamefont {Vaugier}}, \bibinfo
  {author} {\bibfnamefont {P.}~\bibnamefont {Werner}}, \ and\ \bibinfo {author}
  {\bibfnamefont {S.}~\bibnamefont {Biermann}},\ }\bibfield  {title} {\emph
  {\enquote {\bibinfo {title} {Long-Range Coulomb Interactions in Surface
  Systems: A First-Principles Description within Self-Consistently Combined
  $GW$ and Dynamical Mean-Field Theory},}\ }}\href {\doibase
  10.1103/PhysRevLett.110.166401} {\bibfield  {journal} {\bibinfo  {journal}
  {Phys. Rev. Lett.}\ }\textbf {\bibinfo {volume} {110}},\ \bibinfo {pages}
  {166401} (\bibinfo {year} {2013})}\BibitemShut {NoStop}%
\bibitem [{\citenamefont {Hubbard}(1963)}]{Hubbard_1963}%
  \BibitemOpen
  \bibfield  {author} {\bibinfo {author} {\bibfnamefont {J.}~\bibnamefont
  {Hubbard}},\ }\bibfield  {title} {\emph {\enquote {\bibinfo {title} {Electron
  Correlations in Narrow Energy Bands},}\ }}\href {\doibase
  10.1098/rspa.1963.0204} {\bibfield  {journal} {\bibinfo  {journal}
  {Proceedings of the Royal Society of London. Series A, Mathematical and
  Physical Sciences}\ }\textbf {\bibinfo {volume} {276}},\ \bibinfo {pages}
  {238} (\bibinfo {year} {1963})}\BibitemShut {NoStop}%
\bibitem [{\citenamefont {Gutzwiller}(1963)}]{Gutzwiller_1963}%
  \BibitemOpen
  \bibfield  {author} {\bibinfo {author} {\bibfnamefont {M.~C.}\ \bibnamefont
  {Gutzwiller}},\ }\bibfield  {title} {\emph {\enquote {\bibinfo {title}
  {Effect of Correlation on the Ferromagnetism of Transition Metals},}\ }}\href
  {\doibase 10.1103/PhysRevLett.10.159} {\bibfield  {journal} {\bibinfo
  {journal} {Phys. Rev. Lett.}\ }\textbf {\bibinfo {volume} {10}},\ \bibinfo
  {pages} {159} (\bibinfo {year} {1963})}\BibitemShut {NoStop}%
\bibitem [{\citenamefont {Kanamori}(1963)}]{Kanamori_1963}%
  \BibitemOpen
  \bibfield  {author} {\bibinfo {author} {\bibfnamefont {J.}~\bibnamefont
  {Kanamori}},\ }\bibfield  {title} {\emph {\enquote {\bibinfo {title}
  {Electron Correlation and Ferromagnetism of Transition Metals},}\ }}\href
  {\doibase 10.1143/PTP.30.275} {\bibfield  {journal} {\bibinfo  {journal}
  {Progress of Theoretical Physics}\ }\textbf {\bibinfo {volume} {30}},\
  \bibinfo {pages} {275} (\bibinfo {year} {1963})},\ \Eprint
  {http://arxiv.org/abs/https://academic.oup.com/ptp/article-pdf/30/3/275/5278869/30-3-275.pdf}{https://academic.oup.com/ptp/article-pdf/30/3/275/5278869/30-3-275.pdf}\BibitemShut
  {NoStop}%
\bibitem [{\citenamefont {Physics}(2013)}]{Editorial_2013}%
  \BibitemOpen
  \bibfield  {author} {\bibinfo {author} {\bibfnamefont {N.}~\bibnamefont
  {Physics}},\ }\bibfield  {title} {\emph {\enquote {\bibinfo {title} {The
  Hubbard model at half a century},}\ }}\href {\doibase 10.1038/nphys2759}
  {\bibfield  {journal} {\bibinfo  {journal} {Nature Physics}\ }\textbf
  {\bibinfo {volume} {9}},\ \bibinfo {pages} {523} (\bibinfo {year}
  {2013})}\BibitemShut {NoStop}%
\bibitem [{\citenamefont {Bethe}(1931)}]{Bethe_1931}%
  \BibitemOpen
  \bibfield  {author} {\bibinfo {author} {\bibfnamefont {H.}~\bibnamefont
  {Bethe}},\ }\bibfield  {title} {\emph {\enquote {\bibinfo {title} {Zur
  Theorie der Metalle},}\ }}\href {\doibase 10.1007/BF01341708} {\bibfield
  {journal} {\bibinfo  {journal} {Z. Phys.}\ }\textbf {\bibinfo {volume}
  {71}},\ \bibinfo {pages} {205} (\bibinfo {year} {1931})}\BibitemShut
  {NoStop}%
\bibitem [{\citenamefont {Essler}\ \emph {et~al.}(2005)\citenamefont {Essler},
  \citenamefont {Frahm}, \citenamefont {Göhmann}, \citenamefont {Klümper},\
  and\ \citenamefont {Korepin}}]{Essler_Korepin_2005}%
  \BibitemOpen
  \bibfield  {author} {\bibinfo {author} {\bibfnamefont {F.~H.~L.}\
  \bibnamefont {Essler}}, \bibinfo {author} {\bibfnamefont {H.}~\bibnamefont
  {Frahm}}, \bibinfo {author} {\bibfnamefont {F.}~\bibnamefont {Göhmann}},
  \bibinfo {author} {\bibfnamefont {A.}~\bibnamefont {Klümper}}, \ and\
  \bibinfo {author} {\bibfnamefont {V.~E.}\ \bibnamefont {Korepin}},\
  }\href@noop {} {\emph {\bibinfo {title} {The One-Dimensional Hubbard
  Model}}}\ (\bibinfo  {publisher} {Cambridge University Press},\ \bibinfo
  {year} {2005})\BibitemShut {NoStop}%
\bibitem [{\citenamefont {White}(1992)}]{White_1992}%
  \BibitemOpen
  \bibfield  {author} {\bibinfo {author} {\bibfnamefont {S.~R.}\ \bibnamefont
  {White}},\ }\bibfield  {title} {\emph {\enquote {\bibinfo {title} {Density
  matrix formulation for quantum renormalization groups},}\ }}\href {\doibase
  10.1103/PhysRevLett.69.2863} {\bibfield  {journal} {\bibinfo  {journal}
  {Phys. Rev. Lett.}\ }\textbf {\bibinfo {volume} {69}},\ \bibinfo {pages}
  {2863} (\bibinfo {year} {1992})}\BibitemShut {NoStop}%
\bibitem [{\citenamefont {White}(1993)}]{White_1993}%
  \BibitemOpen
  \bibfield  {author} {\bibinfo {author} {\bibfnamefont {S.~R.}\ \bibnamefont
  {White}},\ }\bibfield  {title} {\emph {\enquote {\bibinfo {title}
  {Density-matrix algorithms for quantum renormalization groups},}\ }}\href
  {\doibase 10.1103/PhysRevB.48.10345} {\bibfield  {journal} {\bibinfo
  {journal} {Phys. Rev. B}\ }\textbf {\bibinfo {volume} {48}},\ \bibinfo
  {pages} {10345} (\bibinfo {year} {1993})}\BibitemShut {NoStop}%
\bibitem [{\citenamefont {Schollwöck}(2011)}]{Schollwoeck_2011}%
  \BibitemOpen
  \bibfield  {author} {\bibinfo {author} {\bibfnamefont {U.}~\bibnamefont
  {Schollwöck}},\ }\bibfield  {title} {\emph {\enquote {\bibinfo {title} {The
  density-matrix renormalization group in the age of matrix product states},}\
  }}\href {\doibase https://doi.org/10.1016/j.aop.2010.09.012} {\bibfield
  {journal} {\bibinfo  {journal} {Annals of Physics}\ }\textbf {\bibinfo
  {volume} {326}},\ \bibinfo {pages} {96} (\bibinfo {year} {2011})},\ \bibinfo
  {note} {january 2011 Special Issue}\BibitemShut {NoStop}%
\bibitem [{\citenamefont {Georges}\ \emph {et~al.}(1996)\citenamefont
  {Georges}, \citenamefont {Kotliar}, \citenamefont {Krauth},\ and\
  \citenamefont {Rozenberg}}]{Georges_Rosenberg_1996}%
  \BibitemOpen
  \bibfield  {author} {\bibinfo {author} {\bibfnamefont {A.}~\bibnamefont
  {Georges}}, \bibinfo {author} {\bibfnamefont {G.}~\bibnamefont {Kotliar}},
  \bibinfo {author} {\bibfnamefont {W.}~\bibnamefont {Krauth}}, \ and\ \bibinfo
  {author} {\bibfnamefont {M.~J.}\ \bibnamefont {Rozenberg}},\ }\bibfield
  {title} {\emph {\enquote {\bibinfo {title} {Dynamical mean-field theory of
  strongly correlated fermion systems and the limit of infinite dimensions},}\
  }}\href {\doibase 10.1103/RevModPhys.68.13} {\bibfield  {journal} {\bibinfo
  {journal} {Rev. Mod. Phys.}\ }\textbf {\bibinfo {volume} {68}},\ \bibinfo
  {pages} {13} (\bibinfo {year} {1996})}\BibitemShut {NoStop}%
\bibitem [{\citenamefont {Metzner}\ and\ \citenamefont
  {Vollhardt}(1989)}]{Metzner_Vollhard_1989}%
  \BibitemOpen
  \bibfield  {author} {\bibinfo {author} {\bibfnamefont {W.}~\bibnamefont
  {Metzner}}\ and\ \bibinfo {author} {\bibfnamefont {D.}~\bibnamefont
  {Vollhardt}},\ }\bibfield  {title} {\emph {\enquote {\bibinfo {title}
  {Correlated Lattice Fermions in $d=\ensuremath{\infty}$ Dimensions},}\
  }}\href {\doibase 10.1103/PhysRevLett.62.324} {\bibfield  {journal} {\bibinfo
   {journal} {Phys. Rev. Lett.}\ }\textbf {\bibinfo {volume} {62}},\ \bibinfo
  {pages} {324} (\bibinfo {year} {1989})}\BibitemShut {NoStop}%
\bibitem [{\citenamefont {Vollhardt}\ \emph {et~al.}(2012)\citenamefont
  {Vollhardt}, \citenamefont {Byczuk},\ and\ \citenamefont
  {Kollar}}]{Vollhardt_Kollar_2012}%
  \BibitemOpen
  \bibfield  {author} {\bibinfo {author} {\bibfnamefont {D.}~\bibnamefont
  {Vollhardt}}, \bibinfo {author} {\bibfnamefont {K.}~\bibnamefont {Byczuk}}, \
  and\ \bibinfo {author} {\bibfnamefont {M.}~\bibnamefont {Kollar}},\ }\enquote
  {\bibinfo {title} {Dynamical Mean-Field Theory},}\ in\ \href {\doibase
  10.1007/978-3-642-21831-6_7} {\emph {\bibinfo {booktitle} {Strongly
  Correlated Systems: Theoretical Methods}}},\ \bibinfo {editor} {edited by\
  \bibinfo {editor} {\bibfnamefont {A.}~\bibnamefont {Avella}}\ and\ \bibinfo
  {editor} {\bibfnamefont {F.}~\bibnamefont {Mancini}}}\ (\bibinfo  {publisher}
  {Springer Berlin Heidelberg},\ \bibinfo {address} {Berlin, Heidelberg},\
  \bibinfo {year} {2012})\ pp.\ \bibinfo {pages} {203--236}\BibitemShut
  {NoStop}%
\bibitem [{\citenamefont {Imada}\ \emph {et~al.}(1998)\citenamefont {Imada},
  \citenamefont {Fujimori},\ and\ \citenamefont {Tokura}}]{Imada_1998}%
  \BibitemOpen
  \bibfield  {author} {\bibinfo {author} {\bibfnamefont {M.}~\bibnamefont
  {Imada}}, \bibinfo {author} {\bibfnamefont {A.}~\bibnamefont {Fujimori}}, \
  and\ \bibinfo {author} {\bibfnamefont {Y.}~\bibnamefont {Tokura}},\
  }\bibfield  {title} {\emph {\enquote {\bibinfo {title} {Metal-insulator
  transitions},}\ }}\href {\doibase 10.1103/RevModPhys.70.1039} {\bibfield
  {journal} {\bibinfo  {journal} {Rev. Mod. Phys.}\ }\textbf {\bibinfo {volume}
  {70}},\ \bibinfo {pages} {1039} (\bibinfo {year} {1998})}\BibitemShut
  {NoStop}%
\bibitem [{\citenamefont {Lee}\ \emph {et~al.}(2006)\citenamefont {Lee},
  \citenamefont {Nagaosa},\ and\ \citenamefont {Wen}}]{Lee_2006}%
  \BibitemOpen
  \bibfield  {author} {\bibinfo {author} {\bibfnamefont {P.~A.}\ \bibnamefont
  {Lee}}, \bibinfo {author} {\bibfnamefont {N.}~\bibnamefont {Nagaosa}}, \ and\
  \bibinfo {author} {\bibfnamefont {X.-G.}\ \bibnamefont {Wen}},\ }\bibfield
  {title} {\emph {\enquote {\bibinfo {title} {Doping a Mott insulator: Physics
  of high-temperature superconductivity},}\ }}\href {\doibase
  10.1103/RevModPhys.78.17} {\bibfield  {journal} {\bibinfo  {journal} {Rev.
  Mod. Phys.}\ }\textbf {\bibinfo {volume} {78}},\ \bibinfo {pages} {17}
  (\bibinfo {year} {2006})}\BibitemShut {NoStop}%
\bibitem [{\citenamefont {Kotliar}\ \emph {et~al.}(2002)\citenamefont
  {Kotliar}, \citenamefont {Murthy},\ and\ \citenamefont
  {Rozenberg}}]{Kotliar_Rozenberg_2002}%
  \BibitemOpen
  \bibfield  {author} {\bibinfo {author} {\bibfnamefont {G.}~\bibnamefont
  {Kotliar}}, \bibinfo {author} {\bibfnamefont {S.}~\bibnamefont {Murthy}}, \
  and\ \bibinfo {author} {\bibfnamefont {M.~J.}\ \bibnamefont {Rozenberg}},\
  }\bibfield  {title} {\emph {\enquote {\bibinfo {title} {Compressibility
  Divergence and the Finite Temperature Mott Transition},}\ }}\href {\doibase
  10.1103/PhysRevLett.89.046401} {\bibfield  {journal} {\bibinfo  {journal}
  {Phys. Rev. Lett.}\ }\textbf {\bibinfo {volume} {89}},\ \bibinfo {pages}
  {046401} (\bibinfo {year} {2002})}\BibitemShut {NoStop}%
\bibitem [{\citenamefont {Nourafkan}\ \emph {et~al.}(2019)\citenamefont
  {Nourafkan}, \citenamefont {Côté},\ and\ \citenamefont
  {Tremblay}}]{Nourafkan_Tremblay_2019}%
  \BibitemOpen
  \bibfield  {author} {\bibinfo {author} {\bibfnamefont {R.}~\bibnamefont
  {Nourafkan}}, \bibinfo {author} {\bibfnamefont {M.}~\bibnamefont {Côté}}, \
  and\ \bibinfo {author} {\bibfnamefont {A.-M.~S.}\ \bibnamefont {Tremblay}},\
  }\bibfield  {title} {\emph {\enquote {\bibinfo {title} {Charge-fluctuations
  in lightly hole-doped cuprates: effect of vertex corrections},}\ }}\href
  {\doibase 10.1103/PhysRevB.99.035161} {\bibfield  {journal} {\bibinfo
  {journal} {Physical Review B}\ }\textbf {\bibinfo {volume} {99}},\ \bibinfo
  {pages} {035161} (\bibinfo {year} {2019})},\ \Eprint
  {http://arxiv.org/abs/1807.03855 [cond-mat]}{1807.03855
  [cond-mat]}\BibitemShut {NoStop}%
\bibitem [{\citenamefont {Reitner}\ \emph {et~al.}(2020)\citenamefont
  {Reitner}, \citenamefont {Chalupa}, \citenamefont {Del~Re}, \citenamefont
  {Springer}, \citenamefont {Ciuchi}, \citenamefont {Sangiovanni},\ and\
  \citenamefont {Toschi}}]{Reitner_2020}%
  \BibitemOpen
  \bibfield  {author} {\bibinfo {author} {\bibfnamefont {M.}~\bibnamefont
  {Reitner}}, \bibinfo {author} {\bibfnamefont {P.}~\bibnamefont {Chalupa}},
  \bibinfo {author} {\bibfnamefont {L.}~\bibnamefont {Del~Re}}, \bibinfo
  {author} {\bibfnamefont {D.}~\bibnamefont {Springer}}, \bibinfo {author}
  {\bibfnamefont {S.}~\bibnamefont {Ciuchi}}, \bibinfo {author} {\bibfnamefont
  {G.}~\bibnamefont {Sangiovanni}}, \ and\ \bibinfo {author} {\bibfnamefont
  {A.}~\bibnamefont {Toschi}},\ }\bibfield  {title} {\emph {\enquote {\bibinfo
  {title} {Attractive Effect of a Strong Electronic Repulsion: The Physics of
  Vertex Divergences},}\ }}\href {\doibase 10.1103/PhysRevLett.125.196403}
  {\bibfield  {journal} {\bibinfo  {journal} {Phys. Rev. Lett.}\ }\textbf
  {\bibinfo {volume} {125}},\ \bibinfo {pages} {196403} (\bibinfo {year}
  {2020})}\BibitemShut {NoStop}%
\bibitem [{\citenamefont {Kowalski}\ \emph {et~al.}(2024)\citenamefont
  {Kowalski}, \citenamefont {Reitner}, \citenamefont {Del~Re}, \citenamefont
  {Chatzieleftheriou}, \citenamefont {Amaricci}, \citenamefont {Toschi},
  \citenamefont {de' Medici}, \citenamefont {Sangiovanni},\ and\ \citenamefont
  {Sch\"afer}}]{Kowalski_2024}%
  \BibitemOpen
  \bibfield  {author} {\bibinfo {author} {\bibfnamefont {A.}~\bibnamefont
  {Kowalski}}, \bibinfo {author} {\bibfnamefont {M.}~\bibnamefont {Reitner}},
  \bibinfo {author} {\bibfnamefont {L.}~\bibnamefont {Del~Re}}, \bibinfo
  {author} {\bibfnamefont {M.}~\bibnamefont {Chatzieleftheriou}}, \bibinfo
  {author} {\bibfnamefont {A.}~\bibnamefont {Amaricci}}, \bibinfo {author}
  {\bibfnamefont {A.}~\bibnamefont {Toschi}}, \bibinfo {author} {\bibfnamefont
  {L.}~\bibnamefont {de' Medici}}, \bibinfo {author} {\bibfnamefont
  {G.}~\bibnamefont {Sangiovanni}}, \ and\ \bibinfo {author} {\bibfnamefont
  {T.}~\bibnamefont {Sch\"afer}},\ }\bibfield  {title} {\emph {\enquote
  {\bibinfo {title} {Thermodynamic Stability at the Two-Particle Level},}\
  }}\href {\doibase 10.1103/PhysRevLett.133.066502} {\bibfield  {journal}
  {\bibinfo  {journal} {Phys. Rev. Lett.}\ }\textbf {\bibinfo {volume} {133}},\
  \bibinfo {pages} {066502} (\bibinfo {year} {2024})}\BibitemShut {NoStop}%
\bibitem [{\citenamefont {Lichtenstein}\ and\ \citenamefont
  {Katsnelson}(2000)}]{Lichtenstein_Katsnelson_2000}%
  \BibitemOpen
  \bibfield  {author} {\bibinfo {author} {\bibfnamefont {A.~I.}\ \bibnamefont
  {Lichtenstein}}\ and\ \bibinfo {author} {\bibfnamefont {M.~I.}\ \bibnamefont
  {Katsnelson}},\ }\bibfield  {title} {\emph {\enquote {\bibinfo {title}
  {Antiferromagnetism and d-wave superconductivity in cuprates: A cluster
  dynamical mean-field theory},}\ }}\href {\doibase 10.1103/PhysRevB.62.R9283}
  {\bibfield  {journal} {\bibinfo  {journal} {Phys. Rev. B}\ }\textbf {\bibinfo
  {volume} {62}},\ \bibinfo {pages} {R9283} (\bibinfo {year}
  {2000})}\BibitemShut {NoStop}%
\bibitem [{\citenamefont {Kotliar}\ \emph {et~al.}(2001)\citenamefont
  {Kotliar}, \citenamefont {Savrasov}, \citenamefont {P\'alsson},\ and\
  \citenamefont {Biroli}}]{Kotliar_Biroli_2001}%
  \BibitemOpen
  \bibfield  {author} {\bibinfo {author} {\bibfnamefont {G.}~\bibnamefont
  {Kotliar}}, \bibinfo {author} {\bibfnamefont {S.~Y.}\ \bibnamefont
  {Savrasov}}, \bibinfo {author} {\bibfnamefont {G.}~\bibnamefont {P\'alsson}},
  \ and\ \bibinfo {author} {\bibfnamefont {G.}~\bibnamefont {Biroli}},\
  }\bibfield  {title} {\emph {\enquote {\bibinfo {title} {Cellular Dynamical
  Mean Field Approach to Strongly Correlated Systems},}\ }}\href {\doibase
  10.1103/PhysRevLett.87.186401} {\bibfield  {journal} {\bibinfo  {journal}
  {Phys. Rev. Lett.}\ }\textbf {\bibinfo {volume} {87}},\ \bibinfo {pages}
  {186401} (\bibinfo {year} {2001})}\BibitemShut {NoStop}%
\bibitem [{\citenamefont {Maier}\ \emph {et~al.}(2005)\citenamefont {Maier},
  \citenamefont {Jarrell}, \citenamefont {Pruschke},\ and\ \citenamefont
  {Hettler}}]{Maier_Matthias_2005}%
  \BibitemOpen
  \bibfield  {author} {\bibinfo {author} {\bibfnamefont {T.}~\bibnamefont
  {Maier}}, \bibinfo {author} {\bibfnamefont {M.}~\bibnamefont {Jarrell}},
  \bibinfo {author} {\bibfnamefont {T.}~\bibnamefont {Pruschke}}, \ and\
  \bibinfo {author} {\bibfnamefont {M.~H.}\ \bibnamefont {Hettler}},\
  }\bibfield  {title} {\emph {\enquote {\bibinfo {title} {Quantum cluster
  theories},}\ }}\href {\doibase 10.1103/RevModPhys.77.1027} {\bibfield
  {journal} {\bibinfo  {journal} {Rev. Mod. Phys.}\ }\textbf {\bibinfo {volume}
  {77}},\ \bibinfo {pages} {1027} (\bibinfo {year} {2005})}\BibitemShut
  {NoStop}%
\bibitem [{\citenamefont {Hettler}\ \emph {et~al.}(1998)\citenamefont
  {Hettler}, \citenamefont {Tahvildar-Zadeh}, \citenamefont {Jarrell},
  \citenamefont {Pruschke},\ and\ \citenamefont
  {Krishnamurthy}}]{Hettler_Krishnamurthy_1998}%
  \BibitemOpen
  \bibfield  {author} {\bibinfo {author} {\bibfnamefont {M.~H.}\ \bibnamefont
  {Hettler}}, \bibinfo {author} {\bibfnamefont {A.~N.}\ \bibnamefont
  {Tahvildar-Zadeh}}, \bibinfo {author} {\bibfnamefont {M.}~\bibnamefont
  {Jarrell}}, \bibinfo {author} {\bibfnamefont {T.}~\bibnamefont {Pruschke}}, \
  and\ \bibinfo {author} {\bibfnamefont {H.~R.}\ \bibnamefont
  {Krishnamurthy}},\ }\bibfield  {title} {\emph {\enquote {\bibinfo {title}
  {Nonlocal dynamical correlations of strongly interacting electron systems},}\
  }}\href {\doibase 10.1103/PhysRevB.58.R7475} {\bibfield  {journal} {\bibinfo
  {journal} {Phys. Rev. B}\ }\textbf {\bibinfo {volume} {58}},\ \bibinfo
  {pages} {R7475} (\bibinfo {year} {1998})}\BibitemShut {NoStop}%
\bibitem [{\citenamefont {Hettler}\ \emph {et~al.}(2000)\citenamefont
  {Hettler}, \citenamefont {Mukherjee}, \citenamefont {Jarrell},\ and\
  \citenamefont {Krishnamurthy}}]{Hettler_Krishnamurthy_2000}%
  \BibitemOpen
  \bibfield  {author} {\bibinfo {author} {\bibfnamefont {M.~H.}\ \bibnamefont
  {Hettler}}, \bibinfo {author} {\bibfnamefont {M.}~\bibnamefont {Mukherjee}},
  \bibinfo {author} {\bibfnamefont {M.}~\bibnamefont {Jarrell}}, \ and\
  \bibinfo {author} {\bibfnamefont {H.~R.}\ \bibnamefont {Krishnamurthy}},\
  }\bibfield  {title} {\emph {\enquote {\bibinfo {title} {Dynamical cluster
  approximation: Nonlocal dynamics of correlated electron systems},}\ }}\href
  {\doibase 10.1103/PhysRevB.61.12739} {\bibfield  {journal} {\bibinfo
  {journal} {Phys. Rev. B}\ }\textbf {\bibinfo {volume} {61}},\ \bibinfo
  {pages} {12739} (\bibinfo {year} {2000})}\BibitemShut {NoStop}%
\bibitem [{\citenamefont {Rubtsov}\ \emph {et~al.}(2008)\citenamefont
  {Rubtsov}, \citenamefont {Katsnelson},\ and\ \citenamefont
  {Lichtenstein}}]{Rubtsov_Lichtenstein_2008}%
  \BibitemOpen
  \bibfield  {author} {\bibinfo {author} {\bibfnamefont {A.~N.}\ \bibnamefont
  {Rubtsov}}, \bibinfo {author} {\bibfnamefont {M.~I.}\ \bibnamefont
  {Katsnelson}}, \ and\ \bibinfo {author} {\bibfnamefont {A.~I.}\ \bibnamefont
  {Lichtenstein}},\ }\bibfield  {title} {\emph {\enquote {\bibinfo {title}
  {Dual fermion approach to nonlocal correlations in the Hubbard model},}\
  }}\href {\doibase 10.1103/PhysRevB.77.033101} {\bibfield  {journal} {\bibinfo
   {journal} {Phys. Rev. B}\ }\textbf {\bibinfo {volume} {77}},\ \bibinfo
  {pages} {033101} (\bibinfo {year} {2008})}\BibitemShut {NoStop}%
\bibitem [{\citenamefont {Hafermann}\ \emph {et~al.}(2009)\citenamefont
  {Hafermann}, \citenamefont {Li}, \citenamefont {Rubtsov}, \citenamefont
  {Katsnelson}, \citenamefont {Lichtenstein},\ and\ \citenamefont
  {Monien}}]{Hafermann_Monien_2009}%
  \BibitemOpen
  \bibfield  {author} {\bibinfo {author} {\bibfnamefont {H.}~\bibnamefont
  {Hafermann}}, \bibinfo {author} {\bibfnamefont {G.}~\bibnamefont {Li}},
  \bibinfo {author} {\bibfnamefont {A.~N.}\ \bibnamefont {Rubtsov}}, \bibinfo
  {author} {\bibfnamefont {M.~I.}\ \bibnamefont {Katsnelson}}, \bibinfo
  {author} {\bibfnamefont {A.~I.}\ \bibnamefont {Lichtenstein}}, \ and\
  \bibinfo {author} {\bibfnamefont {H.}~\bibnamefont {Monien}},\ }\bibfield
  {title} {\emph {\enquote {\bibinfo {title} {Efficient Perturbation Theory for
  Quantum Lattice Models},}\ }}\href {\doibase 10.1103/PhysRevLett.102.206401}
  {\bibfield  {journal} {\bibinfo  {journal} {Phys. Rev. Lett.}\ }\textbf
  {\bibinfo {volume} {102}},\ \bibinfo {pages} {206401} (\bibinfo {year}
  {2009})}\BibitemShut {NoStop}%
\bibitem [{\citenamefont {Rohringer}\ \emph {et~al.}(2018)\citenamefont
  {Rohringer}, \citenamefont {Hafermann}, \citenamefont {Toschi}, \citenamefont
  {Katanin}, \citenamefont {Antipov}, \citenamefont {Katsnelson}, \citenamefont
  {Lichtenstein}, \citenamefont {Rubtsov},\ and\ \citenamefont
  {Held}}]{Rohringer_Held_2018}%
  \BibitemOpen
  \bibfield  {author} {\bibinfo {author} {\bibfnamefont {G.}~\bibnamefont
  {Rohringer}}, \bibinfo {author} {\bibfnamefont {H.}~\bibnamefont
  {Hafermann}}, \bibinfo {author} {\bibfnamefont {A.}~\bibnamefont {Toschi}},
  \bibinfo {author} {\bibfnamefont {A.~A.}\ \bibnamefont {Katanin}}, \bibinfo
  {author} {\bibfnamefont {A.~E.}\ \bibnamefont {Antipov}}, \bibinfo {author}
  {\bibfnamefont {M.~I.}\ \bibnamefont {Katsnelson}}, \bibinfo {author}
  {\bibfnamefont {A.~I.}\ \bibnamefont {Lichtenstein}}, \bibinfo {author}
  {\bibfnamefont {A.~N.}\ \bibnamefont {Rubtsov}}, \ and\ \bibinfo {author}
  {\bibfnamefont {K.}~\bibnamefont {Held}},\ }\bibfield  {title} {\emph
  {\enquote {\bibinfo {title} {Diagrammatic routes to nonlocal correlations
  beyond dynamical mean field theory},}\ }}\href {\doibase
  10.1103/RevModPhys.90.025003} {\bibfield  {journal} {\bibinfo  {journal}
  {Rev. Mod. Phys.}\ }\textbf {\bibinfo {volume} {90}},\ \bibinfo {pages}
  {025003} (\bibinfo {year} {2018})}\BibitemShut {NoStop}%
\bibitem [{\citenamefont {Stepanov}\ \emph {et~al.}(2016)\citenamefont
  {Stepanov}, \citenamefont {van Loon}, \citenamefont {Katanin}, \citenamefont
  {Lichtenstein}, \citenamefont {Katsnelson},\ and\ \citenamefont
  {Rubtsov}}]{Stepanov_Rubtsov_2016}%
  \BibitemOpen
  \bibfield  {author} {\bibinfo {author} {\bibfnamefont {E.~A.}\ \bibnamefont
  {Stepanov}}, \bibinfo {author} {\bibfnamefont {E.~G. C.~P.}\ \bibnamefont
  {van Loon}}, \bibinfo {author} {\bibfnamefont {A.~A.}\ \bibnamefont
  {Katanin}}, \bibinfo {author} {\bibfnamefont {A.~I.}\ \bibnamefont
  {Lichtenstein}}, \bibinfo {author} {\bibfnamefont {M.~I.}\ \bibnamefont
  {Katsnelson}}, \ and\ \bibinfo {author} {\bibfnamefont {A.~N.}\ \bibnamefont
  {Rubtsov}},\ }\bibfield  {title} {\emph {\enquote {\bibinfo {title}
  {Self-consistent dual boson approach to single-particle and collective
  excitations in correlated systems},}\ }}\href {\doibase
  10.1103/PhysRevB.93.045107} {\bibfield  {journal} {\bibinfo  {journal} {Phys.
  Rev. B}\ }\textbf {\bibinfo {volume} {93}},\ \bibinfo {pages} {045107}
  (\bibinfo {year} {2016})}\BibitemShut {NoStop}%
\bibitem [{\citenamefont {Peters}\ \emph {et~al.}(2019)\citenamefont {Peters},
  \citenamefont {van Loon}, \citenamefont {Rubtsov}, \citenamefont
  {Lichtenstein}, \citenamefont {Katsnelson},\ and\ \citenamefont
  {Stepanov}}]{Peters_Stepanov_2019}%
  \BibitemOpen
  \bibfield  {author} {\bibinfo {author} {\bibfnamefont {L.}~\bibnamefont
  {Peters}}, \bibinfo {author} {\bibfnamefont {E.~G. C.~P.}\ \bibnamefont {van
  Loon}}, \bibinfo {author} {\bibfnamefont {A.~N.}\ \bibnamefont {Rubtsov}},
  \bibinfo {author} {\bibfnamefont {A.~I.}\ \bibnamefont {Lichtenstein}},
  \bibinfo {author} {\bibfnamefont {M.~I.}\ \bibnamefont {Katsnelson}}, \ and\
  \bibinfo {author} {\bibfnamefont {E.~A.}\ \bibnamefont {Stepanov}},\
  }\bibfield  {title} {\emph {\enquote {\bibinfo {title} {Dual boson approach
  with instantaneous interaction},}\ }}\href {\doibase
  10.1103/PhysRevB.100.165128} {\bibfield  {journal} {\bibinfo  {journal}
  {Phys. Rev. B}\ }\textbf {\bibinfo {volume} {100}},\ \bibinfo {pages}
  {165128} (\bibinfo {year} {2019})}\BibitemShut {NoStop}%
\bibitem [{\citenamefont {Rohringer}\ \emph {et~al.}(2013)\citenamefont
  {Rohringer}, \citenamefont {Toschi}, \citenamefont {Hafermann}, \citenamefont
  {Held}, \citenamefont {Anisimov},\ and\ \citenamefont
  {Katanin}}]{Rohringer_Katanin_2013}%
  \BibitemOpen
  \bibfield  {author} {\bibinfo {author} {\bibfnamefont {G.}~\bibnamefont
  {Rohringer}}, \bibinfo {author} {\bibfnamefont {A.}~\bibnamefont {Toschi}},
  \bibinfo {author} {\bibfnamefont {H.}~\bibnamefont {Hafermann}}, \bibinfo
  {author} {\bibfnamefont {K.}~\bibnamefont {Held}}, \bibinfo {author}
  {\bibfnamefont {V.~I.}\ \bibnamefont {Anisimov}}, \ and\ \bibinfo {author}
  {\bibfnamefont {A.~A.}\ \bibnamefont {Katanin}},\ }\bibfield  {title} {\emph
  {\enquote {\bibinfo {title} {One-particle irreducible functional approach: A
  route to diagrammatic extensions of the dynamical mean-field theory},}\
  }}\href {\doibase 10.1103/PhysRevB.88.115112} {\bibfield  {journal} {\bibinfo
   {journal} {Phys. Rev. B}\ }\textbf {\bibinfo {volume} {88}},\ \bibinfo
  {pages} {115112} (\bibinfo {year} {2013})}\BibitemShut {NoStop}%
\bibitem [{\citenamefont {Ayral}\ and\ \citenamefont
  {Parcollet}(2015)}]{Ayral2015}%
  \BibitemOpen
  \bibfield  {author} {\bibinfo {author} {\bibfnamefont {T.}~\bibnamefont
  {Ayral}}\ and\ \bibinfo {author} {\bibfnamefont {O.}~\bibnamefont
  {Parcollet}},\ }\bibfield  {title} {\emph {\enquote {\bibinfo {title} {Mott
  physics and spin fluctuations: A unified framework},}\ }}\href {\doibase
  10.1103/PhysRevB.92.115109} {\bibfield  {journal} {\bibinfo  {journal} {Phys.
  Rev. B}\ }\textbf {\bibinfo {volume} {92}},\ \bibinfo {pages} {115109}
  (\bibinfo {year} {2015})}\BibitemShut {NoStop}%
\bibitem [{\citenamefont {Ayral}\ and\ \citenamefont
  {Parcollet}(2016{\natexlab{a}})}]{Ayral2016a}%
  \BibitemOpen
  \bibfield  {author} {\bibinfo {author} {\bibfnamefont {T.}~\bibnamefont
  {Ayral}}\ and\ \bibinfo {author} {\bibfnamefont {O.}~\bibnamefont
  {Parcollet}},\ }\bibfield  {title} {\emph {\enquote {\bibinfo {title} {Mott
  physics and spin fluctuations: A functional viewpoint},}\ }}\href {\doibase
  10.1103/PhysRevB.93.235124} {\bibfield  {journal} {\bibinfo  {journal} {Phys.
  Rev. B}\ }\textbf {\bibinfo {volume} {93}},\ \bibinfo {pages} {235124}
  (\bibinfo {year} {2016}{\natexlab{a}})}\BibitemShut {NoStop}%
\bibitem [{\citenamefont {Ayral}\ and\ \citenamefont
  {Parcollet}(2016{\natexlab{b}})}]{Ayral2016}%
  \BibitemOpen
  \bibfield  {author} {\bibinfo {author} {\bibfnamefont {T.}~\bibnamefont
  {Ayral}}\ and\ \bibinfo {author} {\bibfnamefont {O.}~\bibnamefont
  {Parcollet}},\ }\bibfield  {title} {\emph {\enquote {\bibinfo {title} {Mott
  physics and collective modes: An atomic approximation of the four-particle
  irreducible functional},}\ }}\href {\doibase 10.1103/PhysRevB.94.075159}
  {\bibfield  {journal} {\bibinfo  {journal} {Phys. Rev. B}\ }\textbf {\bibinfo
  {volume} {94}},\ \bibinfo {pages} {075159} (\bibinfo {year}
  {2016}{\natexlab{b}})}\BibitemShut {NoStop}%
\bibitem [{\citenamefont {Katanin}\ \emph {et~al.}(2009)\citenamefont
  {Katanin}, \citenamefont {Toschi},\ and\ \citenamefont
  {Held}}]{Katanin_Held_2009}%
  \BibitemOpen
  \bibfield  {author} {\bibinfo {author} {\bibfnamefont {A.~A.}\ \bibnamefont
  {Katanin}}, \bibinfo {author} {\bibfnamefont {A.}~\bibnamefont {Toschi}}, \
  and\ \bibinfo {author} {\bibfnamefont {K.}~\bibnamefont {Held}},\ }\bibfield
  {title} {\emph {\enquote {\bibinfo {title} {Comparing pertinent effects of
  antiferromagnetic fluctuations in the two- and three-dimensional Hubbard
  model},}\ }}\href {\doibase 10.1103/PhysRevB.80.075104} {\bibfield  {journal}
  {\bibinfo  {journal} {Phys. Rev. B}\ }\textbf {\bibinfo {volume} {80}},\
  \bibinfo {pages} {075104} (\bibinfo {year} {2009})}\BibitemShut {NoStop}%
\bibitem [{\citenamefont {Toschi}\ \emph {et~al.}(2007)\citenamefont {Toschi},
  \citenamefont {Katanin},\ and\ \citenamefont {Held}}]{Toschi_Held_2007}%
  \BibitemOpen
  \bibfield  {author} {\bibinfo {author} {\bibfnamefont {A.}~\bibnamefont
  {Toschi}}, \bibinfo {author} {\bibfnamefont {A.~A.}\ \bibnamefont {Katanin}},
  \ and\ \bibinfo {author} {\bibfnamefont {K.}~\bibnamefont {Held}},\
  }\bibfield  {title} {\emph {\enquote {\bibinfo {title} {Dynamical vertex
  approximation: A step beyond dynamical mean-field theory},}\ }}\href
  {\doibase 10.1103/PhysRevB.75.045118} {\bibfield  {journal} {\bibinfo
  {journal} {Phys. Rev. B}\ }\textbf {\bibinfo {volume} {75}},\ \bibinfo
  {pages} {045118} (\bibinfo {year} {2007})}\BibitemShut {NoStop}%
\bibitem [{\citenamefont {Rohringer}\ and\ \citenamefont
  {Toschi}(2016)}]{Rohringer_Toschi_2016}%
  \BibitemOpen
  \bibfield  {author} {\bibinfo {author} {\bibfnamefont {G.}~\bibnamefont
  {Rohringer}}\ and\ \bibinfo {author} {\bibfnamefont {A.}~\bibnamefont
  {Toschi}},\ }\bibfield  {title} {\emph {\enquote {\bibinfo {title} {Impact of
  nonlocal correlations over different energy scales: A dynamical vertex
  approximation study},}\ }}\href {\doibase 10.1103/PhysRevB.94.125144}
  {\bibfield  {journal} {\bibinfo  {journal} {Phys. Rev. B}\ }\textbf {\bibinfo
  {volume} {94}},\ \bibinfo {pages} {125144} (\bibinfo {year}
  {2016})}\BibitemShut {NoStop}%
\bibitem [{\citenamefont {Stobbe}\ and\ \citenamefont
  {Rohringer}(2022)}]{Stobbe_Rohringer_2022}%
  \BibitemOpen
  \bibfield  {author} {\bibinfo {author} {\bibfnamefont {J.}~\bibnamefont
  {Stobbe}}\ and\ \bibinfo {author} {\bibfnamefont {G.}~\bibnamefont
  {Rohringer}},\ }\bibfield  {title} {\emph {\enquote {\bibinfo {title}
  {Consistency of potential energy in the dynamical vertex approximation},}\
  }}\href {\doibase 10.1103/PhysRevB.106.205101} {\bibfield  {journal}
  {\bibinfo  {journal} {Phys. Rev. B}\ }\textbf {\bibinfo {volume} {106}},\
  \bibinfo {pages} {205101} (\bibinfo {year} {2022})}\BibitemShut {NoStop}%
\bibitem [{\citenamefont {Tam}\ \emph {et~al.}(2013)\citenamefont {Tam},
  \citenamefont {Fotso}, \citenamefont {Yang}, \citenamefont {Lee},
  \citenamefont {Moreno}, \citenamefont {Ramanujam},\ and\ \citenamefont
  {Jarrell}}]{Tam_Jarrell_2013}%
  \BibitemOpen
  \bibfield  {author} {\bibinfo {author} {\bibfnamefont {K.-M.}\ \bibnamefont
  {Tam}}, \bibinfo {author} {\bibfnamefont {H.}~\bibnamefont {Fotso}}, \bibinfo
  {author} {\bibfnamefont {S.-X.}\ \bibnamefont {Yang}}, \bibinfo {author}
  {\bibfnamefont {T.-W.}\ \bibnamefont {Lee}}, \bibinfo {author} {\bibfnamefont
  {J.}~\bibnamefont {Moreno}}, \bibinfo {author} {\bibfnamefont
  {J.}~\bibnamefont {Ramanujam}}, \ and\ \bibinfo {author} {\bibfnamefont
  {M.}~\bibnamefont {Jarrell}},\ }\bibfield  {title} {\emph {\enquote {\bibinfo
  {title} {Solving the parquet equations for the Hubbard model beyond weak
  coupling},}\ }}\href {\doibase 10.1103/PhysRevE.87.013311} {\bibfield
  {journal} {\bibinfo  {journal} {Phys. Rev. E}\ }\textbf {\bibinfo {volume}
  {87}},\ \bibinfo {pages} {013311} (\bibinfo {year} {2013})}\BibitemShut
  {NoStop}%
\bibitem [{\citenamefont {Chen}\ and\ \citenamefont
  {Bickers}(1992)}]{Chen_Bickers_1992}%
  \BibitemOpen
  \bibfield  {author} {\bibinfo {author} {\bibfnamefont {C.-X.}\ \bibnamefont
  {Chen}}\ and\ \bibinfo {author} {\bibfnamefont {N.}~\bibnamefont {Bickers}},\
  }\bibfield  {title} {\emph {\enquote {\bibinfo {title} {Numerical solution of
  parquet equations for the Anderson impurity model},}\ }}\href {\doibase
  https://doi.org/10.1016/0038-1098(92)90358-G} {\bibfield  {journal} {\bibinfo
   {journal} {Solid State Communications}\ }\textbf {\bibinfo {volume} {82}},\
  \bibinfo {pages} {311} (\bibinfo {year} {1992})}\BibitemShut {NoStop}%
\bibitem [{\citenamefont {Valli}\ \emph {et~al.}(2015)\citenamefont {Valli},
  \citenamefont {Sch\"afer}, \citenamefont {Thunstr\"om}, \citenamefont
  {Rohringer}, \citenamefont {Andergassen}, \citenamefont {Sangiovanni},
  \citenamefont {Held},\ and\ \citenamefont {Toschi}}]{Vali_Toschi_2015}%
  \BibitemOpen
  \bibfield  {author} {\bibinfo {author} {\bibfnamefont {A.}~\bibnamefont
  {Valli}}, \bibinfo {author} {\bibfnamefont {T.}~\bibnamefont {Sch\"afer}},
  \bibinfo {author} {\bibfnamefont {P.}~\bibnamefont {Thunstr\"om}}, \bibinfo
  {author} {\bibfnamefont {G.}~\bibnamefont {Rohringer}}, \bibinfo {author}
  {\bibfnamefont {S.}~\bibnamefont {Andergassen}}, \bibinfo {author}
  {\bibfnamefont {G.}~\bibnamefont {Sangiovanni}}, \bibinfo {author}
  {\bibfnamefont {K.}~\bibnamefont {Held}}, \ and\ \bibinfo {author}
  {\bibfnamefont {A.}~\bibnamefont {Toschi}},\ }\bibfield  {title} {\emph
  {\enquote {\bibinfo {title} {Dynamical vertex approximation in its parquet
  implementation: Application to Hubbard nanorings},}\ }}\href {\doibase
  10.1103/PhysRevB.91.115115} {\bibfield  {journal} {\bibinfo  {journal} {Phys.
  Rev. B}\ }\textbf {\bibinfo {volume} {91}},\ \bibinfo {pages} {115115}
  (\bibinfo {year} {2015})}\BibitemShut {NoStop}%
\bibitem [{\citenamefont {Li}\ \emph {et~al.}(2016)\citenamefont {Li},
  \citenamefont {Wentzell}, \citenamefont {Pudleiner}, \citenamefont
  {Thunstr\"om},\ and\ \citenamefont {Held}}]{Li_Held_2016}%
  \BibitemOpen
  \bibfield  {author} {\bibinfo {author} {\bibfnamefont {G.}~\bibnamefont
  {Li}}, \bibinfo {author} {\bibfnamefont {N.}~\bibnamefont {Wentzell}},
  \bibinfo {author} {\bibfnamefont {P.}~\bibnamefont {Pudleiner}}, \bibinfo
  {author} {\bibfnamefont {P.}~\bibnamefont {Thunstr\"om}}, \ and\ \bibinfo
  {author} {\bibfnamefont {K.}~\bibnamefont {Held}},\ }\bibfield  {title}
  {\emph {\enquote {\bibinfo {title} {Efficient implementation of the parquet
  equations: Role of the reducible vertex function and its kernel
  approximation},}\ }}\href {\doibase 10.1103/PhysRevB.93.165103} {\bibfield
  {journal} {\bibinfo  {journal} {Phys. Rev. B}\ }\textbf {\bibinfo {volume}
  {93}},\ \bibinfo {pages} {165103} (\bibinfo {year} {2016})}\BibitemShut
  {NoStop}%
\bibitem [{\citenamefont {Kozik}\ \emph {et~al.}(2015)\citenamefont {Kozik},
  \citenamefont {Ferrero},\ and\ \citenamefont {Georges}}]{Kozik2015}%
  \BibitemOpen
  \bibfield  {author} {\bibinfo {author} {\bibfnamefont {E.}~\bibnamefont
  {Kozik}}, \bibinfo {author} {\bibfnamefont {M.}~\bibnamefont {Ferrero}}, \
  and\ \bibinfo {author} {\bibfnamefont {A.}~\bibnamefont {Georges}},\
  }\bibfield  {title} {\emph {\enquote {\bibinfo {title} {Nonexistence of the
  Luttinger-Ward Functional and Misleading Convergence of Skeleton Diagrammatic
  Series for Hubbard-Like Models},}\ }}\href {\doibase
  10.1103/PhysRevLett.114.156402} {\bibfield  {journal} {\bibinfo  {journal}
  {Phys. Rev. Lett.}\ }\textbf {\bibinfo {volume} {114}},\ \bibinfo {pages}
  {156402} (\bibinfo {year} {2015})}\BibitemShut {NoStop}%
\bibitem [{\citenamefont {Gunnarsson}\ \emph {et~al.}(2017)\citenamefont
  {Gunnarsson}, \citenamefont {Rohringer}, \citenamefont {Sch\"afer},
  \citenamefont {Sangiovanni},\ and\ \citenamefont {Toschi}}]{Gunnarsson2017}%
  \BibitemOpen
  \bibfield  {author} {\bibinfo {author} {\bibfnamefont {O.}~\bibnamefont
  {Gunnarsson}}, \bibinfo {author} {\bibfnamefont {G.}~\bibnamefont
  {Rohringer}}, \bibinfo {author} {\bibfnamefont {T.}~\bibnamefont
  {Sch\"afer}}, \bibinfo {author} {\bibfnamefont {G.}~\bibnamefont
  {Sangiovanni}}, \ and\ \bibinfo {author} {\bibfnamefont {A.}~\bibnamefont
  {Toschi}},\ }\bibfield  {title} {\emph {\enquote {\bibinfo {title} {Breakdown
  of Traditional Many-Body Theories for Correlated Electrons},}\ }}\href
  {\doibase 10.1103/PhysRevLett.119.056402} {\bibfield  {journal} {\bibinfo
  {journal} {Phys. Rev. Lett.}\ }\textbf {\bibinfo {volume} {119}},\ \bibinfo
  {pages} {056402} (\bibinfo {year} {2017})}\BibitemShut {NoStop}%
\bibitem [{\citenamefont {Sch\"afer}\ \emph {et~al.}(2013)\citenamefont
  {Sch\"afer}, \citenamefont {Rohringer}, \citenamefont {Gunnarsson},
  \citenamefont {Ciuchi}, \citenamefont {Sangiovanni},\ and\ \citenamefont
  {Toschi}}]{Schaefer_2013}%
  \BibitemOpen
  \bibfield  {author} {\bibinfo {author} {\bibfnamefont {T.}~\bibnamefont
  {Sch\"afer}}, \bibinfo {author} {\bibfnamefont {G.}~\bibnamefont
  {Rohringer}}, \bibinfo {author} {\bibfnamefont {O.}~\bibnamefont
  {Gunnarsson}}, \bibinfo {author} {\bibfnamefont {S.}~\bibnamefont {Ciuchi}},
  \bibinfo {author} {\bibfnamefont {G.}~\bibnamefont {Sangiovanni}}, \ and\
  \bibinfo {author} {\bibfnamefont {A.}~\bibnamefont {Toschi}},\ }\bibfield
  {title} {\emph {\enquote {\bibinfo {title} {Divergent Precursors of the
  Mott-Hubbard Transition at the Two-Particle Level},}\ }}\href {\doibase
  10.1103/PhysRevLett.110.246405} {\bibfield  {journal} {\bibinfo  {journal}
  {Phys. Rev. Lett.}\ }\textbf {\bibinfo {volume} {110}},\ \bibinfo {pages}
  {246405} (\bibinfo {year} {2013})}\BibitemShut {NoStop}%
\bibitem [{\citenamefont {Sch\"afer}\ \emph {et~al.}(2016)\citenamefont
  {Sch\"afer}, \citenamefont {Ciuchi}, \citenamefont {Wallerberger},
  \citenamefont {Thunstr\"om}, \citenamefont {Gunnarsson}, \citenamefont
  {Sangiovanni}, \citenamefont {Rohringer},\ and\ \citenamefont
  {Toschi}}]{Schaefer_2016}%
  \BibitemOpen
  \bibfield  {author} {\bibinfo {author} {\bibfnamefont {T.}~\bibnamefont
  {Sch\"afer}}, \bibinfo {author} {\bibfnamefont {S.}~\bibnamefont {Ciuchi}},
  \bibinfo {author} {\bibfnamefont {M.}~\bibnamefont {Wallerberger}}, \bibinfo
  {author} {\bibfnamefont {P.}~\bibnamefont {Thunstr\"om}}, \bibinfo {author}
  {\bibfnamefont {O.}~\bibnamefont {Gunnarsson}}, \bibinfo {author}
  {\bibfnamefont {G.}~\bibnamefont {Sangiovanni}}, \bibinfo {author}
  {\bibfnamefont {G.}~\bibnamefont {Rohringer}}, \ and\ \bibinfo {author}
  {\bibfnamefont {A.}~\bibnamefont {Toschi}},\ }\bibfield  {title} {\emph
  {\enquote {\bibinfo {title} {Nonperturbative landscape of the Mott-Hubbard
  transition: Multiple divergence lines around the critical endpoint},}\
  }}\href {\doibase 10.1103/PhysRevB.94.235108} {\bibfield  {journal} {\bibinfo
   {journal} {Phys. Rev. B}\ }\textbf {\bibinfo {volume} {94}},\ \bibinfo
  {pages} {235108} (\bibinfo {year} {2016})}\BibitemShut {NoStop}%
\bibitem [{\citenamefont {Thunstr\"om}\ \emph {et~al.}(2018)\citenamefont
  {Thunstr\"om}, \citenamefont {Gunnarsson}, \citenamefont {Ciuchi},\ and\
  \citenamefont {Rohringer}}]{Thunstroem_2018}%
  \BibitemOpen
  \bibfield  {author} {\bibinfo {author} {\bibfnamefont {P.}~\bibnamefont
  {Thunstr\"om}}, \bibinfo {author} {\bibfnamefont {O.}~\bibnamefont
  {Gunnarsson}}, \bibinfo {author} {\bibfnamefont {S.}~\bibnamefont {Ciuchi}},
  \ and\ \bibinfo {author} {\bibfnamefont {G.}~\bibnamefont {Rohringer}},\
  }\bibfield  {title} {\emph {\enquote {\bibinfo {title} {Analytical
  investigation of singularities in two-particle irreducible vertex functions
  of the Hubbard atom},}\ }}\href {\doibase 10.1103/PhysRevB.98.235107}
  {\bibfield  {journal} {\bibinfo  {journal} {Phys. Rev. B}\ }\textbf {\bibinfo
  {volume} {98}},\ \bibinfo {pages} {235107} (\bibinfo {year}
  {2018})}\BibitemShut {NoStop}%
\bibitem [{\citenamefont {Chalupa}\ \emph {et~al.}(2018)\citenamefont
  {Chalupa}, \citenamefont {Gunacker}, \citenamefont {Sch\"afer}, \citenamefont
  {Held},\ and\ \citenamefont {Toschi}}]{Chalupa2018}%
  \BibitemOpen
  \bibfield  {author} {\bibinfo {author} {\bibfnamefont {P.}~\bibnamefont
  {Chalupa}}, \bibinfo {author} {\bibfnamefont {P.}~\bibnamefont {Gunacker}},
  \bibinfo {author} {\bibfnamefont {T.}~\bibnamefont {Sch\"afer}}, \bibinfo
  {author} {\bibfnamefont {K.}~\bibnamefont {Held}}, \ and\ \bibinfo {author}
  {\bibfnamefont {A.}~\bibnamefont {Toschi}},\ }\bibfield  {title} {\emph
  {\enquote {\bibinfo {title} {Divergences of the irreducible vertex functions
  in correlated metallic systems: Insights from the Anderson impurity model},}\
  }}\href {\doibase 10.1103/PhysRevB.97.245136} {\bibfield  {journal} {\bibinfo
   {journal} {Phys. Rev. B}\ }\textbf {\bibinfo {volume} {97}},\ \bibinfo
  {pages} {245136} (\bibinfo {year} {2018})}\BibitemShut {NoStop}%
\bibitem [{\citenamefont {Springer}\ \emph {et~al.}(2020)\citenamefont
  {Springer}, \citenamefont {Chalupa}, \citenamefont {Ciuchi}, \citenamefont
  {Sangiovanni},\ and\ \citenamefont {Toschi}}]{Springer2020}%
  \BibitemOpen
  \bibfield  {author} {\bibinfo {author} {\bibfnamefont {D.}~\bibnamefont
  {Springer}}, \bibinfo {author} {\bibfnamefont {P.}~\bibnamefont {Chalupa}},
  \bibinfo {author} {\bibfnamefont {S.}~\bibnamefont {Ciuchi}}, \bibinfo
  {author} {\bibfnamefont {G.}~\bibnamefont {Sangiovanni}}, \ and\ \bibinfo
  {author} {\bibfnamefont {A.}~\bibnamefont {Toschi}},\ }\bibfield  {title}
  {\emph {\enquote {\bibinfo {title} {Interplay between local response and
  vertex divergences in many-fermion systems with on-site attraction},}\
  }}\href {\doibase 10.1103/PhysRevB.101.155148} {\bibfield  {journal}
  {\bibinfo  {journal} {Phys. Rev. B}\ }\textbf {\bibinfo {volume} {101}},\
  \bibinfo {pages} {155148} (\bibinfo {year} {2020})}\BibitemShut {NoStop}%
\bibitem [{\citenamefont {Lihm}\ \emph {et~al.}(2025)\citenamefont {Lihm},
  \citenamefont {Kiese}, \citenamefont {Lee},\ and\ \citenamefont
  {Kugler}}]{Lihm_2025}%
  \BibitemOpen
  \bibfield  {author} {\bibinfo {author} {\bibfnamefont {J.-M.}\ \bibnamefont
  {Lihm}}, \bibinfo {author} {\bibfnamefont {D.}~\bibnamefont {Kiese}},
  \bibinfo {author} {\bibfnamefont {S.-S.~B.}\ \bibnamefont {Lee}}, \ and\
  \bibinfo {author} {\bibfnamefont {F.~B.}\ \bibnamefont {Kugler}},\ }\href
  {https://arxiv.org/abs/2505.20116} {\enquote {\bibinfo {title} {The
  finite-difference parquet method: Enhanced electron-paramagnon scattering
  opens a pseudogap},}\ } (\bibinfo {year} {2025}),\ \Eprint
  {http://arxiv.org/abs/2505.20116}{arXiv:2505.20116
  [cond-mat.str-el]}\BibitemShut {NoStop}%
\bibitem [{\citenamefont {Tagliavini}\ \emph {et~al.}(2018)\citenamefont
  {Tagliavini}, \citenamefont {Hummel}, \citenamefont {Wentzell}, \citenamefont
  {Andergassen}, \citenamefont {Toschi},\ and\ \citenamefont
  {Rohringer}}]{Tagliavini2018}%
  \BibitemOpen
  \bibfield  {author} {\bibinfo {author} {\bibfnamefont {A.}~\bibnamefont
  {Tagliavini}}, \bibinfo {author} {\bibfnamefont {S.}~\bibnamefont {Hummel}},
  \bibinfo {author} {\bibfnamefont {N.}~\bibnamefont {Wentzell}}, \bibinfo
  {author} {\bibfnamefont {S.}~\bibnamefont {Andergassen}}, \bibinfo {author}
  {\bibfnamefont {A.}~\bibnamefont {Toschi}}, \ and\ \bibinfo {author}
  {\bibfnamefont {G.}~\bibnamefont {Rohringer}},\ }\bibfield  {title} {\emph
  {\enquote {\bibinfo {title} {Efficient Bethe-Salpeter equation treatment in
  dynamical mean-field theory},}\ }}\href {\doibase 10.1103/PhysRevB.97.235140}
  {\bibfield  {journal} {\bibinfo  {journal} {Phys. Rev. B}\ }\textbf {\bibinfo
  {volume} {97}},\ \bibinfo {pages} {235140} (\bibinfo {year}
  {2018})}\BibitemShut {NoStop}%
\bibitem [{\citenamefont {Wentzell}\ \emph {et~al.}(2020)\citenamefont
  {Wentzell}, \citenamefont {Li}, \citenamefont {Tagliavini}, \citenamefont
  {Taranto}, \citenamefont {Rohringer}, \citenamefont {Held}, \citenamefont
  {Toschi},\ and\ \citenamefont {Andergassen}}]{Wentzell2020}%
  \BibitemOpen
  \bibfield  {author} {\bibinfo {author} {\bibfnamefont {N.}~\bibnamefont
  {Wentzell}}, \bibinfo {author} {\bibfnamefont {G.}~\bibnamefont {Li}},
  \bibinfo {author} {\bibfnamefont {A.}~\bibnamefont {Tagliavini}}, \bibinfo
  {author} {\bibfnamefont {C.}~\bibnamefont {Taranto}}, \bibinfo {author}
  {\bibfnamefont {G.}~\bibnamefont {Rohringer}}, \bibinfo {author}
  {\bibfnamefont {K.}~\bibnamefont {Held}}, \bibinfo {author} {\bibfnamefont
  {A.}~\bibnamefont {Toschi}}, \ and\ \bibinfo {author} {\bibfnamefont
  {S.}~\bibnamefont {Andergassen}},\ }\bibfield  {title} {\emph {\enquote
  {\bibinfo {title} {High-frequency asymptotics of the vertex function:
  Diagrammatic parametrization and algorithmic implementation},}\ }}\href
  {\doibase 10.1103/PhysRevB.102.085106} {\bibfield  {journal} {\bibinfo
  {journal} {Phys. Rev. B}\ }\textbf {\bibinfo {volume} {102}},\ \bibinfo
  {pages} {085106} (\bibinfo {year} {2020})}\BibitemShut {NoStop}%
\bibitem [{\citenamefont {Krien}\ \emph {et~al.}(2019)\citenamefont {Krien},
  \citenamefont {Valli},\ and\ \citenamefont {Capone}}]{Krien2019}%
  \BibitemOpen
  \bibfield  {author} {\bibinfo {author} {\bibfnamefont {F.}~\bibnamefont
  {Krien}}, \bibinfo {author} {\bibfnamefont {A.}~\bibnamefont {Valli}}, \ and\
  \bibinfo {author} {\bibfnamefont {M.}~\bibnamefont {Capone}},\ }\bibfield
  {title} {\emph {\enquote {\bibinfo {title} {Single-boson exchange
  decomposition of the vertex function},}\ }}\href {\doibase
  10.1103/PhysRevB.100.155149} {\bibfield  {journal} {\bibinfo  {journal}
  {Phys. Rev. B}\ }\textbf {\bibinfo {volume} {100}},\ \bibinfo {pages}
  {155149} (\bibinfo {year} {2019})}\BibitemShut {NoStop}%
\bibitem [{\citenamefont {Krien}\ and\ \citenamefont
  {Valli}(2019)}]{Krien2019a}%
  \BibitemOpen
  \bibfield  {author} {\bibinfo {author} {\bibfnamefont {F.}~\bibnamefont
  {Krien}}\ and\ \bibinfo {author} {\bibfnamefont {A.}~\bibnamefont {Valli}},\
  }\bibfield  {title} {\emph {\enquote {\bibinfo {title} {Parquetlike equations
  for the Hedin three-leg vertex},}\ }}\href {\doibase
  10.1103/PhysRevB.100.245147} {\bibfield  {journal} {\bibinfo  {journal}
  {Phys. Rev. B}\ }\textbf {\bibinfo {volume} {100}},\ \bibinfo {pages}
  {245147} (\bibinfo {year} {2019})}\BibitemShut {NoStop}%
\bibitem [{\citenamefont {Krien}\ \emph
  {et~al.}(2020{\natexlab{a}})\citenamefont {Krien}, \citenamefont
  {Lichtenstein},\ and\ \citenamefont {Rohringer}}]{Krien_2020}%
  \BibitemOpen
  \bibfield  {author} {\bibinfo {author} {\bibfnamefont {F.}~\bibnamefont
  {Krien}}, \bibinfo {author} {\bibfnamefont {A.~I.}\ \bibnamefont
  {Lichtenstein}}, \ and\ \bibinfo {author} {\bibfnamefont {G.}~\bibnamefont
  {Rohringer}},\ }\bibfield  {title} {\emph {\enquote {\bibinfo {title}
  {Fluctuation diagnostic of the nodal/antinodal dichotomy in the Hubbard model
  at weak coupling: A parquet dual fermion approach},}\ }}\href {\doibase
  10.1103/PhysRevB.102.235133} {\bibfield  {journal} {\bibinfo  {journal}
  {Phys. Rev. B}\ }\textbf {\bibinfo {volume} {102}},\ \bibinfo {pages}
  {235133} (\bibinfo {year} {2020}{\natexlab{a}})}\BibitemShut {NoStop}%
\bibitem [{\citenamefont {Krien}\ \emph
  {et~al.}(2020{\natexlab{b}})\citenamefont {Krien}, \citenamefont {Valli},
  \citenamefont {Chalupa}, \citenamefont {Capone}, \citenamefont
  {Lichtenstein},\ and\ \citenamefont {Toschi}}]{Krien_2020a}%
  \BibitemOpen
  \bibfield  {author} {\bibinfo {author} {\bibfnamefont {F.}~\bibnamefont
  {Krien}}, \bibinfo {author} {\bibfnamefont {A.}~\bibnamefont {Valli}},
  \bibinfo {author} {\bibfnamefont {P.}~\bibnamefont {Chalupa}}, \bibinfo
  {author} {\bibfnamefont {M.}~\bibnamefont {Capone}}, \bibinfo {author}
  {\bibfnamefont {A.~I.}\ \bibnamefont {Lichtenstein}}, \ and\ \bibinfo
  {author} {\bibfnamefont {A.}~\bibnamefont {Toschi}},\ }\bibfield  {title}
  {\emph {\enquote {\bibinfo {title} {Boson-exchange parquet solver for dual
  fermions},}\ }}\href {\doibase 10.1103/PhysRevB.102.195131} {\bibfield
  {journal} {\bibinfo  {journal} {Phys. Rev. B}\ }\textbf {\bibinfo {volume}
  {102}},\ \bibinfo {pages} {195131} (\bibinfo {year}
  {2020}{\natexlab{b}})}\BibitemShut {NoStop}%
\bibitem [{\citenamefont {Krien}\ \emph {et~al.}(2022)\citenamefont {Krien},
  \citenamefont {Worm}, \citenamefont {Chalupa-Gantner}, \citenamefont
  {Toschi},\ and\ \citenamefont {Held}}]{Krien_2022}%
  \BibitemOpen
  \bibfield  {author} {\bibinfo {author} {\bibfnamefont {F.}~\bibnamefont
  {Krien}}, \bibinfo {author} {\bibfnamefont {P.}~\bibnamefont {Worm}},
  \bibinfo {author} {\bibfnamefont {P.}~\bibnamefont {Chalupa-Gantner}},
  \bibinfo {author} {\bibfnamefont {A.}~\bibnamefont {Toschi}}, \ and\ \bibinfo
  {author} {\bibfnamefont {K.}~\bibnamefont {Held}},\ }\bibfield  {title}
  {\emph {\enquote {\bibinfo {title} {Explaining the pseudogap through damping
  and antidamping on the Fermi surface by imaginary spin scattering},}\ }}\href
  {\doibase 10.1038/s42005-022-01117-5} {\bibfield  {journal} {\bibinfo
  {journal} {Communications Physics}\ }\textbf {\bibinfo {volume} {5}},\
  \bibinfo {pages} {336} (\bibinfo {year} {2022})}\BibitemShut {NoStop}%
\bibitem [{\citenamefont {Eckhardt}\ \emph {et~al.}(2018)\citenamefont
  {Eckhardt}, \citenamefont {Schober}, \citenamefont {Ehrlich},\ and\
  \citenamefont {Honerkamp}}]{Eckhardt2018}%
  \BibitemOpen
  \bibfield  {author} {\bibinfo {author} {\bibfnamefont {C.~J.}\ \bibnamefont
  {Eckhardt}}, \bibinfo {author} {\bibfnamefont {G.~A.~H.}\ \bibnamefont
  {Schober}}, \bibinfo {author} {\bibfnamefont {J.}~\bibnamefont {Ehrlich}}, \
  and\ \bibinfo {author} {\bibfnamefont {C.}~\bibnamefont {Honerkamp}},\
  }\bibfield  {title} {\emph {\enquote {\bibinfo {title} {Truncated-unity
  parquet equations: Application to the repulsive Hubbard model},}\ }}\href
  {\doibase 10.1103/PhysRevB.98.075143} {\bibfield  {journal} {\bibinfo
  {journal} {Phys. Rev. B}\ }\textbf {\bibinfo {volume} {98}},\ \bibinfo
  {pages} {075143} (\bibinfo {year} {2018})}\BibitemShut {NoStop}%
\bibitem [{\citenamefont {Eckhardt}\ \emph {et~al.}(2020)\citenamefont
  {Eckhardt}, \citenamefont {Honerkamp}, \citenamefont {Held},\ and\
  \citenamefont {Kauch}}]{Eckhardt2020}%
  \BibitemOpen
  \bibfield  {author} {\bibinfo {author} {\bibfnamefont {C.~J.}\ \bibnamefont
  {Eckhardt}}, \bibinfo {author} {\bibfnamefont {C.}~\bibnamefont {Honerkamp}},
  \bibinfo {author} {\bibfnamefont {K.}~\bibnamefont {Held}}, \ and\ \bibinfo
  {author} {\bibfnamefont {A.}~\bibnamefont {Kauch}},\ }\bibfield  {title}
  {\emph {\enquote {\bibinfo {title} {Truncated unity parquet solver},}\
  }}\href {\doibase 10.1103/PhysRevB.101.155104} {\bibfield  {journal}
  {\bibinfo  {journal} {Phys. Rev. B}\ }\textbf {\bibinfo {volume} {101}},\
  \bibinfo {pages} {155104} (\bibinfo {year} {2020})}\BibitemShut {NoStop}%
\bibitem [{\citenamefont {Wallerberger}\ \emph {et~al.}(2021)\citenamefont
  {Wallerberger}, \citenamefont {Shinaoka},\ and\ \citenamefont
  {Kauch}}]{Wallerberger2021}%
  \BibitemOpen
  \bibfield  {author} {\bibinfo {author} {\bibfnamefont {M.}~\bibnamefont
  {Wallerberger}}, \bibinfo {author} {\bibfnamefont {H.}~\bibnamefont
  {Shinaoka}}, \ and\ \bibinfo {author} {\bibfnamefont {A.}~\bibnamefont
  {Kauch}},\ }\bibfield  {title} {\emph {\enquote {\bibinfo {title} {Solving
  the Bethe-Salpeter equation with exponential convergence},}\ }}\href
  {\doibase 10.1103/PhysRevResearch.3.033168} {\bibfield  {journal} {\bibinfo
  {journal} {Phys. Rev. Res.}\ }\textbf {\bibinfo {volume} {3}},\ \bibinfo
  {pages} {033168} (\bibinfo {year} {2021})}\BibitemShut {NoStop}%
\bibitem [{\citenamefont {Wallerberger}\ \emph {et~al.}(2023)\citenamefont
  {Wallerberger}, \citenamefont {Badr}, \citenamefont {Hoshino}, \citenamefont
  {Huber}, \citenamefont {Kakizawa}, \citenamefont {Koretsune}, \citenamefont
  {Nagai}, \citenamefont {Nogaki}, \citenamefont {Nomoto}, \citenamefont
  {Mori}, \citenamefont {Otsuki}, \citenamefont {Ozaki}, \citenamefont
  {Plaikner}, \citenamefont {Sakurai}, \citenamefont {Vogel}, \citenamefont
  {Witt}, \citenamefont {Yoshimi},\ and\ \citenamefont
  {Shinaoka}}]{Wallerberger2023}%
  \BibitemOpen
  \bibfield  {author} {\bibinfo {author} {\bibfnamefont {M.}~\bibnamefont
  {Wallerberger}}, \bibinfo {author} {\bibfnamefont {S.}~\bibnamefont {Badr}},
  \bibinfo {author} {\bibfnamefont {S.}~\bibnamefont {Hoshino}}, \bibinfo
  {author} {\bibfnamefont {S.}~\bibnamefont {Huber}}, \bibinfo {author}
  {\bibfnamefont {F.}~\bibnamefont {Kakizawa}}, \bibinfo {author}
  {\bibfnamefont {T.}~\bibnamefont {Koretsune}}, \bibinfo {author}
  {\bibfnamefont {Y.}~\bibnamefont {Nagai}}, \bibinfo {author} {\bibfnamefont
  {K.}~\bibnamefont {Nogaki}}, \bibinfo {author} {\bibfnamefont
  {T.}~\bibnamefont {Nomoto}}, \bibinfo {author} {\bibfnamefont
  {H.}~\bibnamefont {Mori}}, \bibinfo {author} {\bibfnamefont {J.}~\bibnamefont
  {Otsuki}}, \bibinfo {author} {\bibfnamefont {S.}~\bibnamefont {Ozaki}},
  \bibinfo {author} {\bibfnamefont {T.}~\bibnamefont {Plaikner}}, \bibinfo
  {author} {\bibfnamefont {R.}~\bibnamefont {Sakurai}}, \bibinfo {author}
  {\bibfnamefont {C.}~\bibnamefont {Vogel}}, \bibinfo {author} {\bibfnamefont
  {N.}~\bibnamefont {Witt}}, \bibinfo {author} {\bibfnamefont {K.}~\bibnamefont
  {Yoshimi}}, \ and\ \bibinfo {author} {\bibfnamefont {H.}~\bibnamefont
  {Shinaoka}},\ }\bibfield  {title} {\emph {\enquote {\bibinfo {title}
  {sparse-ir: Optimal compression and sparse sampling of many-body
  propagators},}\ }}\href {\doibase
  https://doi.org/10.1016/j.softx.2022.101266} {\bibfield  {journal} {\bibinfo
  {journal} {SoftwareX}\ }\textbf {\bibinfo {volume} {21}},\ \bibinfo {pages}
  {101266} (\bibinfo {year} {2023})}\BibitemShut {NoStop}%
\bibitem [{\citenamefont {Rohshap}\ \emph {et~al.}(2025)\citenamefont
  {Rohshap}, \citenamefont {Ritter}, \citenamefont {Shinaoka}, \citenamefont
  {von Delft}, \citenamefont {Wallerberger},\ and\ \citenamefont
  {Kauch}}]{Rohshap_2025}%
  \BibitemOpen
  \bibfield  {author} {\bibinfo {author} {\bibfnamefont {S.}~\bibnamefont
  {Rohshap}}, \bibinfo {author} {\bibfnamefont {M.~K.}\ \bibnamefont {Ritter}},
  \bibinfo {author} {\bibfnamefont {H.}~\bibnamefont {Shinaoka}}, \bibinfo
  {author} {\bibfnamefont {J.}~\bibnamefont {von Delft}}, \bibinfo {author}
  {\bibfnamefont {M.}~\bibnamefont {Wallerberger}}, \ and\ \bibinfo {author}
  {\bibfnamefont {A.}~\bibnamefont {Kauch}},\ }\bibfield  {title} {\emph
  {\enquote {\bibinfo {title} {Two-particle calculations with quantics tensor
  trains: Solving the parquet equations},}\ }}\href {\doibase
  10.1103/PhysRevResearch.7.023087} {\bibfield  {journal} {\bibinfo  {journal}
  {Phys. Rev. Res.}\ }\textbf {\bibinfo {volume} {7}},\ \bibinfo {pages}
  {023087} (\bibinfo {year} {2025})}\BibitemShut {NoStop}%
\bibitem [{\citenamefont {Sordi}\ \emph {et~al.}(2011)\citenamefont {Sordi},
  \citenamefont {Haule},\ and\ \citenamefont {Tremblay}}]{Sordi_Tremblay_2011}%
  \BibitemOpen
  \bibfield  {author} {\bibinfo {author} {\bibfnamefont {G.}~\bibnamefont
  {Sordi}}, \bibinfo {author} {\bibfnamefont {K.}~\bibnamefont {Haule}}, \ and\
  \bibinfo {author} {\bibfnamefont {A.-M.~S.}\ \bibnamefont {Tremblay}},\
  }\bibfield  {title} {\emph {\enquote {\bibinfo {title} {Mott physics and
  first-order transition between two metals in the normal-state phase diagram
  of the two-dimensional Hubbard model},}\ }}\href {\doibase
  10.1103/PhysRevB.84.075161} {\bibfield  {journal} {\bibinfo  {journal} {Phys.
  Rev. B}\ }\textbf {\bibinfo {volume} {84}},\ \bibinfo {pages} {075161}
  (\bibinfo {year} {2011})}\BibitemShut {NoStop}%
\bibitem [{\citenamefont {Werner}\ and\ \citenamefont
  {Millis}(2007)}]{Werner_Millis_2007}%
  \BibitemOpen
  \bibfield  {author} {\bibinfo {author} {\bibfnamefont {P.}~\bibnamefont
  {Werner}}\ and\ \bibinfo {author} {\bibfnamefont {A.~J.}\ \bibnamefont
  {Millis}},\ }\bibfield  {title} {\emph {\enquote {\bibinfo {title}
  {Doping-driven Mott transition in the one-band Hubbard model},}\ }}\href
  {\doibase 10.1103/PhysRevB.75.085108} {\bibfield  {journal} {\bibinfo
  {journal} {Phys. Rev. B}\ }\textbf {\bibinfo {volume} {75}},\ \bibinfo
  {pages} {085108} (\bibinfo {year} {2007})}\BibitemShut {NoStop}%
\bibitem [{\citenamefont {Liebsch}(2008)}]{Liebsch_2008}%
  \BibitemOpen
  \bibfield  {author} {\bibinfo {author} {\bibfnamefont {A.}~\bibnamefont
  {Liebsch}},\ }\bibfield  {title} {\emph {\enquote {\bibinfo {title}
  {Doping-driven Mott transition in
  ${\mathrm{La}}_{1\ensuremath{-}x}{\mathrm{Sr}}_{x}\mathrm{Ti}{\mathrm{O}}_{3}$
  via simultaneous electron and hole doping of ${t}_{2g}$ subbands as predicted
  by $\mathrm{LDA}+\mathrm{DMFT}$ calculations},}\ }}\href {\doibase
  10.1103/PhysRevB.77.115115} {\bibfield  {journal} {\bibinfo  {journal} {Phys.
  Rev. B}\ }\textbf {\bibinfo {volume} {77}},\ \bibinfo {pages} {115115}
  (\bibinfo {year} {2008})}\BibitemShut {NoStop}%
\bibitem [{\citenamefont {Steinbauer}\ \emph {et~al.}(2019)\citenamefont
  {Steinbauer}, \citenamefont {de' Medici},\ and\ \citenamefont
  {Biermann}}]{Steinbauer_Biermann_2019}%
  \BibitemOpen
  \bibfield  {author} {\bibinfo {author} {\bibfnamefont {J.}~\bibnamefont
  {Steinbauer}}, \bibinfo {author} {\bibfnamefont {L.}~\bibnamefont {de'
  Medici}}, \ and\ \bibinfo {author} {\bibfnamefont {S.}~\bibnamefont
  {Biermann}},\ }\bibfield  {title} {\emph {\enquote {\bibinfo {title}
  {Doping-driven metal-insulator transition in correlated electron systems with
  strong Hund's exchange coupling},}\ }}\href {\doibase
  10.1103/PhysRevB.100.085104} {\bibfield  {journal} {\bibinfo  {journal}
  {Phys. Rev. B}\ }\textbf {\bibinfo {volume} {100}},\ \bibinfo {pages}
  {085104} (\bibinfo {year} {2019})}\BibitemShut {NoStop}%
\bibitem [{\citenamefont {Mermin}\ and\ \citenamefont
  {Wagner}(1966)}]{Mermin1966}%
  \BibitemOpen
  \bibfield  {author} {\bibinfo {author} {\bibfnamefont {N.}~\bibnamefont
  {Mermin}}\ and\ \bibinfo {author} {\bibfnamefont {H.}~\bibnamefont
  {Wagner}},\ }\bibfield  {title} {\emph {\enquote {\bibinfo {title} {Absence
  of Ferromagnetism or Antiferromagnetism in One- or Two-Dimensional Isotropic
  Heisenberg Models},}\ }}\href@noop {} {\bibfield  {journal} {\bibinfo
  {journal} {Phys. Rev. Lett.}\ }\textbf {\bibinfo {volume} {17}},\ \bibinfo
  {pages} {1133} (\bibinfo {year} {1966})}\BibitemShut {NoStop}%
\bibitem [{\citenamefont {Huang}\ \emph {et~al.}(2003)\citenamefont {Huang},
  \citenamefont {Hanke}, \citenamefont {Arrigoni},\ and\ \citenamefont
  {Scalapino}}]{Huang_2003}%
  \BibitemOpen
  \bibfield  {author} {\bibinfo {author} {\bibfnamefont {Z.~B.}\ \bibnamefont
  {Huang}}, \bibinfo {author} {\bibfnamefont {W.}~\bibnamefont {Hanke}},
  \bibinfo {author} {\bibfnamefont {E.}~\bibnamefont {Arrigoni}}, \ and\
  \bibinfo {author} {\bibfnamefont {D.~J.}\ \bibnamefont {Scalapino}},\
  }\bibfield  {title} {\emph {\enquote {\bibinfo {title} {Electron-phonon
  vertex in the two-dimensional one-band Hubbard model},}\ }}\href {\doibase
  10.1103/PhysRevB.68.220507} {\bibfield  {journal} {\bibinfo  {journal} {Phys.
  Rev. B}\ }\textbf {\bibinfo {volume} {68}},\ \bibinfo {pages} {220507}
  (\bibinfo {year} {2003})}\BibitemShut {NoStop}%
\bibitem [{\citenamefont {Coulter}\ and\ \citenamefont
  {Millis}(2025)}]{Coulter_2025}%
  \BibitemOpen
  \bibfield  {author} {\bibinfo {author} {\bibfnamefont {J.}~\bibnamefont
  {Coulter}}\ and\ \bibinfo {author} {\bibfnamefont {A.~J.}\ \bibnamefont
  {Millis}},\ }\href {https://arxiv.org/abs/2505.08081} {\enquote {\bibinfo
  {title} {Electron-phonon coupling in correlated materials: insights from the
  Hubbard-Holstein model},}\ } (\bibinfo {year} {2025}),\ \Eprint
  {http://arxiv.org/abs/2505.08081}{arXiv:2505.08081
  [cond-mat.str-el]}\BibitemShut {NoStop}%
\bibitem [{\citenamefont {Moghadas}\ \emph {et~al.}(2025)\citenamefont
  {Moghadas}, \citenamefont {Reitner}, \citenamefont {Wehling}, \citenamefont
  {Sangiovanni}, \citenamefont {Ciuchi},\ and\ \citenamefont
  {Toschi}}]{Moghadas_2025}%
  \BibitemOpen
  \bibfield  {author} {\bibinfo {author} {\bibfnamefont {E.}~\bibnamefont
  {Moghadas}}, \bibinfo {author} {\bibfnamefont {M.}~\bibnamefont {Reitner}},
  \bibinfo {author} {\bibfnamefont {T.}~\bibnamefont {Wehling}}, \bibinfo
  {author} {\bibfnamefont {G.}~\bibnamefont {Sangiovanni}}, \bibinfo {author}
  {\bibfnamefont {S.}~\bibnamefont {Ciuchi}}, \ and\ \bibinfo {author}
  {\bibfnamefont {A.}~\bibnamefont {Toschi}},\ }\href
  {https://arxiv.org/abs/2503.12113} {\enquote {\bibinfo {title} {Effective
  enhancement of the electron-phonon coupling driven by nonperturbative
  electronic density fluctuations},}\ } (\bibinfo {year} {2025}),\ \Eprint
  {http://arxiv.org/abs/2503.12113}{arXiv:2503.12113
  [cond-mat.str-el]}\BibitemShut {NoStop}%
\bibitem [{\citenamefont {Ornstein}\ and\ \citenamefont
  {Zernike}(1914)}]{Ornstein1914}%
  \BibitemOpen
  \bibfield  {author} {\bibinfo {author} {\bibfnamefont {L.~S.}\ \bibnamefont
  {Ornstein}}\ and\ \bibinfo {author} {\bibfnamefont {F.}~\bibnamefont
  {Zernike}},\ }\bibfield  {title} {\emph {\enquote {\bibinfo {title} {Integral
  equation in liquid state theory},}\ }}\href
  {https://dwc.knaw.nl/DL/publications/PU00012727.pdf} {\bibfield  {journal}
  {\bibinfo  {journal} {Proc. K. Ned. Akad. Wet.}\ }\textbf {\bibinfo {volume}
  {17}},\ \bibinfo {pages} {793} (\bibinfo {year} {1914})}\BibitemShut
  {NoStop}%
\bibitem [{\citenamefont {Rojo}\ \emph {et~al.}(1993)\citenamefont {Rojo},
  \citenamefont {Kotliar},\ and\ \citenamefont {Canright}}]{Rojo_1993}%
  \BibitemOpen
  \bibfield  {author} {\bibinfo {author} {\bibfnamefont {A.~G.}\ \bibnamefont
  {Rojo}}, \bibinfo {author} {\bibfnamefont {G.}~\bibnamefont {Kotliar}}, \
  and\ \bibinfo {author} {\bibfnamefont {G.~S.}\ \bibnamefont {Canright}},\
  }\bibfield  {title} {\emph {\enquote {\bibinfo {title} {Sign of equilibrium
  Hall conductivity in strongly correlated systems},}\ }}\href {\doibase
  10.1103/PhysRevB.47.9140} {\bibfield  {journal} {\bibinfo  {journal} {Phys.
  Rev. B}\ }\textbf {\bibinfo {volume} {47}},\ \bibinfo {pages} {9140}
  (\bibinfo {year} {1993})}\BibitemShut {NoStop}%
\bibitem [{\citenamefont {Markov}\ \emph {et~al.}(2019)\citenamefont {Markov},
  \citenamefont {Rohringer},\ and\ \citenamefont {Rubtsov}}]{Markov_2019}%
  \BibitemOpen
  \bibfield  {author} {\bibinfo {author} {\bibfnamefont {A.~A.}\ \bibnamefont
  {Markov}}, \bibinfo {author} {\bibfnamefont {G.}~\bibnamefont {Rohringer}}, \
  and\ \bibinfo {author} {\bibfnamefont {A.~N.}\ \bibnamefont {Rubtsov}},\
  }\bibfield  {title} {\emph {\enquote {\bibinfo {title} {Robustness of the
  topological quantization of the Hall conductivity for correlated lattice
  electrons at finite temperatures},}\ }}\href {\doibase
  10.1103/PhysRevB.100.115102} {\bibfield  {journal} {\bibinfo  {journal}
  {Phys. Rev. B}\ }\textbf {\bibinfo {volume} {100}},\ \bibinfo {pages}
  {115102} (\bibinfo {year} {2019})}\BibitemShut {NoStop}%
\bibitem [{\citenamefont {S\'en\'echal}\ and\ \citenamefont
  {Tremblay}(2004)}]{Senechal_Tremblay_2004}%
  \BibitemOpen
  \bibfield  {author} {\bibinfo {author} {\bibfnamefont {D.}~\bibnamefont
  {S\'en\'echal}}\ and\ \bibinfo {author} {\bibfnamefont {A.-M.~S.}\
  \bibnamefont {Tremblay}},\ }\bibfield  {title} {\emph {\enquote {\bibinfo
  {title} {Hot Spots and Pseudogaps for Hole- and Electron-Doped
  High-Temperature Superconductors},}\ }}\href {\doibase
  10.1103/PhysRevLett.92.126401} {\bibfield  {journal} {\bibinfo  {journal}
  {Phys. Rev. Lett.}\ }\textbf {\bibinfo {volume} {92}},\ \bibinfo {pages}
  {126401} (\bibinfo {year} {2004})}\BibitemShut {NoStop}%
\bibitem [{\citenamefont {Worm}\ \emph {et~al.}(2024)\citenamefont {Worm},
  \citenamefont {Reitner}, \citenamefont {Held},\ and\ \citenamefont
  {Toschi}}]{Worm_Toschi_2024}%
  \BibitemOpen
  \bibfield  {author} {\bibinfo {author} {\bibfnamefont {P.}~\bibnamefont
  {Worm}}, \bibinfo {author} {\bibfnamefont {M.}~\bibnamefont {Reitner}},
  \bibinfo {author} {\bibfnamefont {K.}~\bibnamefont {Held}}, \ and\ \bibinfo
  {author} {\bibfnamefont {A.}~\bibnamefont {Toschi}},\ }\bibfield  {title}
  {\emph {\enquote {\bibinfo {title} {Fermi and Luttinger Arcs: Two Concepts,
  Realized on One Surface},}\ }}\href {\doibase 10.1103/PhysRevLett.133.166501}
  {\bibfield  {journal} {\bibinfo  {journal} {Phys. Rev. Lett.}\ }\textbf
  {\bibinfo {volume} {133}},\ \bibinfo {pages} {166501} (\bibinfo {year}
  {2024})}\BibitemShut {NoStop}%
\bibitem [{\citenamefont {Stobbe}\ \emph {et~al.}(2025)\citenamefont {Stobbe},
  \citenamefont {Leusch}, \citenamefont {Titvinidze},\ and\ \citenamefont
  {Rohringer}}]{Stobbe_2025prep}%
  \BibitemOpen
  \bibfield  {author} {\bibinfo {author} {\bibfnamefont {J.}~\bibnamefont
  {Stobbe}}, \bibinfo {author} {\bibfnamefont {M.}~\bibnamefont {Leusch}},
  \bibinfo {author} {\bibfnamefont {I.}~\bibnamefont {Titvinidze}}, \ and\
  \bibinfo {author} {\bibfnamefont {G.}~\bibnamefont {Rohringer}},\ }\href@noop
  {} {\enquote {\bibinfo {title} {Tail correction of the self-energy},}\ }
  (\bibinfo {year} {2025}),\ \bibinfo {note} {in preparation}\BibitemShut
  {NoStop}%
\bibitem [{\citenamefont {van Loon}\ \emph {et~al.}(2016)\citenamefont {van
  Loon}, \citenamefont {Krien}, \citenamefont {Hafermann}, \citenamefont
  {Stepanov}, \citenamefont {Lichtenstein},\ and\ \citenamefont
  {Katsnelson}}]{vanLoon_2016}%
  \BibitemOpen
  \bibfield  {author} {\bibinfo {author} {\bibfnamefont {E.~G. C.~P.}\
  \bibnamefont {van Loon}}, \bibinfo {author} {\bibfnamefont {F.}~\bibnamefont
  {Krien}}, \bibinfo {author} {\bibfnamefont {H.}~\bibnamefont {Hafermann}},
  \bibinfo {author} {\bibfnamefont {E.~A.}\ \bibnamefont {Stepanov}}, \bibinfo
  {author} {\bibfnamefont {A.~I.}\ \bibnamefont {Lichtenstein}}, \ and\
  \bibinfo {author} {\bibfnamefont {M.~I.}\ \bibnamefont {Katsnelson}},\
  }\bibfield  {title} {\emph {\enquote {\bibinfo {title} {Double occupancy in
  dynamical mean-field theory and the dual boson approach},}\ }}\href {\doibase
  10.1103/PhysRevB.93.155162} {\bibfield  {journal} {\bibinfo  {journal} {Phys.
  Rev. B}\ }\textbf {\bibinfo {volume} {93}},\ \bibinfo {pages} {155162}
  (\bibinfo {year} {2016})}\BibitemShut {NoStop}%
\bibitem [{\citenamefont {{J.
  Stobbe}}(2025{\natexlab{a}})}]{VertexPostprocessing}%
  \BibitemOpen
  \bibfield  {author} {\bibinfo {author} {\bibnamefont {{J. Stobbe}}},\
  }\href@noop {} {\enquote {\bibinfo {title} {Vertex Postprocessing},}\
  }\bibinfo {howpublished}
  {\url{https://github.com/Atomtomate/VertexPostprocessing.jl}} (\bibinfo
  {year} {2025}{\natexlab{a}}),\ \bibinfo {note} {commit:
  f12318160}\BibitemShut {NoStop}%
\bibitem [{\citenamefont {{J. Stobbe}}(2025{\natexlab{b}})}]{jED}%
  \BibitemOpen
  \bibfield  {author} {\bibinfo {author} {\bibnamefont {{J. Stobbe}}},\
  }\href@noop {} {\enquote {\bibinfo {title} {Exact Diagonalization},}\
  }\bibinfo {howpublished} {\url{https://github.com/Atomtomate/jED.jl}}
  (\bibinfo {year} {2025}{\natexlab{b}}),\ \bibinfo {note} {commit:
  4d61874c}\BibitemShut {NoStop}%
\bibitem [{\citenamefont {{J.
  Stobbe}}(2025{\natexlab{c}})}]{EquivalenceClassesConstructor}%
  \BibitemOpen
  \bibfield  {author} {\bibinfo {author} {\bibnamefont {{J. Stobbe}}},\
  }\href@noop {} {\enquote {\bibinfo {title} {Equivalence Classes
  Constructor},}\ }\bibinfo {howpublished} {\url{
  https://github.com/Atomtomate/EquivalenceClassesConstructor.jl}} (\bibinfo
  {year} {2025}{\natexlab{c}}),\ \bibinfo {note} {commit: 33aa5216}\BibitemShut
  {NoStop}%
\bibitem [{\citenamefont {{J. Stobbe}}(2025{\natexlab{d}})}]{LadderDGA}%
  \BibitemOpen
  \bibfield  {author} {\bibinfo {author} {\bibnamefont {{J. Stobbe}}},\
  }\href@noop {} {\enquote {\bibinfo {title} {lD$\Gamma$A},}\ }\bibinfo
  {howpublished} {\url{https://github.com/Atomtomate/LadderDGA.jl.}} (\bibinfo
  {year} {2025}{\natexlab{d}}),\ \bibinfo {note} {commit: d6c2854c}\BibitemShut
  {NoStop}%
\bibitem [{\citenamefont {{J. Stobbe and M. Leusch}}(2025)}]{Dispersions}%
  \BibitemOpen
  \bibfield  {author} {\bibinfo {author} {\bibnamefont {{J. Stobbe and M.
  Leusch}}},\ }\href@noop {} {\enquote {\bibinfo {title} {Dispersion},}\
  }\bibinfo {howpublished} {\url{
  https://github.com/Atomtomate/Dispersions.jl}} (\bibinfo {year} {2025}),\
  \bibinfo {note} {commit: 3be4597a}\BibitemShut {NoStop}%
\bibitem [{\citenamefont {{J. Stobbe}}(2025{\natexlab{e}})}]{BSE_Asymptotics}%
  \BibitemOpen
  \bibfield  {author} {\bibinfo {author} {\bibnamefont {{J. Stobbe}}},\
  }\href@noop {} {\enquote {\bibinfo {title} {BSE Asymptotics},}\ }\bibinfo
  {howpublished} {\url{ https://github.com/Atomtomate/BSE_Asymptotics.jl}}
  (\bibinfo {year} {2025}{\natexlab{e}}),\ \bibinfo {note} {commit:
  0d342c48}\BibitemShut {NoStop}%
\bibitem [{\citenamefont {{I.Titvinidze, J. Stobbe, M. Leusch, and G.
  Rohringer}}(2025)}]{Our_Data}%
  \BibitemOpen
  \bibfield  {author} {\bibinfo {author} {\bibnamefont {{I.Titvinidze, J.
  Stobbe, M. Leusch, and G. Rohringer}}},\ }\href@noop {} {\enquote {\bibinfo
  {title} {Our Data},}\ }\bibinfo {howpublished} {\url{
  https://www.fdr.uni-hamburg.de/record/17952}} (\bibinfo {year}
  {2025})\BibitemShut {NoStop}%
\end{thebibliography}
%

\end{document}